\documentclass[preprint,12pt,authoryear,nopreprintline]{elsarticle}

\usepackage{amssymb}
\usepackage{amsmath}

\usepackage{graphicx}
\usepackage{cuted}
\usepackage{booktabs, siunitx, multirow, longtable}
\usepackage[breaklinks,colorlinks]{hyperref}
\usepackage[color=gray!20,textsize=tiny]{todonotes}
\usepackage{subcaption}
\usepackage[most]{tcolorbox}
\usepackage{tabularx}
\usepackage[labelfont=bf, labelsep=period]{caption}
\usepackage[utf8]{inputenc}
\usepackage[top=2.5cm, bottom=2.5cm, left=2.5cm, right=2.5cm]{geometry}
\usepackage{titlesec}
\titleformat{\section}
  {\normalfont\large\bfseries}
  {\thesection}
  {0.5em}
  {}

\titleformat{\subsection}
  {\normalfont\normalsize\bfseries}
  {\thesubsection}{0.5em}{}

\titleformat{\subsubsection}
  {\normalfont\normalsize\bfseries}
  {\thesubsubsection}{0.5em}{}

\titleformat{\paragraph}[runin]
  {\normalfont\normalsize\bfseries}
  {\theparagraph}{0.5em}{}[. ]

\titlespacing*{\paragraph}
  {0pt}
  {1ex plus .1ex minus .2ex}
  {1em}

\usepackage{xspace}
\newcommand{\myTitle}{A Vision-Language Framework for Measuring Social Life on Sidewalks}

\usepackage[most]{tcolorbox}
\tcbuselibrary{breakable}
\usepackage{fvextra}

\newcounter{promptbox}

\begin{document}

\begin{frontmatter}

\title{\myTitle}

\author[mit,cfl]{Liu Liu}
\ead{lyons66@mit.edu}

\author[mit,cfl]{Andres Sevtsuk}
\ead{sevtsuk@mit.edu}

\affiliation[mit]{
  organization={Massachusetts Institute of Technology},
  addressline={77 Massachusetts Avenue},
  city={Cambridge},
  postcode={02139},
  state={MA},
  country={USA}
}

\affiliation[cfl]{
  organization={City Form Lab}
}

\begin{abstract}
While a number of methods exist for counting pedestrians in street-view imagery, these mostly ignore the social dimensions of pedestrian activity. A street traversed by a high volume of pedestrians has the same headcount as a street where people linger, sit, and socialize. This paper presents a vision-language framework for extracting social indicators from street-level imagery. Panoramic street-level imagery is reprojected to sidewalk-facing sideviews with preserved timestamps. A vision-language model (VLM)-based activity detection system codes each person across ten independent observable dimensions, resolving a systematic failure mode in which models prompted with high-level social categories conflate observable states with contextual inferences. The resulting social indicator system produces a Social Dwelling Index (SDI) that jointly considers pedestrian grouping and dwelling, provides activity labels documenting behavioral diversity, and issues binary flags for the presence of accessibility-sensitive populations. We apply the framework to 102,514 sideviews in New York City, revealing that pedestrian volume and SDI are only weakly associated ($r = 0.168$): streets with the highest foot traffic are not where social activity is most intense. The framework provides a scalable method for measuring not only how many people are on city sidewalks, but also their grouping, posture, and activity type, summarizing the non-transient activities that occur on city sidewalks.
\end{abstract}




\end{frontmatter}


\section{Introduction} \label{sec:intro}

The pedestrian realm serves as the stage for urban social life. Sidewalks, plazas, parks, and public spaces are not merely conduits for movement; they are places where people encounter one another and where everyday public life takes shape \citep{Sennett2024}. Observation of human behavior in public space has long occupied urban planning, design, and sociology \citep{Whyte1943, Park1925, Jacobs1961, Gehl1987LifeSpace, Leyden2003, Talen2008, Salazar-Miranda2025}. As Park wrote in 1925, ``the clairvoyant, the vaudeville performer, the quack doctor, the bartender, the ward boss, the strike-breaker, the labor agitator, the school teacher, the reporter, the stockbroker, the pawnbroker; all of these are characteristic products of the conditions of city life'' \citep{Park1925}.

However, research on social behavior in public space has persistently faced two problems. The first is scale. The in-person observational method that produces the richest account of social life is labor-intensive to apply at scale. The second is measurement. When researchers have sought to study pedestrian activities at scale, they have primarily relied on pedestrian counts, which do not distinguish between sitting, standing, or walking and cannot tell apart groups of people who happen to be near each other from those who are actually interacting.

Recent advances in computer vision and analysis of street-view imagery have addressed the scale problem. Yet in these studies, pedestrians are typically counted, but their behavior is not distinguished \citep{liu2024clarity}. This paper addresses this gap. We present a computer-vision framework for extracting social indicators from street-level imagery, not simply counting pedestrians but observing whether they are alone or together, moving or lingering, sitting, waiting, talking, or carrying things. These distinctions have not been measured at scale previously.

This paper presents a vision-language framework for measuring sidewalk social life from street-level imagery. Operating on timestamped sidewalk-facing sideviews extracted from panoramic street-level imagery (Appendix~\ref{sec:data-processing}), the framework combines revised social group detection, per-person VLM query, and composite indicator construction to measure grouping, dwelling, activity, and spatial context at the city scale.

\begin{figure}[htbp]
\centering
\begin{minipage}{0.95\textwidth}
    \includegraphics[width=\linewidth]{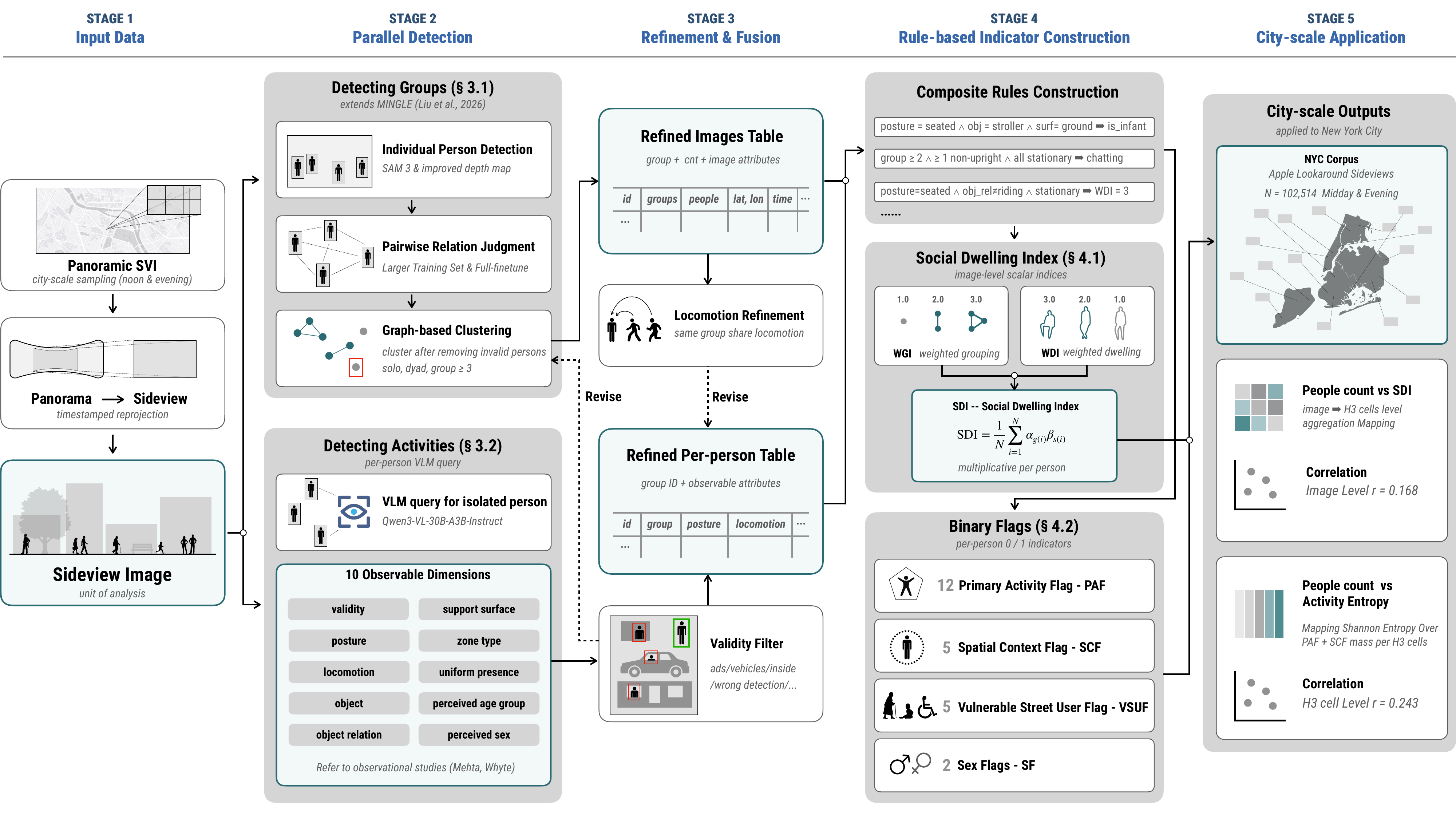}
    \caption{Overview of the framework for measuring sidewalk social life from street-level imagery, organized across five stages. \textbf{Stage 1}: Panoramic imagery is sampled at the city scale and reprojected into timestamped, sidewalk-facing sideview images, the unit of analysis. \textbf{Stage 2}: Each sideview passes through two parallel detection branches. Group detection (Section~\ref{ssec:social-group-detection}, extending MINGLE; \citealt{liu2026mingle}) applies person detection (SAM3 with improved depth estimation) and pairwise social-relation judgment (Qwen2.5-VL-3B, full fine-tuning on an expanded training set). Activity detection (Section~\ref{ssec:activity-detection}) queries a VLM (Qwen3-VL-30B-A3B-Instruct) for each isolated person across ten independent observable dimensions: validity, posture, locomotion, object, object relation, support surface, zone type, uniform presence, perceived age group, and perceived sex. \textbf{Stage 3}: A validity filter removes spurious detections (figures on advertisements, vehicle occupants, indoor figures); graph-based clustering is re-run on the filtered person set; and locomotion states are propagated uniformly within each social group, producing a refined images table and a refined per-person attribute table. Dashed arrows indicate feedback from refinement to both detection branches. \textbf{Stage 4}: A composite rule layer (Section~\ref{sec:social-indicators}) derives two complementary indicator families: the Social Dwelling Index (SDI), combining the Weighted Grouping Index (WGI) and Weighted Dwelling Index (WDI) multiplicatively at the individual level, and the Binary Flag system comprising 12 Primary Activity Flags (PAF), 5 Spatial Context Flags (SCF), 5 Vulnerable Street User Flags (VSUF), and 2 Sex Flags (SF). \textbf{Stage 5}: The framework is applied to 102,514 Apple Lookaround sideviews in New York City (midday and evening peak hours), demonstrating the decoupling between pedestrian volume and SDI ($r = 0.168$ at the image level; $r = 0.243$ between pedestrian volume and activity entropy at the H3 cell level) and supporting city-scale activity-entropy mapping.}
    \label{fig:framework}
\end{minipage}
\end{figure}

The framework makes three methodological contributions. First, building on MINGLE \citep{liu2026mingle}, we develop an integrated sidewalk social detection framework that combines revised social group detection with per-person observable query using vision-language models (VLMs). Second, we identify and resolve a fundamental problem in applying VLMs to urban behavioral observation: models prompted with high-level social categories systematically conflate observable physical states with contextual inferences. To address this issue, we redesign activity detection around ten independent observable dimensions grounded in urban observational studies, substantially improving detection reliability. Third, we construct a composite social indicator system that integrates grouping structure and per-person observations into image-level measurements of sidewalk social life, including the Social Dwelling Index (SDI) and binary indicators describing primary activities, spatial context, vulnerable street users, and perceived sex. Figure~\ref{fig:framework} provides an overview of the complete framework.

We demonstrate the framework on 102,514 sideviews in New York City, showing that pedestrian volume and sidewalk social activity are only weakly associated. Streets with the highest pedestrian counts are not necessarily where people most often gather or linger.

\section{Related Work} \label{sec:background}

\subsection{Observational Studies of Urban Public Space} \label{ssec:urban-behavior}

The study of human behavior in public space has a long empirical tradition in urban design research. Whyte \citeyearpar{whyte1980social} spent more than a decade recording how people sat, congregated, and lingered in New York City plazas, producing findings that influenced zoning regulations. Gehl \citeyearpar{Gehl1987LifeSpace, gehl_cities_2010} documented how small design decisions affect whether people linger or pass through, distinguishing necessary, optional, and social activities as the key categories for evaluating built environments. Jacobs \citeyearpar{Jacobs1961} argued that sidewalks function as social infrastructure, and that the temporal rhythm and mix of their uses determine whether street life is sustained or eroded. Appleyard \citeyearpar{appleyard_livable_1981} compared streets with different traffic volumes over three years, showing how vehicle traffic displaces social activity. Gehl and Svarre \citeyearpar{gehl_how_2013} systematized these approaches into a set of observational protocols, including counting, mapping, tracing, and behavioral coding, that have since been applied in cities worldwide. Mehta \citeyearpar{mehta_lively_2007} extended this tradition to street-level behavioral taxonomy, identifying environmental characteristics that support social behavior. Hampton et al. \citeyearpar{hampton_change_2015} revisited Whyte's original sites thirty years later using the same methods, finding that public spaces had become more socially diverse.

These empirical studies draw on sociological frameworks for understanding public interaction. Goffman \citeyearpar{goffman1963behavior} distinguished focused from unfocused interaction: the former involves mutual engagement between co-present individuals; the latter governs how strangers share space without direct engagement. Hall \citeyearpar{hall1966hidden} documented how interpersonal distance structures public behavior. Lofland \citeyearpar{lofland1998public} classified public sociability into fleeting, routinized, and enduring forms, each associated with distinct physical arrangements.

The ethnographic and sociological depth of this research has been constrained by the scale of its application. Whyte's research required more than a decade of fieldwork at a small number of NYC sites; Gehl's city-wide studies required large teams of trained observers. The methods are rich in behavioral detail, but necessarily local and labor-intensive. Examining social behavior across hundreds of streets requires a different approach.

\subsection{Street-Level Imagery in Urban Research} \label{ssec:streetview}

The growing availability of street-level imagery from commercial platforms such as Google Street View, Apple Lookaround, Bing Streetside, Baidu Maps, Yandex Maps, and Kakao Road View, as well as open platforms such as Mapillary and KartaView, has opened new possibilities for computational urban analysis. A systematic review of over 300 CV-based urban studies found that more than 80\% focus on simple scene-level metrics with limited connection to human behavior \citep{liu2024clarity}. The most common outputs include the Green View Index, the Sky View Factor, and building facade classifications \citep{long2017green, gong_mapping_2018, ki_analyzing_2021, tang_measuring_2019, ye_visual_2019, zhang_systematic_2019}. A parallel strand of work has used street-level imagery to predict human perceptions of safety, beauty, and liveliness, training models on crowdsourced comparison data \citep{zhang2018measuring}. Zhang et al. \citeyearpar{zhang_urban_2024} provide a broad synthesis of how artificial intelligence and street-level imagery are now used across urban research, covering semantic segmentation, attribute measurement, and visual change detection.

In these studies, pedestrians are counted without distinguishing their activities. Studies estimate pedestrian volume from street-level scenes using person detectors \citep{chen_examining_2022, zhao_people_2023}, and others examine associations between streetscape features and walking behavior \citep{li_investigating_2018, koo_how_2022, basu_how_2022, goel_estimating_2018}. Neither line of work addresses whether people present in a scene are alone or together, moving or stationary, seated or standing. Those distinctions are central to the observational tradition established by Whyte, Gehl, and Jacobs, but have no scalable computational analog to date.

\subsection{Detecting Human Behavior with Computer Vision} \label{ssec:detection-vlm}

Computer vision research has developed several approaches for extracting behavioral information from images. Semantic and instance segmentation models, trained on urban scene benchmarks such as Cityscapes \citep{cordts2016cityscapes}, enable the classification of scene elements, including pedestrians, in street-level imagery. Pose estimation methods \citep{cao2019openpose} recover per-person body keypoints from single images, providing proxies for posture and locomotion without requiring video. Crowd counting approaches use CNN-based density maps to estimate the number of people in a scene \citep{zhang2016single}, while pedestrian attribute recognition models classify clothing, posture, and carried objects from cropped person images \citep{zheng2020pedestrian}. These capabilities have been applied to urban imagery for pedestrian flow estimation \citep{zhao_people_2023, chen_examining_2022} and transit demand inference \citep{abdelhalim_computer_2024}. What they share is that they treat behavioral information as a byproduct of detection: individuals are located, counted, or attributed, but not observed in relation to one another or to their spatial context.

Social group detection requires procedures beyond individual detection: inferring whether co-located people are engaged in active or passive interaction and constitute a social unit. This problem has been approached through trajectory clustering in video, pairwise relation estimation from static images, and, more recently, graph-based grouping over detected individuals. Liu et al. \citeyearpar{liu2026mingle} introduced MINGLE, a three-stage pipeline that combines person detection, pairwise social judgment via a VLM, and graph-based clustering, specifically applied to urban street-level imagery data. The present paper builds on MINGLE and extends it to per-person activity coding.

Vision-language models have substantially expanded what can be recovered from a single image. Contrastive pretraining on image-text pairs \citep{radford2021learning} provided models with broad semantic grounding; subsequent instruction-tuning approaches \citep{liu_visual_2023} enabled structured question-answering over image content. Models such as GPT-4V \citep{openai2023gpt4} and V-IRL \citep{yang_v-irl_2024} have demonstrated that VLMs can parse complex scenes and reason about spatial relationships in real-world environments. Open-vocabulary detection architectures \citep{liu_grounding_2023} further allow querying for arbitrary categories without task-specific retraining. Despite these capabilities, no prior work has benchmarked VLMs on urban behavioral coding or identified the failure mode this paper addresses. When prompted with high-level social categories, VLMs conflate observable physical states with contextual inferences. Restricting outputs to directly observable attributes, specifically posture, locomotion, object in hand, support surface, and zone type, is the design principle that resolves this problem and makes the resulting framework reliable enough for social indicator construction.

\section{Detection Methods}\label{sec:detection-methods}

The framework operates on sidewalk-facing sideview images reprojected from timestamped panoramic street-level imagery; the extraction pipeline is described in Appendix~\ref{sec:data-processing}. This section describes the two detection components: a revised social group detection model (Section~\ref{ssec:social-group-detection}) and a VLM-based system for coding per-person observable attributes (Section~\ref{ssec:activity-detection}).

\subsection{Detecting Groups: a revised three-step pipeline}\label{ssec:social-group-detection}\label{sec:social-group-detection}

\paragraph{The Three-step Pipeline for Group Detection}
The group detection methodology in this study builds on the three-step pipeline proposed by \citet{liu2026mingle}, which was trained on 76K human-annotated street-view images in which annotators classified whether individuals belonged to the same social group. As shown in Figure~\ref{fig:framework} (Stage 2), the process begins with person detection within each sideview image. The model then evaluates pairwise social relationships by determining whether two detected individuals belong to the same social group. Pairwise social judgments are subsequently refined and integrated through graph-based clustering after cross-branch validity filtering, described in Section~\ref{ssec:cross-branch-refinement}.

\paragraph{Revision to Step 1: Person Detection and Depth Estimation}
Two components of the original Step~1 pipeline are revised. The ATSS-Swin-L-DyHead detector \citep{zhang2020bridging} used in MINGLE is replaced with SAM3 \citep{carion2025sam}, which improves recall for occluded and small-scale figures (Fig.~\ref{fig:refined-person-detection}). EVP depth estimation \citep{lavreniuk2024evp} is also replaced with PatchFusion \citep{li2024patchfusion}, providing higher-resolution and more geometrically continuous depth estimation for interpersonal distance measurement (Fig.~\ref{fig:depth-comparison}).

\paragraph{Revision to Step 2: Pairwise Social Judgment}
The original MINGLE pairwise model embedded bounding box coordinates directly into the prompt. We find that these coordinates introduce inconsistencies across image resolutions and cropping conventions and therefore remove them entirely. We further upgrade the base model from Qwen2-VL-7B to Qwen2.5-VL-7B and replace LoRA adaptation with full fine-tuning. Training proceeded through four iterative updates (Fig.~\ref{fig:updated-f1}), progressively incorporating improved depth maps and an expanded 76K training corpus. These revisions increased F1 from 72\% in the original MINGLE framework to 80.64\% while improving precision--recall balance. A final stage trained a 3B-parameter variant on the same corpus, achieving comparable performance at substantially lower inference cost; this model is used for all pairwise social interaction judgments in this study. Full training details are provided in Appendix~\ref{app:pipeline-revisions}.

\subsection{Detecting Activities: per-person VLM query}\label{ssec:activity-detection}

Per-person activity detection uses the same SAM3 detector described in Section~\ref{ssec:social-group-detection}, with ``person'' as the query prompt. Before querying the model, each detected person is isolated using an adaptive cropping procedure that incorporates proportional context padding (Fig.~\ref{fig:focus-crop}); the formal specification is provided in Appendix~\ref{sec:app-cropping}.

\paragraph{The observable versus interpretive problem}
When prompted with high-level social categories, VLMs systematically conflate directly observable physical states with contextual inferences. Testing an initial taxonomy derived from Mehta~(\citeyear{mehta2019streets}) with Qwen2.5-VL-72B-Instruct revealed a consistent failure mode: a large share of individuals were assigned the label ``waiting for public transport.'' Inspection of the model's intermediate reasoning revealed the cause: the prompt omitted \textit{walking} as a posture option, so the model approximated \textit{standing}, then reasoned forward to a plausible social situation. The action label was a rationalization of an earlier misclassification, not a description of what was observed. A second failure compounded this: although the prompt permitted multiple action labels, the model routinely returned only one, even when behavior clearly combined several. Both failures stem from a mismatch between urban observational vocabulary and VLM pretraining conventions: in corpora such as Kinetics-400 and AVA, \textit{standing} frequently subsumes \textit{walking}; the boundary is not systematically enforced.

\paragraph{Prompt redesign: a decomposed observational vocabulary}
Rather than pursuing fine-tuning, which would have required prohibitive annotation effort at this vocabulary breadth, we redesigned the prompt to operate within the model's existing semantic conventions. The redesigned prompt replaces the posture-action hierarchy with ten independent observational dimensions chosen to align urban observational distinctions with categories that VLMs can identify more reliably from directly visible evidence (Appendix~\ref{sec:app-act-prompt}): \textit{validity}, \textit{posture}, \textit{locomotion}, \textit{object}, \textit{object relation}, \textit{support surface}, \textit{zone type}, \textit{uniform presence}, \textit{perceived age group}, and \textit{perceived sex}. Each dimension is posed as a self-contained question with a constrained set of allowed values. The core design principle is the separation of observable physical state from contextual inference: a person walking while on the phone is encoded as \textit{locomotion: moving}, \textit{object: phone}, \textit{object\_relation: holding}, rather than as a single synthesized action label, thereby eliminating the cascading error above. Fields fall into three reliability tiers on a 999-image held-out benchmark: Tier~1 ($>$85\% accuracy) covers validity, posture, locomotion, zone type, uniform presence, and perceived age group; Tier~2 (reliable on $\geq$8B-parameter models) covers object, object relation, and support surface; Tier~3 (provisional) covers perceived sex, which is structurally constrained by face-blurring practices of US imagery providers and is excluded from all primary analyses. Based on a comparative evaluation across six candidate models (Appendix~\ref{sec:app-activity-model}), \textbf{Qwen3-VL-30B-A3B-Instruct} was selected as the primary coding engine, achieving a mean field accuracy of 88.5\%.

\subsection{Cross-branch Refinement and Fusion}\label{ssec:cross-branch-refinement}

To improve the detection performance, the group detection and activity detection branches are not independent. Outputs from the per-person VLM query are used to refine social group detection, while detected group structure is in turn used to refine activity interpretation. This refinement stage integrates the two branches into a unified observational framework prior to indicator construction.

\paragraph{Validity Filtering and Group Refinement}
The higher-recall SAM3 detector improves recovery of distant and partially occluded pedestrians, but also introduces additional false positives, including figures on advertisement displays, reflections, and vehicle occupants visible through windows. These cases are filtered using the \textit{validity} dimension produced by the per-person VLM query in Section~\ref{ssec:activity-detection}. Detections classified as \textit{advertisement\_or\_image}, \textit{inside\_vehicle}, or \textit{inside\_building} are removed prior to graph-based clustering. Pairwise social judgments are then integrated through the graph-based clustering procedure introduced in MINGLE, producing refined social group assignments from the validity-filtered person set.

\paragraph{Locomotion Harmonization}
Group structure is subsequently used to refine locomotion interpretation. Because members of the same co-present social group typically share a common movement state, a unified \texttt{locomotion\_effective} field is derived by propagating a consistent locomotion label across all members of the same detected group. This harmonized locomotion field replaces the raw \texttt{locomotion} output in all subsequent indicator construction.
\section{Social Measurement System}\label{sec:social-indicators}

The detection methods described in Section~\ref{sec:detection-methods} produce raw per-person outputs: group membership labels from Section~\ref{ssec:social-group-detection} and VLM-coded observable attributes from Section~\ref{ssec:activity-detection}. These outputs cannot be directly compared across images or locations. To characterize the social character of a street environment quantitatively, we define two complementary measurement systems. The Social Dwelling Index (Section~\ref{ssec:sdi}) aggregates grouping and dwelling signals into a single image-level score. The binary flag system (Section~\ref{ssec:binary-flags}) records the presence of specific population groups and behavioral types as independent indicators.

\subsection{Constructing the Social Dwelling Index}\label{ssec:sdi}

SDI integrates two component measures: the Weighted Grouping Index (WGI), derived from the group structure produced by Section~\ref{ssec:social-group-detection}, and the Weighted Dwelling Index (WDI), derived from the \texttt{posture} and \texttt{locomotion} attributes coded in Section~\ref{ssec:activity-detection}. Each captures a distinct dimension of social presence on the sidewalk; SDI combines them multiplicatively at the individual level.

\begin{figure}[htbp]
    \centering
    \begin{subfigure}[b]{0.32\textwidth}
        \includegraphics[width=\linewidth]{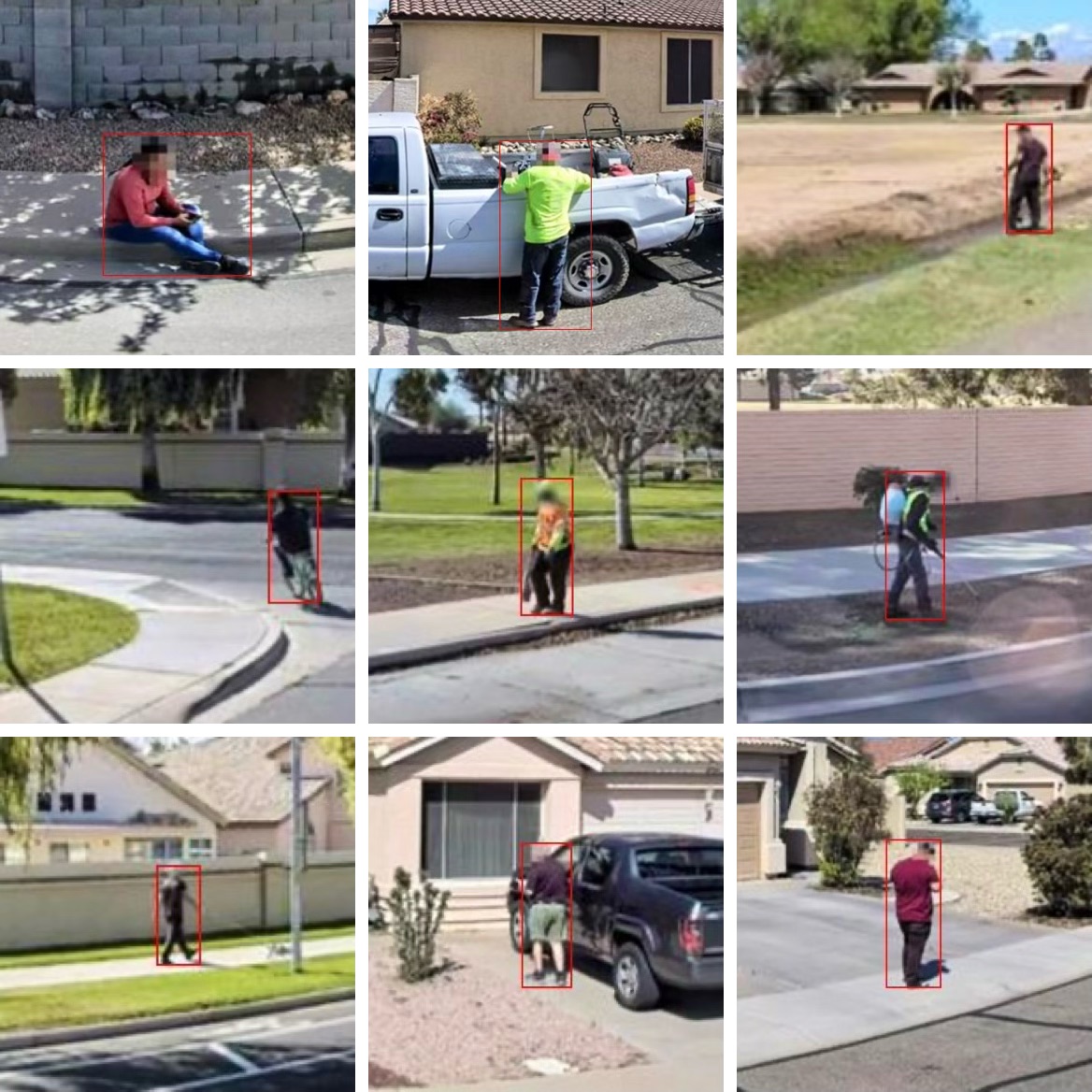}
        \caption{Solo pedestrian ($\alpha = 1$).}
        \label{fig:wgi-g1}
    \end{subfigure}
    \hfill
    \begin{subfigure}[b]{0.32\textwidth}
        \includegraphics[width=\linewidth]{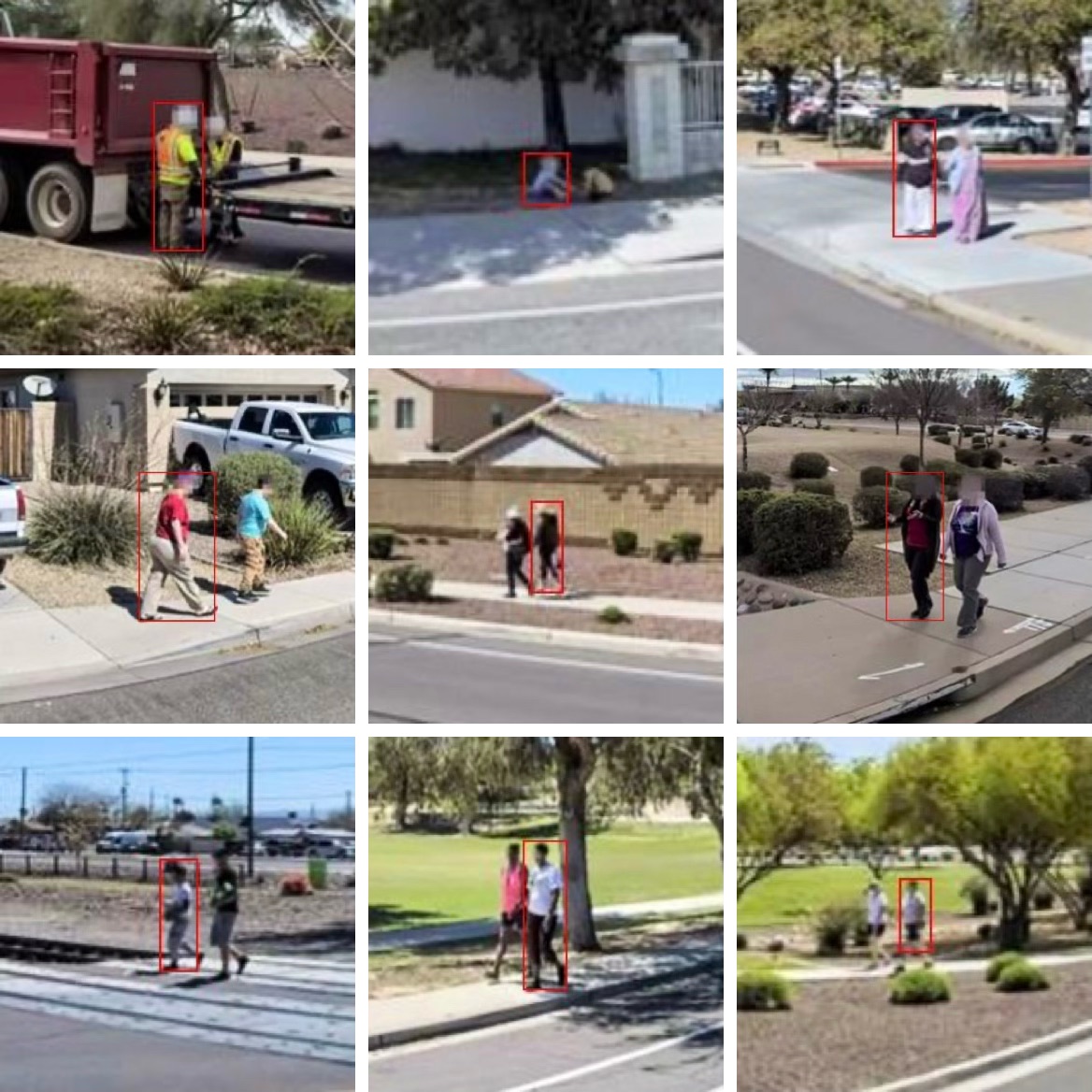}
        \caption{Person in Dyads ($\alpha = 2$).}
        \label{fig:wgi-g2}
    \end{subfigure}
    \hfill
    \begin{subfigure}[b]{0.32\textwidth}
        \includegraphics[width=\linewidth]{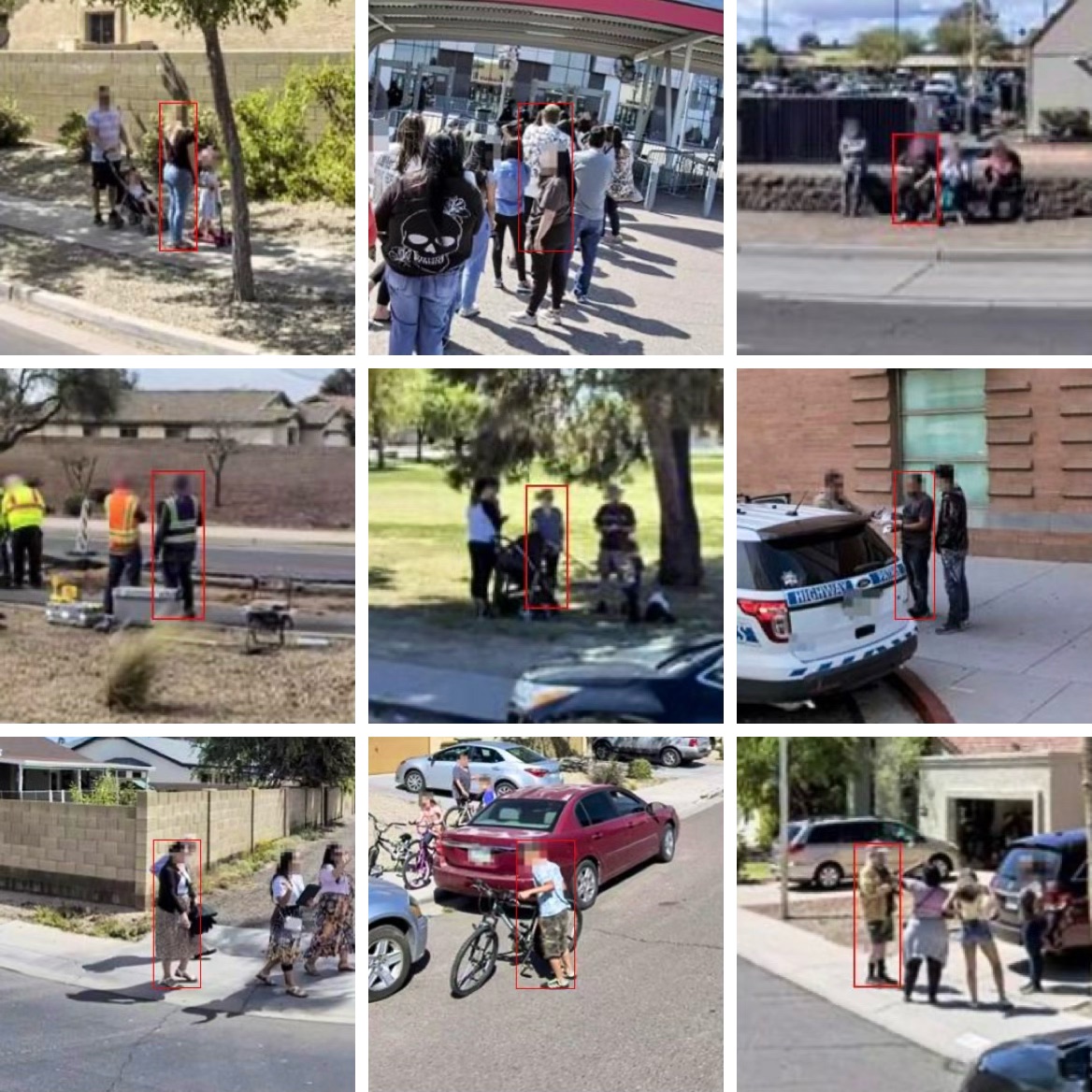}
        \caption{Person in Groups of 3+ ($\alpha = 3$).}
        \label{fig:wgi-g3}
    \end{subfigure}
    \caption{The three grouping tiers used in WGI weight assignment. Each detected person is assigned a weight $\alpha$ equal to the size of their social group: solo pedestrians receive $\alpha = 1$ (a), dyad members receive $\alpha = 2$ (b), and members of groups of three or more receive $\alpha = 3$ (c).}
    \label{fig:wgi-examples}
\end{figure}

\textbf{Weighted Grouping Index (WGI)} measures the average group-size weight across all detected people:
\begin{equation}
    \mathrm{WGI} = \frac{1}{N} \sum_{i=1}^{N} \alpha_{g(i)},
\end{equation}
where $\alpha_{g(i)} = 1$ for solo people (group size $= 1$), $\alpha_{g(i)} = 2$ for dyad members (group size $= 2$), and $\alpha_{g(i)} = 3$ for members of groups of three or more (see Table~\ref{tab:sdi-flags}). These weights should be understood as a simple ordinal representation of increasing levels of observed social grouping rather than empirically estimated parameters. The weighting scheme follows the intuition that larger co-present groups generally indicate higher levels of observed grouping than isolated pedestrians, while preserving a compact three-tier structure suitable for large-scale observational analysis. Figure~\ref{fig:wgi-examples} illustrates the three grouping tiers with street-level examples.

\textbf{Weighted Dwelling Index (WDI)} measures the average dwelling-intensity weight across all detected people:
\begin{equation}
    \mathrm{WDI} = \frac{1}{N} \sum_{i=1}^{N} \beta_{s(i)},
\end{equation}
where $\beta_{s(i)} = 3$ for people who are seated and not in a riding relation, $\beta_{s(i)} = 2$ for people whose locomotion is stationary but who are not seated, and $\beta_{s(i)} = 1$ otherwise. Similar to WGI, the integer ratio $3:2:1$ is intended as a simple ordinal weighting scheme rather than an empirically estimated parameter. It loosely follows the observational distinction between more sustained optional or social uses of public space and more transient movement through space \citep{Gehl1987LifeSpace}. Seated dwelling is therefore assigned the highest weight, stationary upright presence an intermediate weight, and active locomotion the baseline weight. Because the precise magnitudes of these weights are heuristic, we report sensitivity analyses under alternative specifications in Appendix~\ref{sec:app-sensitivity}. Figure~\ref{fig:wdi-examples} illustrates representative cases for each tier.

\begin{figure}[htbp]
    \centering
    \begin{subfigure}[b]{0.32\textwidth}
        \includegraphics[width=\linewidth]{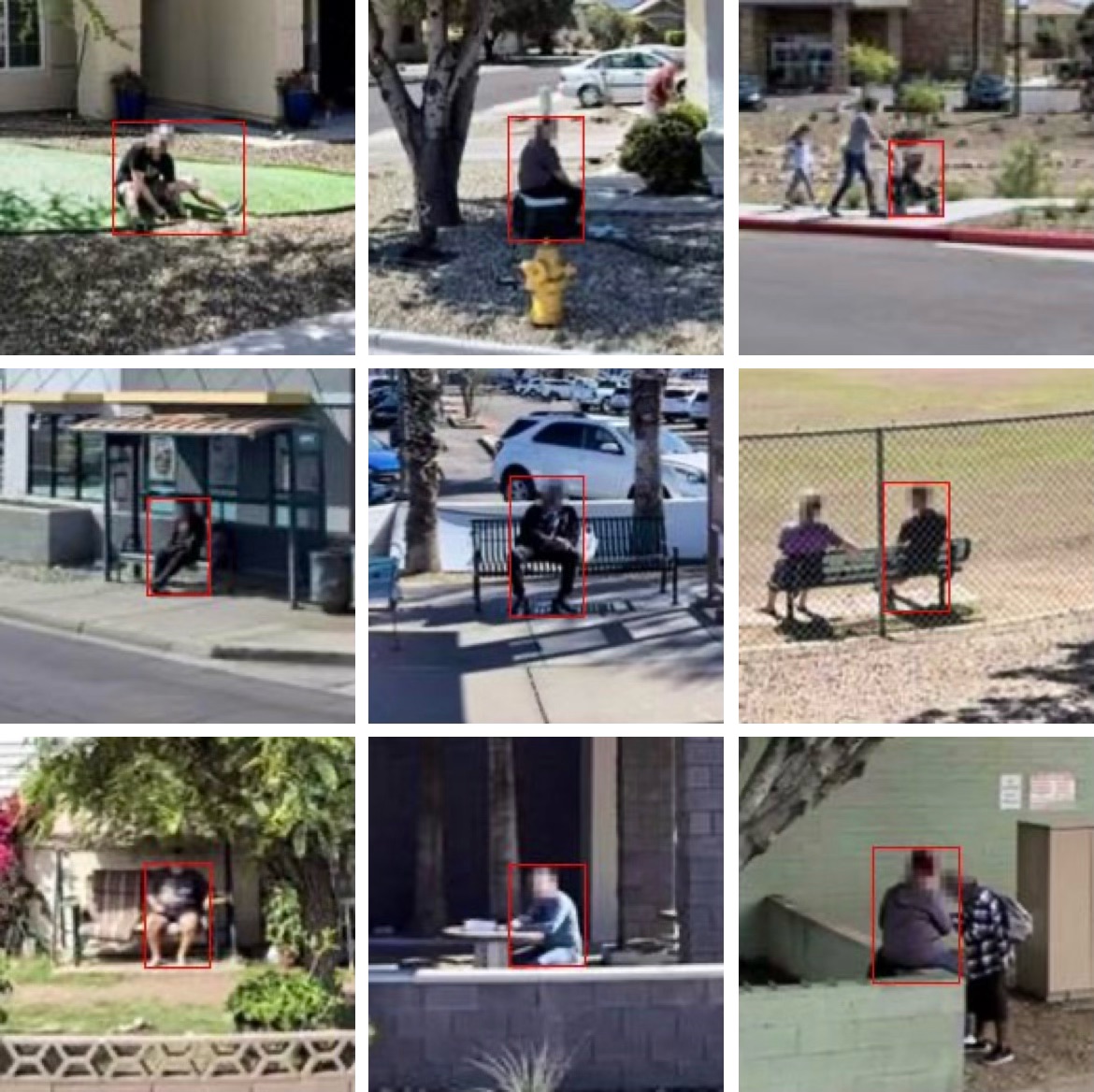}
        \caption{Seated ($\beta = 3$).}
        \label{fig:wdi-s1}
    \end{subfigure}
    \hfill
    \begin{subfigure}[b]{0.32\textwidth}
        \includegraphics[width=\linewidth]{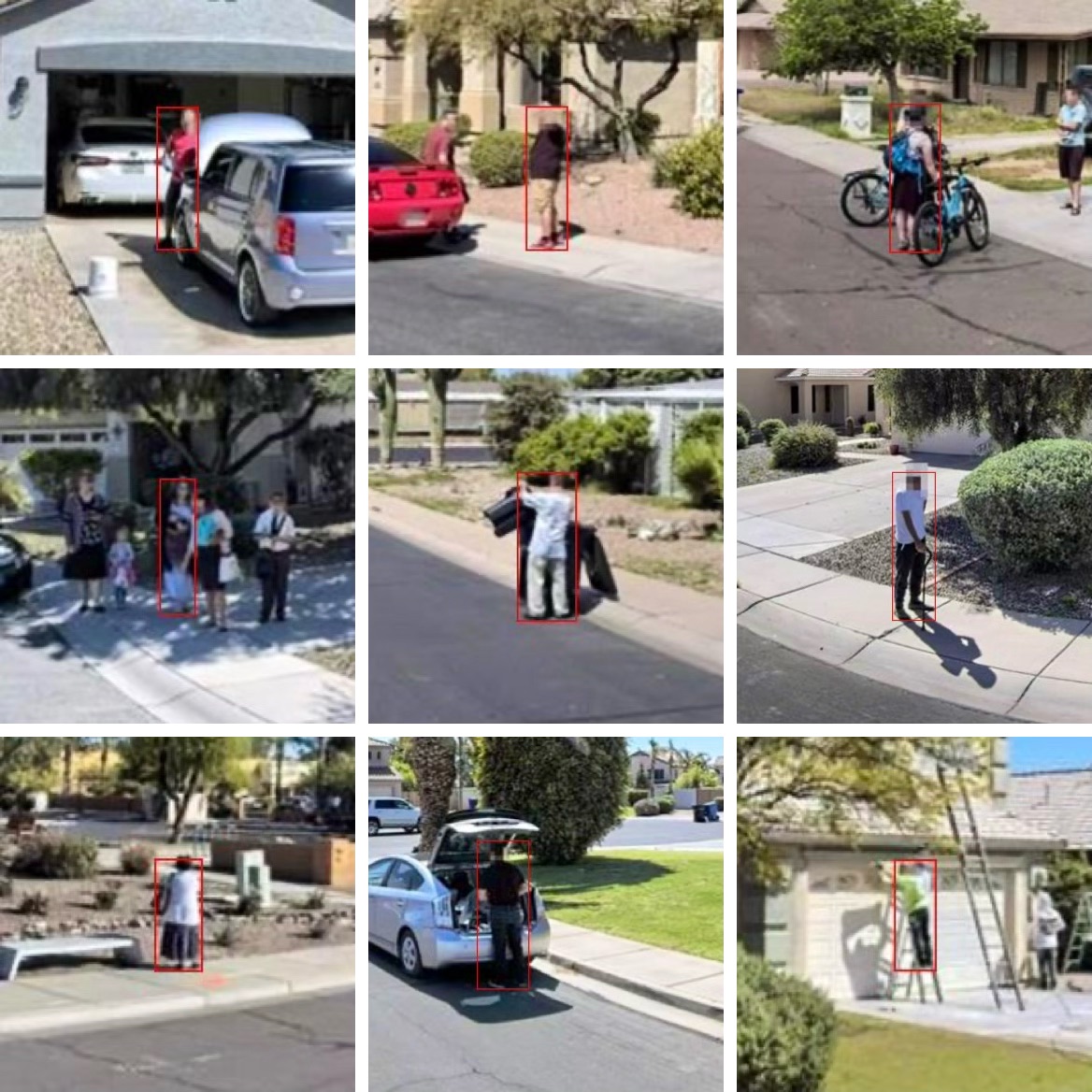}
        \caption{Stationary standing ($\beta = 2$).}
        \label{fig:wdi-s2}
    \end{subfigure}
    \hfill
    \begin{subfigure}[b]{0.32\textwidth}
        \includegraphics[width=\linewidth]{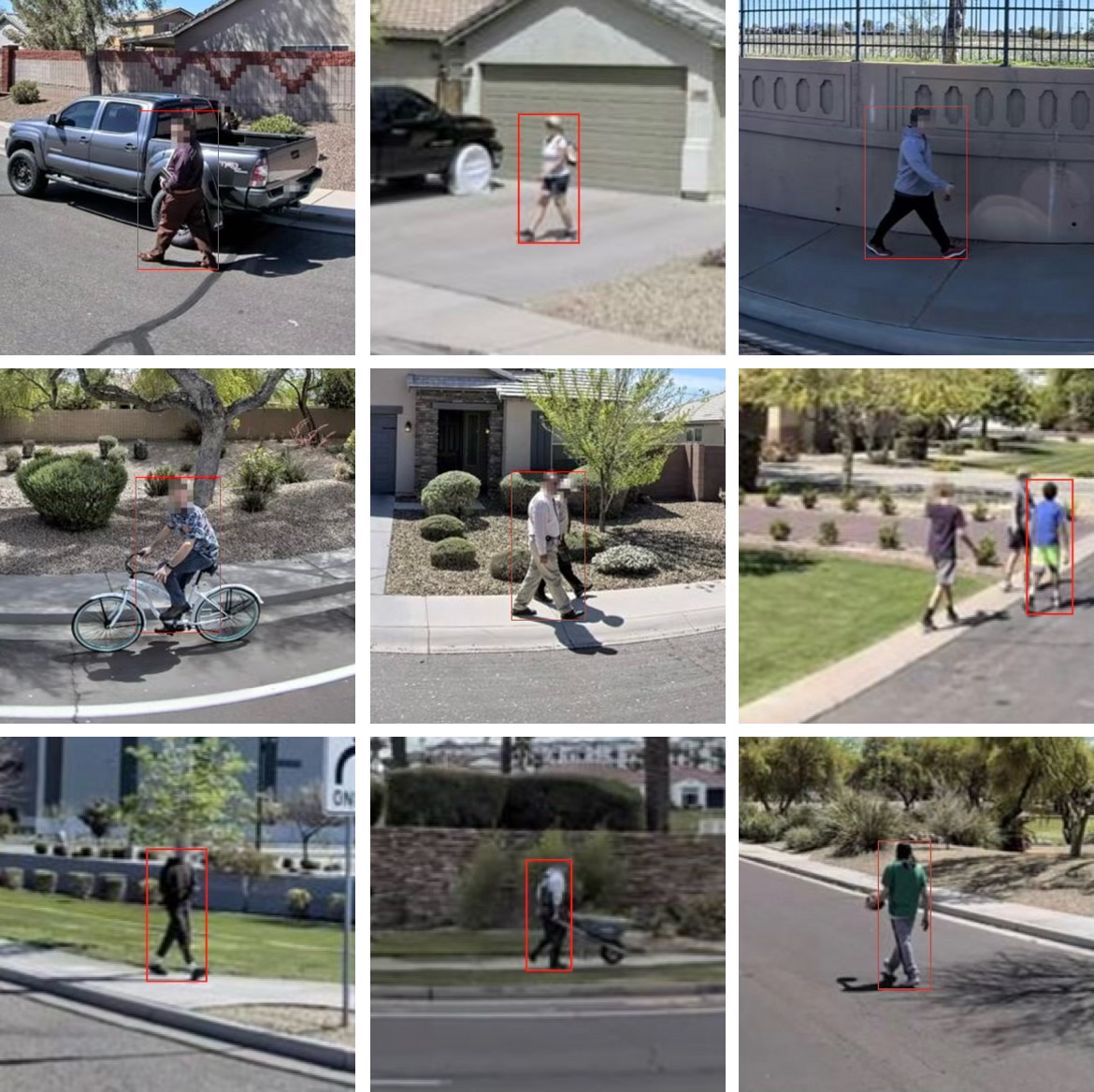}
        \caption{Active locomotion ($\beta = 1$).}
        \label{fig:wdi-s3}
    \end{subfigure}
    \caption{The three dwelling-intensity tiers used in WDI weight assignment. Seated people not in a riding relation receive $\beta = 3$ (a), stationary standing people receive $\beta = 2$ (b), and all others in active locomotion receive $\beta = 1$ (c).}
    \label{fig:wdi-examples}
\end{figure}

\begin{table}[htbp]
\renewcommand{\arraystretch}{1.2}%
\setlength{\tabcolsep}{4pt}%
\scriptsize
\caption{SDI binary flag definitions and weight assignments. Each flag equals 1 when all listed conditions are satisfied simultaneously.}
\label{tab:sdi-flags}
\centering
\begin{tabular}{p{0.8cm} p{2.8cm} p{2.2cm} p{1.2cm} p{5.5cm}}
\toprule
\textbf{Group} & \textbf{Indicator} & \textbf{Variable} & \textbf{Weight} & \textbf{Composite conditions} \\
\midrule
WGI
  & Solo pedestrian
  & \texttt{is\_g1}
  & $\alpha = 1$
  & \texttt{group\_size} $= 1$ \\[4pt]
  & Dyad member
  & \texttt{is\_g2}
  & $\alpha = 2$
  & \texttt{group\_size} $= 2$ \\[4pt]
  & Group of 3 or more
  & \texttt{is\_g3}
  & $\alpha = 3$
  & \texttt{group\_size} $\geq 3$ \\[6pt]
WDI
  & Seated, not riding
  & \texttt{is\_s1}
  & $\beta = 3$
  & \texttt{posture} $=$ \texttt{seated} \textbf{and}\newline
    \texttt{object\_relation} $\neq$ \texttt{riding} \textbf{and}\newline
    \texttt{locomotion} $=$ \texttt{stationary} \newline\\[4pt]
  & Stationary, upright
  & \texttt{is\_s2}
  & $\beta = 2$
  & \texttt{locomotion} $=$ \texttt{stationary} \textbf{and}\newline
    \texttt{posture} $\neq$ \texttt{seated} \\[4pt]
  & All others
  & \texttt{is\_s3}
  & $\beta = 1$
  & \texttt{is\_s1} $= 0$ \textbf{and} \texttt{is\_s2} $= 0$ \\
\bottomrule
\end{tabular}
\end{table}

\begin{figure}[htbp]
\centering
\begin{minipage}{0.85\textwidth}
    \includegraphics[width=1\columnwidth]{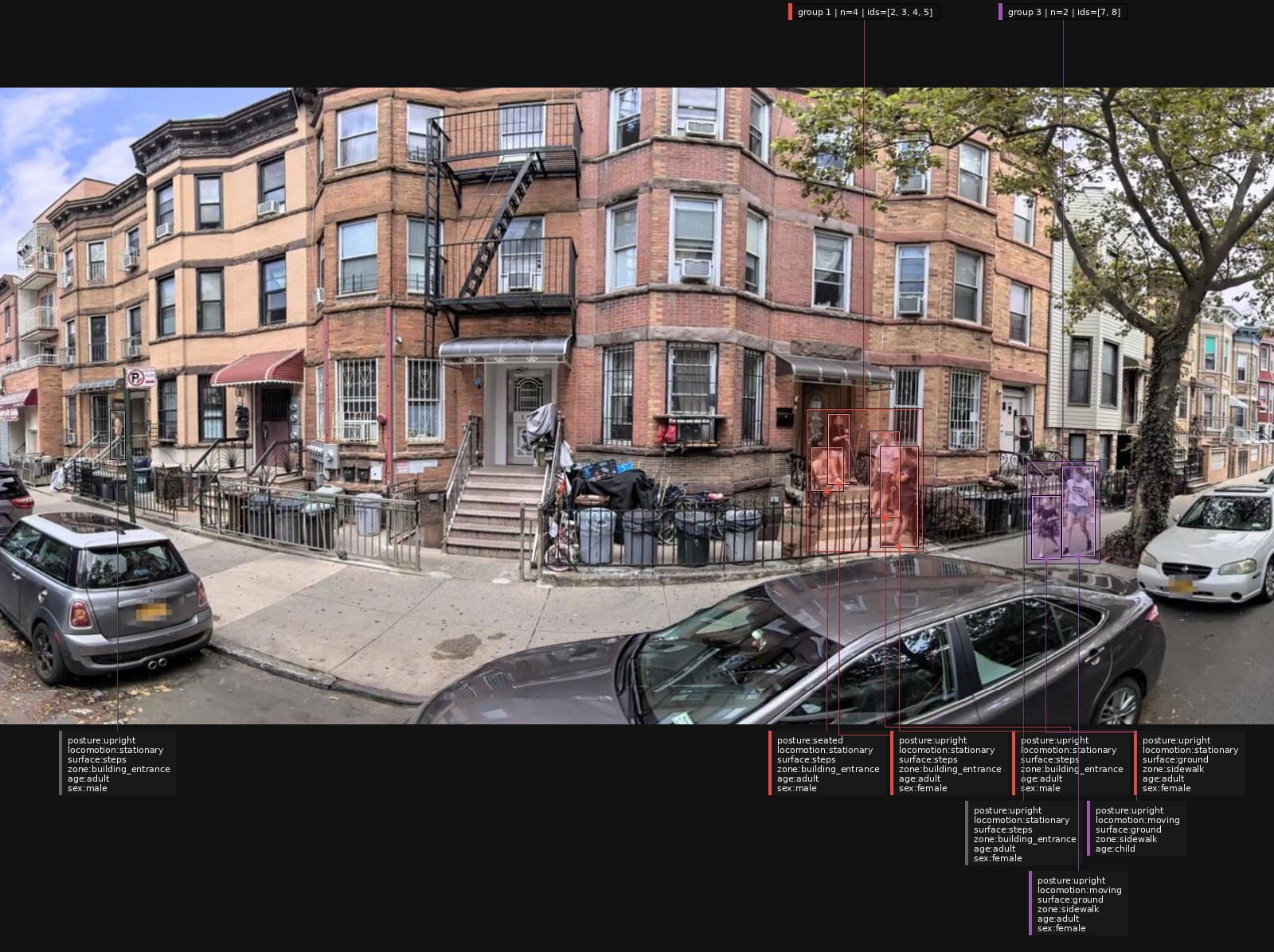}
    \caption{An illustrative sideview containing eight detected people: two solo individuals (both stationary), a dyad (both moving), and a group of
    four (three standing, one seated). The Weighted Grouping Index is $\text{WGI} = (2 \times 1 + 2 \times 2 + 4 \times 3) / 8 = 2.25$, reflecting the group-size composition. The Weighted Dwelling Index is $\text{WDI} = (2 \times 1 + 5 \times 2 + 1 \times 3) / 8 = 1.875$, reflecting the distribution of behavioral states. For the SDI, the two solo stationary individuals each receive a joint weight of $1 \times 2 = 2$; the two dyad members (moving) each receive $2 \times 1 = 2$; the three standing group members each receive $3 \times 2 = 6$; and the seated group member receives $3 \times 3 = 9$. Summing and normalizing yields $\text{SDI} = (2 \times 2 + 2 \times 2 + 3 \times 6 + 1 \times 9) / 8 = 4.375$.
}
\label{fig:sdi-example}
\end{minipage}
\end{figure}

\textbf{Social Dwelling Index (SDI)} combines the same grouping weights used in WGI and the same dwelling-intensity weights used in WDI, but does so at the individual level rather than by multiplying the two aggregate indices:
\begin{equation}
    \mathrm{SDI} = \frac{1}{N}\sum_{i=1}^{N}\alpha_{g(i)}\beta_{s(i)}.
\end{equation}
The multiplicative form is intended as a simple operationalization of the intuition that social presence on sidewalks becomes more pronounced when people are both co-present and lingering in space. Under this formulation, each detected person contributes a joint weight of $\alpha_{g(i)}\beta_{s(i)}/N$ to the final index. For example, a seated person in a group of three receives a higher contribution than a moving solo pedestrian because the formulation jointly prioritizes grouping and dwelling intensity. Because SDI is normalized by $N$, it does not scale mechanically with pedestrian volume and is therefore comparable across images with different numbers of detected people. As with WGI and WDI, the weighting structure should be interpreted as a heuristic observational measure rather than an empirically calibrated behavioral model. Figure~\ref{fig:sdi-example} illustrates the computation for a scene containing eight people distributed across different group configurations and locomotion states.

Sensitivity analyses reported in Appendix~\ref{sec:app-sensitivity} show that image-level SDI rankings are stable across three alternative grouping and dwelling-intensity weighting specifications (Spearman $\rho = 0.949$--$0.998$), and that the weak association between SDI and pedestrian count is preserved across all specifications ($r = 0.163$--$0.212$).

\subsection{Binary Flags}\label{ssec:binary-flags}

The binary flag tier assigns each detected person a set of binary indicators derived from VLM-coded observable attributes, produced by the \texttt{binary\_indicators()} function. Persons whose bounding box falls below a minimum pixel area threshold are excluded prior to flag assignment. The flags are organized into two families. The first two subsections cover sociodemographic flags, which record population composition and group membership. The final two subsections cover activity flags, which record behavioral states and spatial positions.

\textbf{Sex Flag} records the perceived sex of each detected person as coded by the VLM (\texttt{perceived\_sex} $\in$ \{\texttt{male}, \texttt{female}, \texttt{unclear}\}). This attribute is treated as a provisional indicator. Street-level imagery in the United States is subject to systematic face blurring by imagery providers, which removes the primary visual cue to infer sex. As a result, VLM coding of this attribute is structurally unreliable, independent of model capability, and \texttt{perceived\_sex} is excluded from primary analyses. It is retained in the dataset for exploratory use, where blurring rates are known to be low.

\begin{figure}[htbp]
\centering
\begin{minipage}{0.85\textwidth}
    \begin{subfigure}[t]{0.32\textwidth}
        \includegraphics[width=\linewidth]{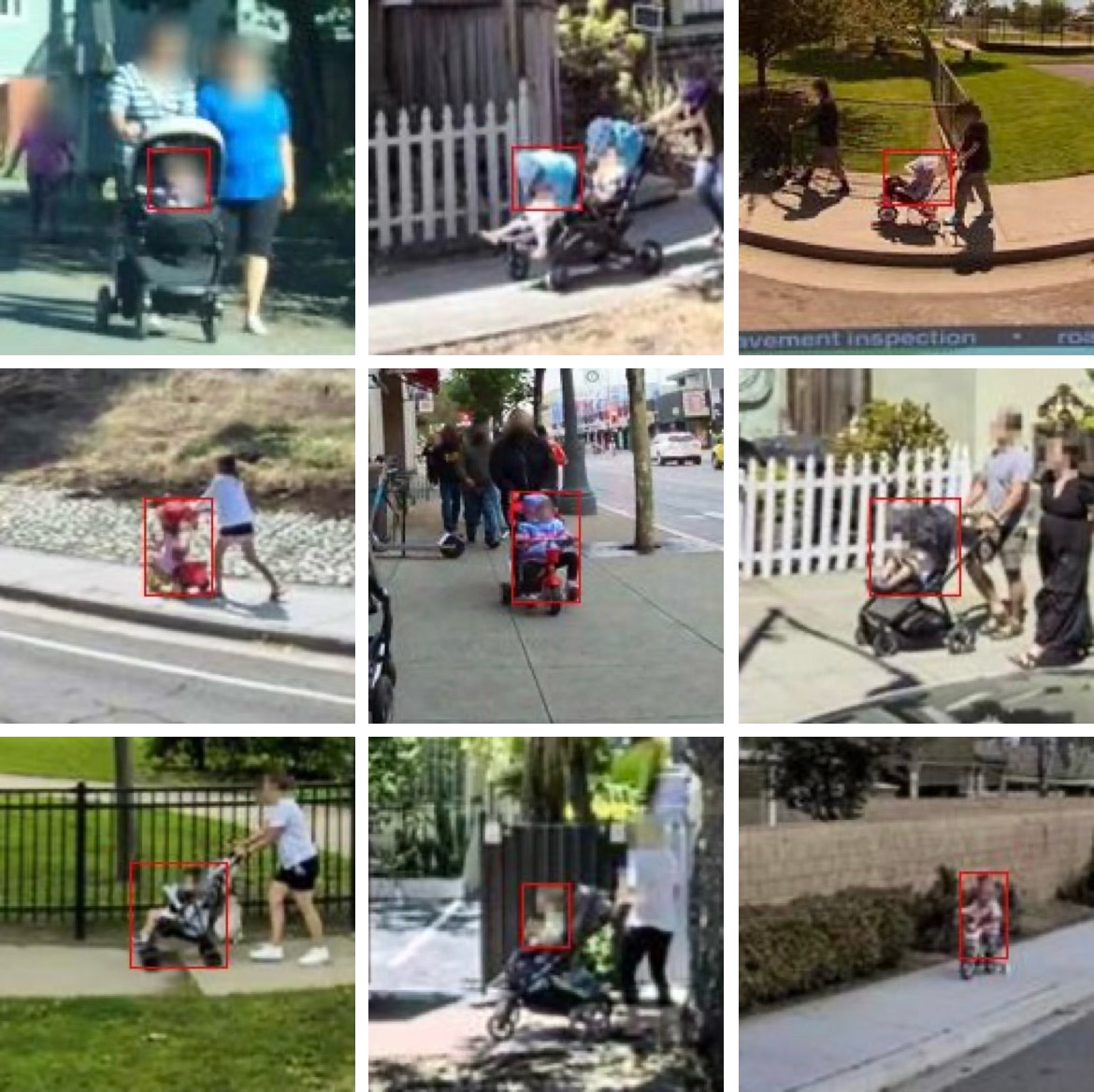}
        \caption{Infant.}
        \label{fig:cpf-infant}
    \end{subfigure}
    \hfill
    \begin{subfigure}[t]{0.32\textwidth}
        \includegraphics[width=\linewidth]{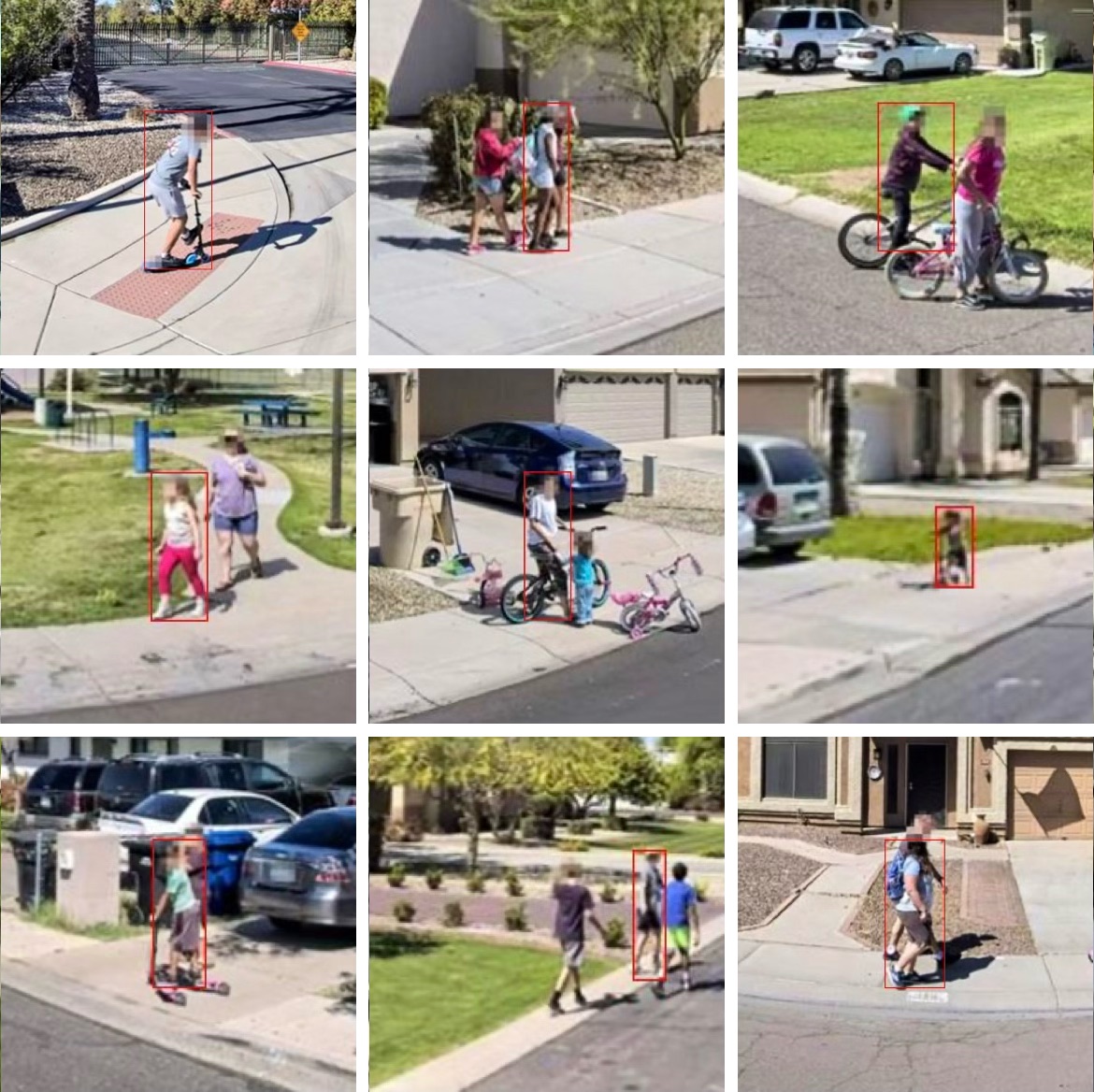}
        \caption{Child.}
        \label{fig:cpf-child}
    \end{subfigure}
    \hfill
    \begin{subfigure}[t]{0.32\textwidth}
        \mbox{} 
    \end{subfigure}
    \medskip
    \begin{subfigure}[t]{0.32\textwidth}
        \includegraphics[width=\linewidth]{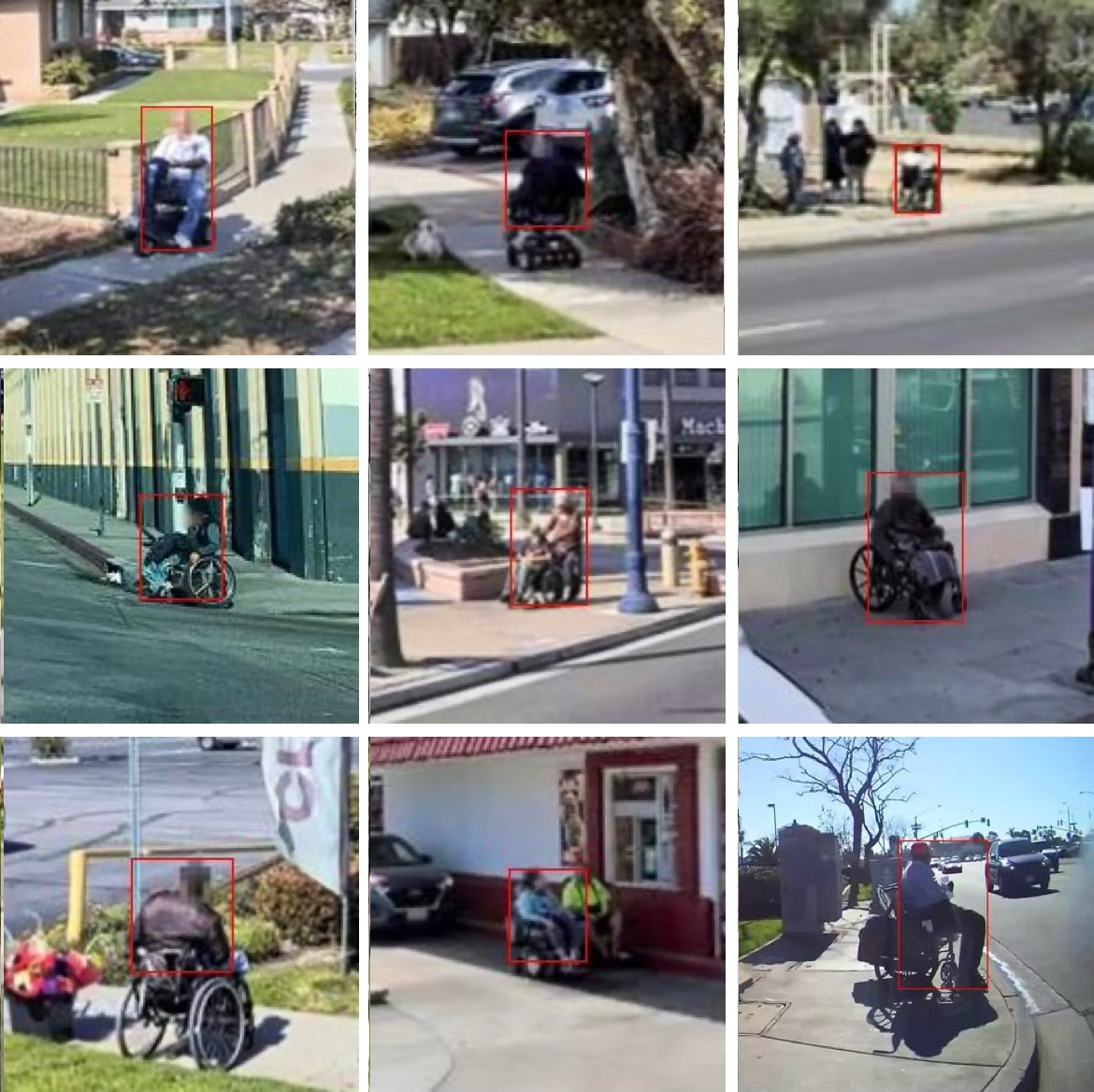}
        \caption{Wheelchair User.}
        \label{fig:cpf-wheelchair}
    \end{subfigure}
    \hfill
    \begin{subfigure}[t]{0.32\textwidth}
        \includegraphics[width=\linewidth]{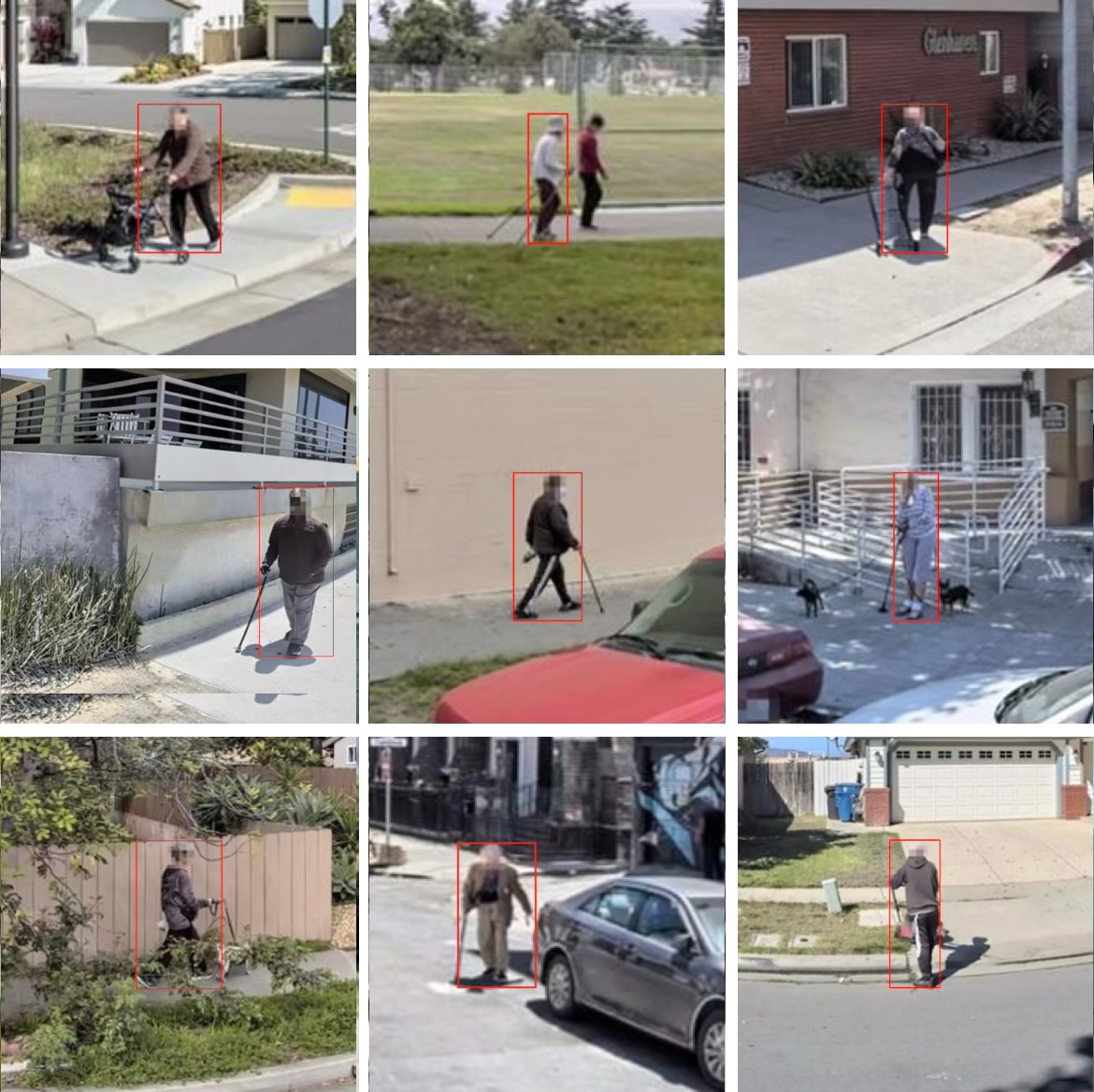}
        \caption{Walker User.}
        \label{fig:cpf-walker}
    \end{subfigure}
    \hfill
    \begin{subfigure}[t]{0.32\textwidth}
        \includegraphics[width=\linewidth]{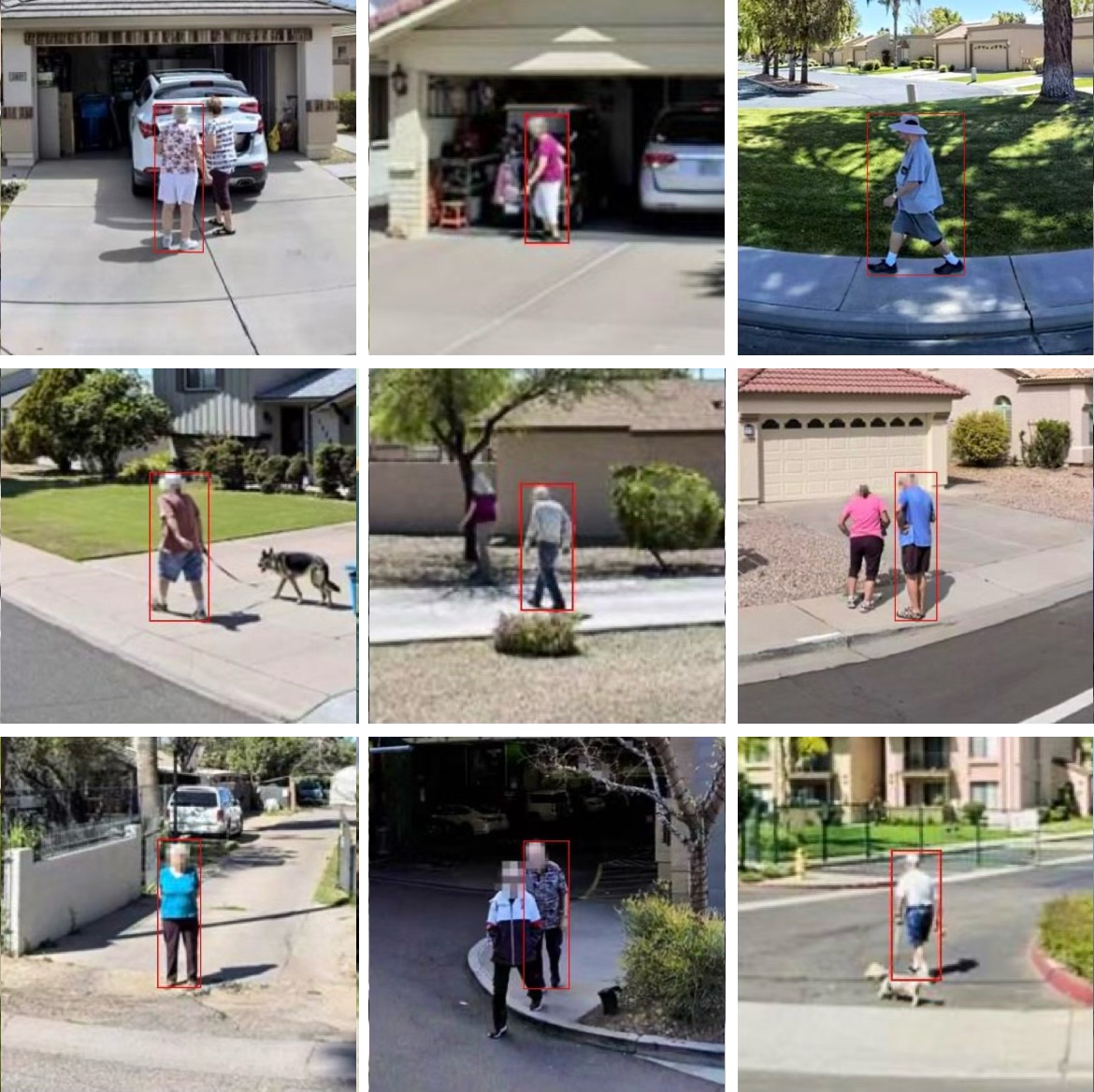}
        \caption{Older Adult.}
        \label{fig:cpf-older-adult}
    \end{subfigure}
    \caption{Representative street-level examples of the five Vulnerable Street User Flag categories, arranged to match the detection conditions listed in Table~\ref{tab:cgf}. Top row: age-based flags (infant, child). Bottom row: mobility-related flags (wheelchair user, walker user, older adult). Each crop corresponds to a single bounding box detection passed to the VLM for attribute coding.}
    \label{fig:cgf-examples}
\end{minipage}
\end{figure}

\textbf{Vulnerable Street User Flag} records whether a person belongs to a population group whose active mobility is sensitive to physical accessibility conditions, including surface continuity, slope, and the presence of curb cuts and resting opportunities. Five binary flags are assigned independently for \textit{infants, children, wheelchair users, walker users, and older adults}. Detection conditions are designed to be conservative: each flag requires a combination of age, posture, and object attributes, which, taken together, reduce false positives in visually ambiguous cases. Table~\ref{tab:cgf} lists the composite detection conditions for each flag; Figure~\ref{fig:cgf-examples} shows a representative example of each category.

\begin{table}[htbp]
\renewcommand{\arraystretch}{1}%
\setlength{\tabcolsep}{4pt}%
\scriptsize
\caption{Vulnerable Street User Flag detection conditions. Each flag equals 1 when all listed conditions are satisfied simultaneously. All flags additionally require that the person's bounding box exceeds the minimum pixel area threshold.}
\label{tab:cgf}
\centering
\begin{tabular}{p{0.5cm} p{3.0cm} p{2.3cm} p{6.2cm}}
\toprule
\textbf{} & \textbf{Flag} & \textbf{Variable} & \textbf{Composite conditions} \\
\midrule
1 & Infant
  & \texttt{is\_infant}
  & \texttt{posture} $=$ \texttt{seated}\newline
    \texttt{object} $=$ \texttt{stroller}\newline
    \texttt{support\_surface} $=$ \texttt{ground}\newline
    \texttt{perceived\_age\_group} $\in$ \{\texttt{child}, \texttt{unclear}\}\newline
    \texttt{uniform\_presence} $=$ \texttt{none} \\[2pt]
2 & Child
  & \texttt{is\_child}
  & \texttt{perceived\_age\_group} $=$ \texttt{child}\newline
    \texttt{posture} $\neq$ \texttt{unclear}\newline
    \texttt{is\_infant} $= 0$\newline
    \texttt{uniform\_presence} $=$ \texttt{none} \\[2pt]
3 & Wheelchair user
  & \texttt{is\_wheelchair\_user}
  & \texttt{posture} $=$ \texttt{seated}\newline
    \texttt{object} $=$ \texttt{wheelchair}\newline
    \texttt{support\_surface} $=$ \texttt{ground}\newline
    \texttt{uniform\_presence} $=$ \texttt{none} \\[2pt]
4 & Walker user
  & \texttt{is\_walker\_user}
  & \texttt{object} $=$ \texttt{walker}\newline
    \texttt{object\_relation} $=$ \texttt{holding}\newline
    \texttt{perceived\_age\_group} $=$ \texttt{older\_adult}\newline
    \texttt{posture} $\neq$ \texttt{unclear}\newline
    \texttt{is\_wheelchair\_user} $= 0$\newline
    \texttt{uniform\_presence} $=$ \texttt{none} \\[2pt]
5 & Older adult
  & \texttt{is\_older\_adult}
  & \texttt{perceived\_age\_group} $=$ \texttt{older\_adult}\newline
    \texttt{posture} $\neq$ \texttt{unclear}\newline
    \texttt{is\_wheelchair\_user} $= 0$\newline
    \texttt{is\_walker\_user} $= 0$\newline
    \texttt{uniform\_presence} $=$ \texttt{none} \\
\bottomrule
\end{tabular}
\end{table}

\textbf{Primary Activity Flag} identifies twelve activity categories, grounded in the pedestrian observational literature reviewed in Section~\ref{ssec:urban-behavior}. Gehl's \citeyearpar{Gehl1987LifeSpace} tripartite distinction between necessary, optional, and social activities provides the primary organizing logic: necessary activities are represented by \texttt{passing} and \texttt{working with tools}; optional activities by \texttt{cycling}, \texttt{dog walking}, \texttt{sports}, \texttt{running}, \texttt{scooter/skateboard riding}, \texttt{eating}, \texttt{phone use}, and \texttt{resting}; and social activities by \texttt{chatting}. \texttt{Maintenance working} captures the institutional street presence that Whyte \citeyearpar{whyte1980social} documented as a persistent feature of active sidewalks. The category boundaries follow Mehta's \citeyearpar{mehta2019streets} behavioral taxonomy, which enumerates sidewalk activities at a granularity compatible with visual observation, covering postures, object-mediated behaviors, and social engagements.

Two constraints shaped the final selection. First, each category must be inferable from directly observable physical attributes, specifically posture, locomotion, held object, object relation, and zone type, without requiring inference about intent or social meaning. This is the observable-versus-interpretive principle described in Section~\ref{ssec:activity-detection}: categories that depend on facial expression, speech, or fine-grained contextual judgment, such as window-shopping, panhandling, or sunbathing, are excluded on these grounds. Second, each category must produce a sufficiently distinct visual signature to support reliable VLM coding. Goffman's \citeyearpar{goffman1963behavior} distinction between focused and unfocused interaction motivates the \texttt{chatting} category specifically: rather than coding interaction in general, which is ambiguous from a static image, the detection conditions require a group of two or more people who are all stationary and include at least one non-upright member, a configuration that reliably signals sustained focused engagement rather than incidental co-presence.

\begin{figure}[htbp]
\centering
\begin{minipage}{0.85\textwidth}
    \begin{subfigure}[t]{0.32\textwidth}
        \includegraphics[width=\linewidth]{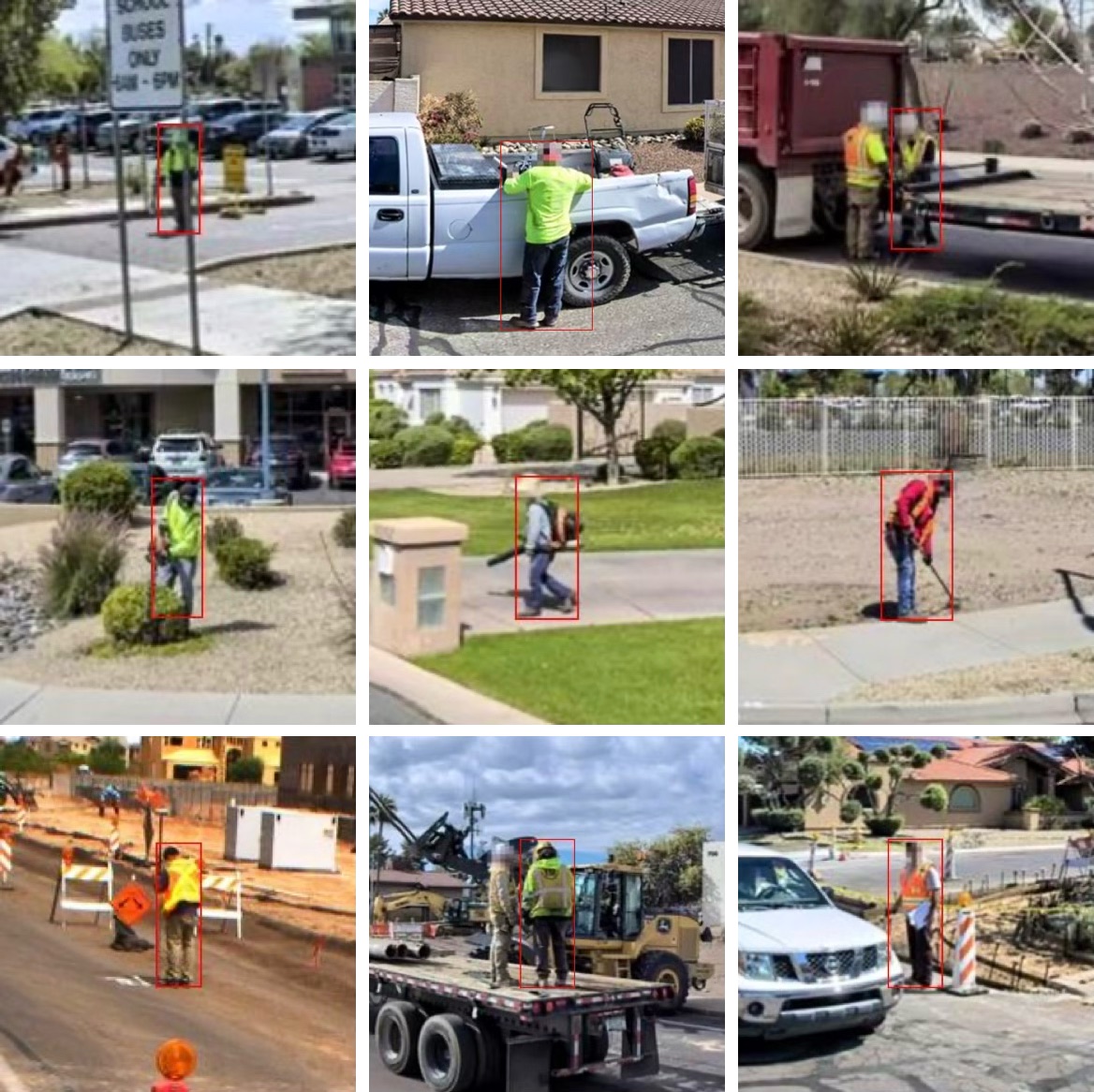}
        \caption{Maintenance Working.}
        \label{fig:paf-maintenance-working}
    \end{subfigure}
    \hfill
    \begin{subfigure}[t]{0.32\textwidth}
        \includegraphics[width=\linewidth]{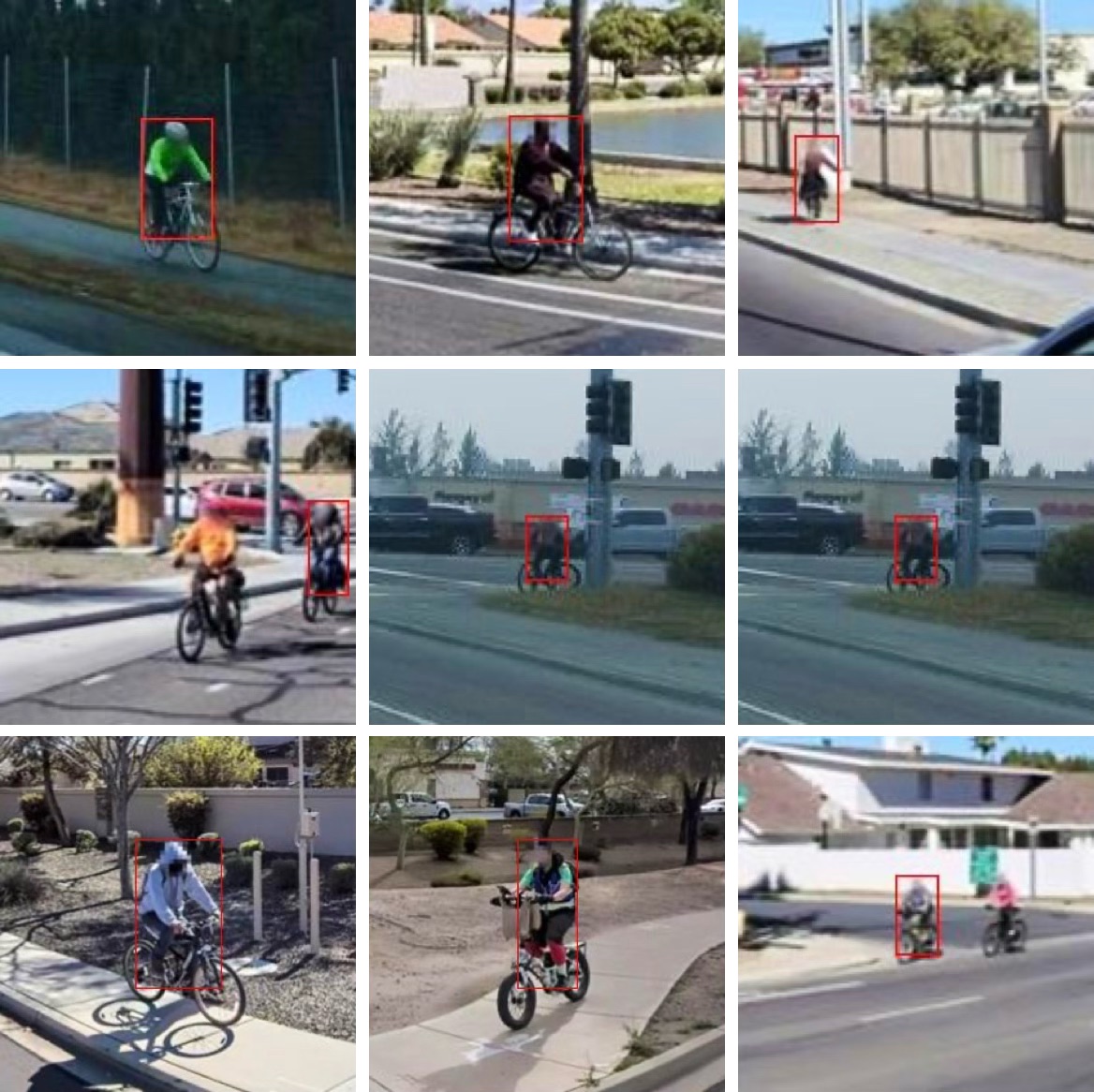}
        \caption{Cycling.}
        \label{fig:paf-cycling}
    \end{subfigure}
    \hfill
    \begin{subfigure}[t]{0.32\textwidth}
        \includegraphics[width=\linewidth]{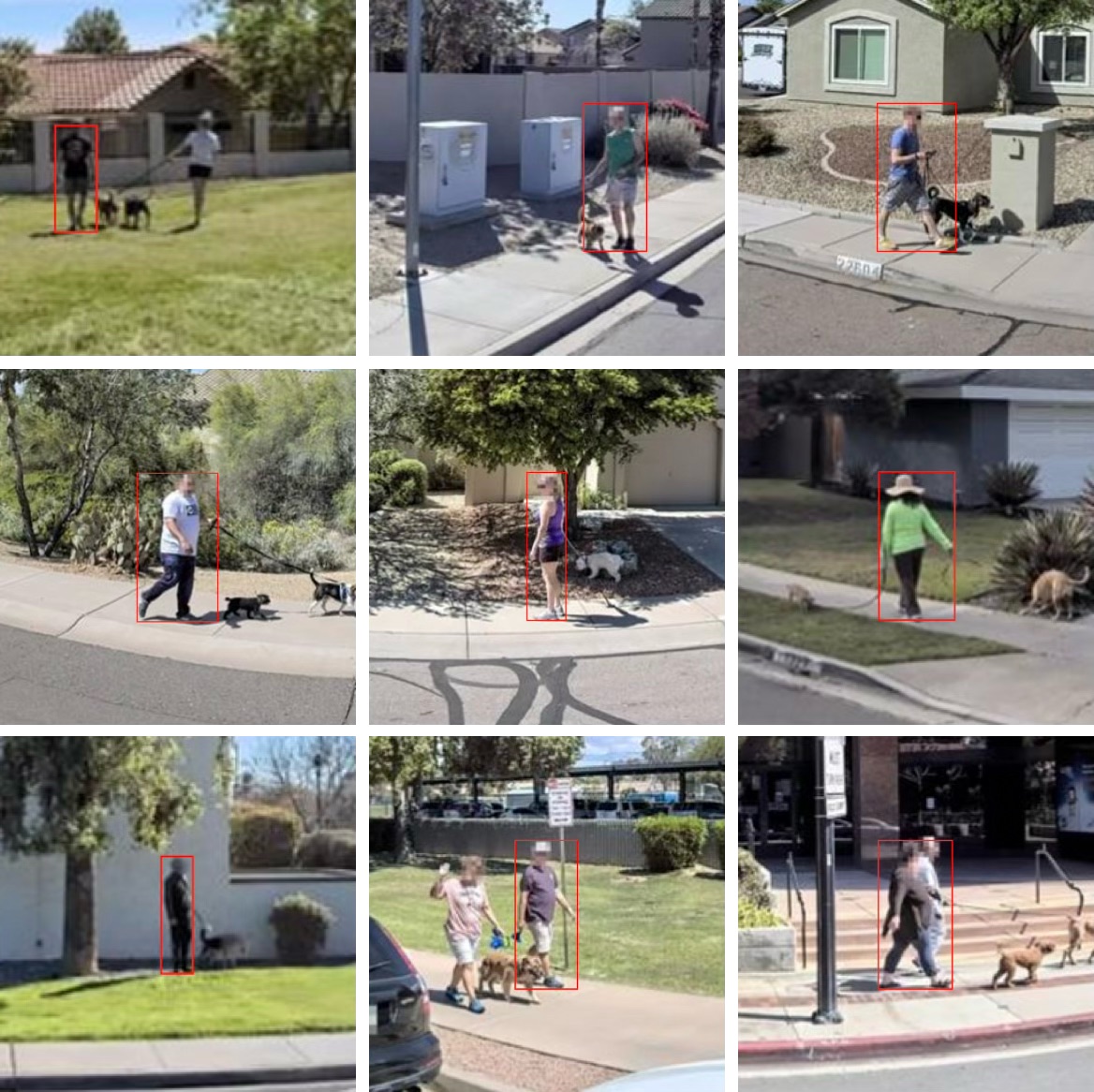}
        \caption{Dog Walking.}
        \label{fig:paf-dog-walking}
    \end{subfigure}

    \begin{subfigure}[t]{0.32\textwidth}
        \includegraphics[width=\linewidth]{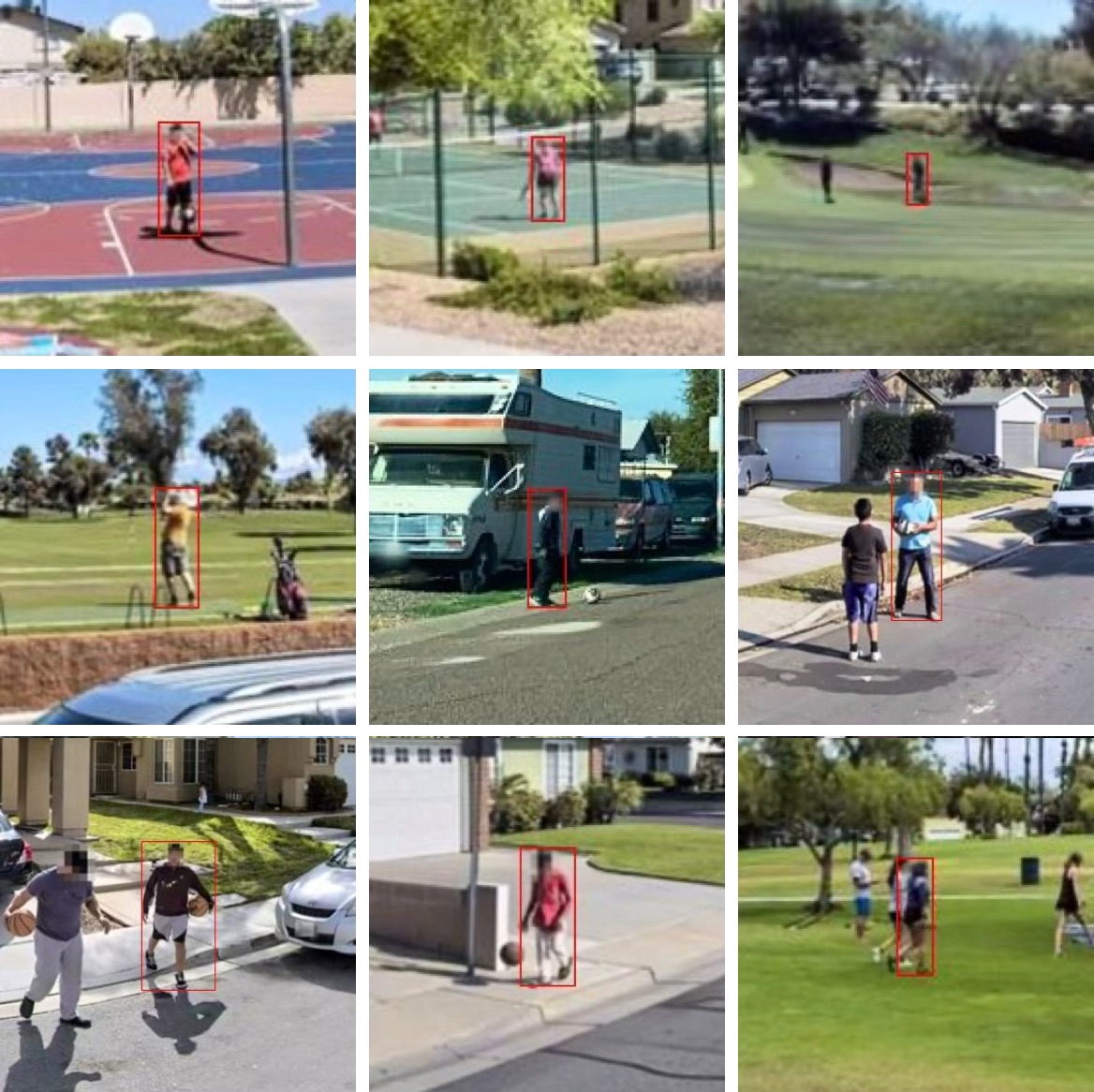}
        \caption{Sports.}
        \label{fig:paf-sports}
    \end{subfigure}
    \hfill
    \begin{subfigure}[t]{0.32\textwidth}
        \includegraphics[width=\linewidth]{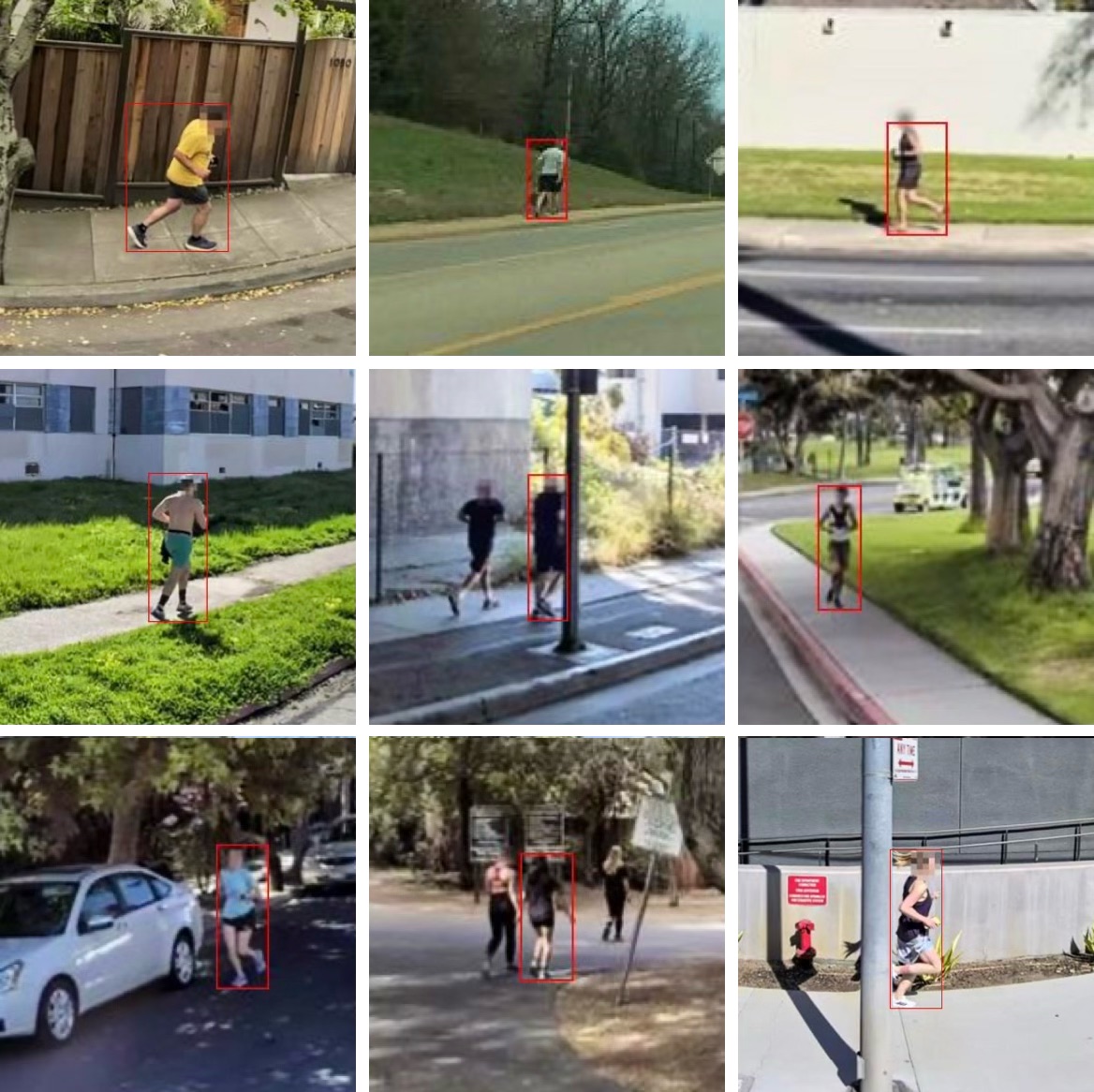}
        \caption{Running.}
        \label{fig:paf-running}
    \end{subfigure}
    \hfill
    \begin{subfigure}[t]{0.32\textwidth}
        \includegraphics[width=\linewidth]{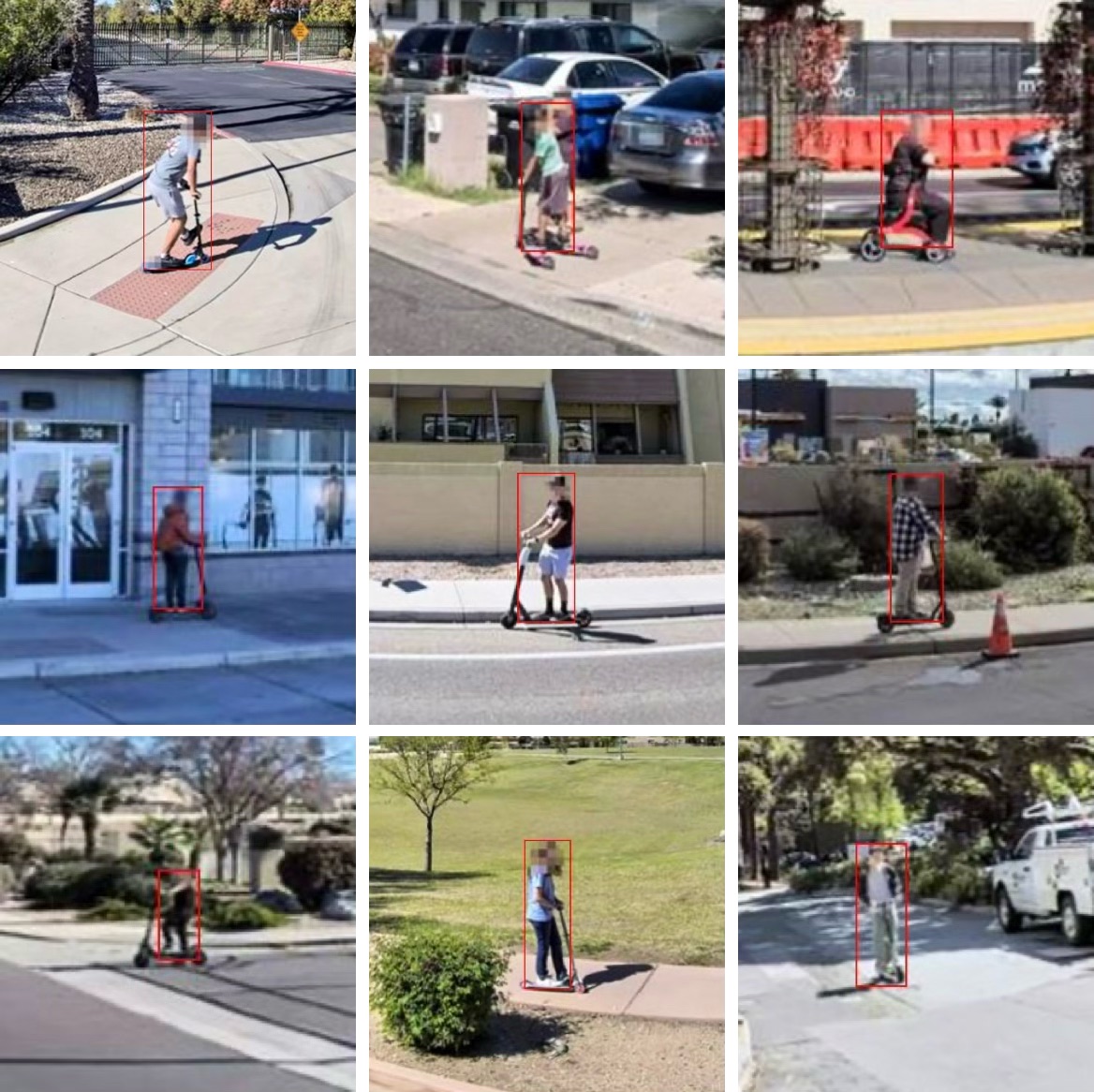}
        \caption{Scooter/Skateboard Riding.}
        \label{fig:paf-scooter-skateboard-riding}
    \end{subfigure}

    \begin{subfigure}[t]{0.32\textwidth}
        \includegraphics[width=\linewidth]{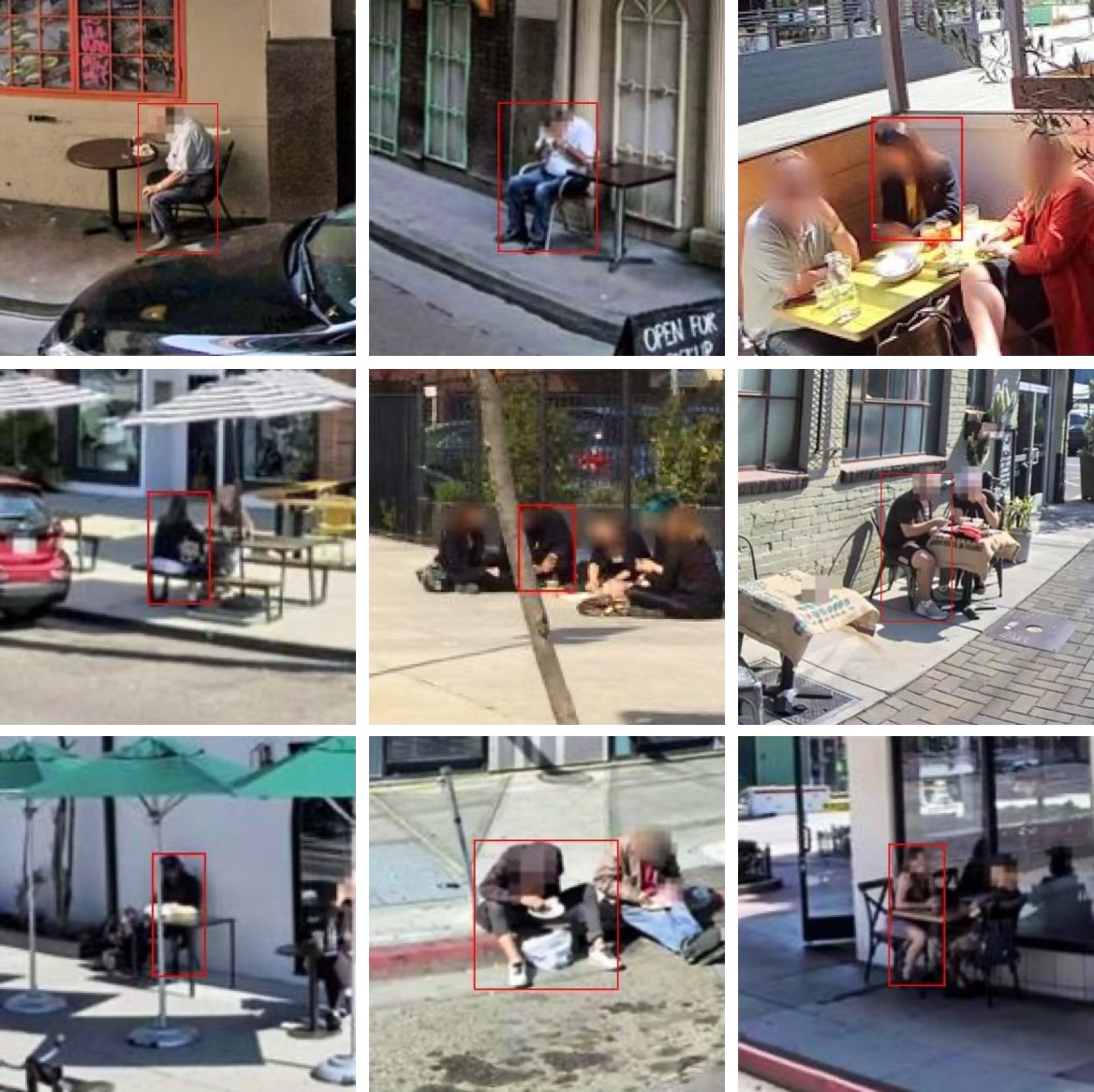}
        \caption{Eating.}
        \label{fig:paf-eating}
    \end{subfigure}
    \hfill
    \begin{subfigure}[t]{0.32\textwidth}
        \includegraphics[width=\linewidth]{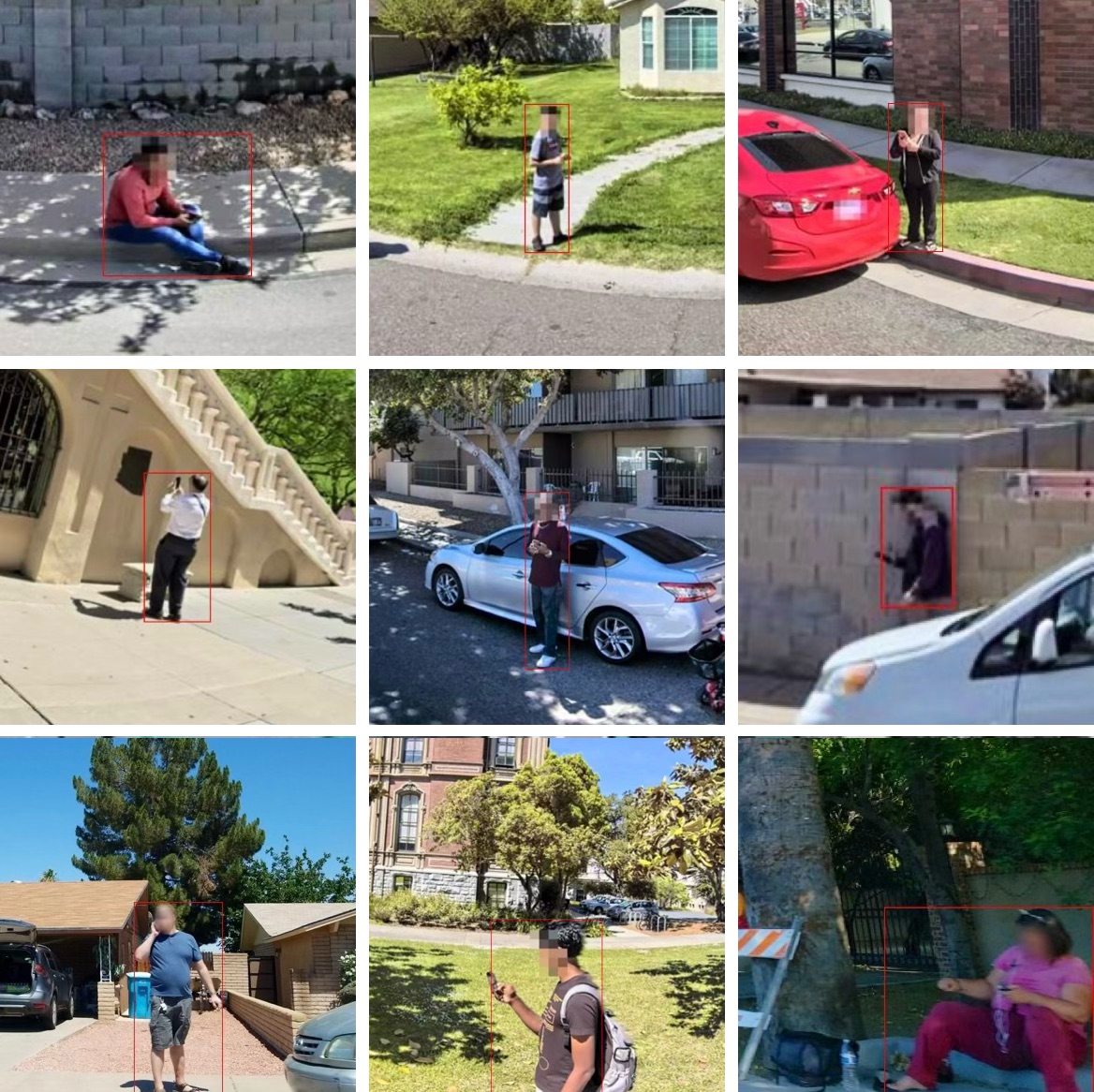}
        \caption{Phone Using.}
        \label{fig:paf-phone-using}
    \end{subfigure}
    \hfill
    \begin{subfigure}[t]{0.32\textwidth}
        \includegraphics[width=\linewidth]{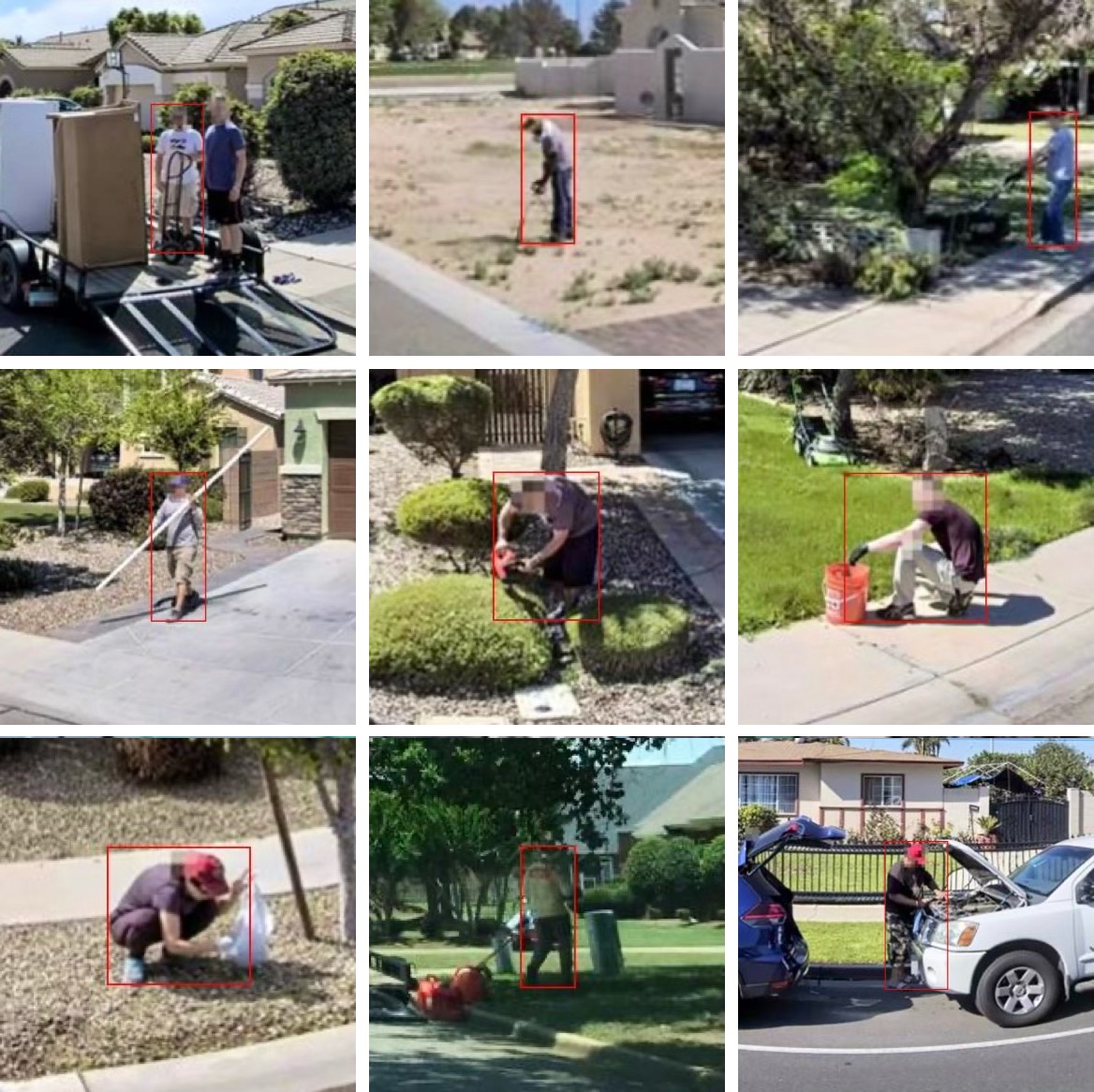}
        \caption{Working with Tools.}
        \label{fig:paf-tool-working}
    \end{subfigure}

    \begin{subfigure}[t]{0.32\textwidth}
        \includegraphics[width=\linewidth]{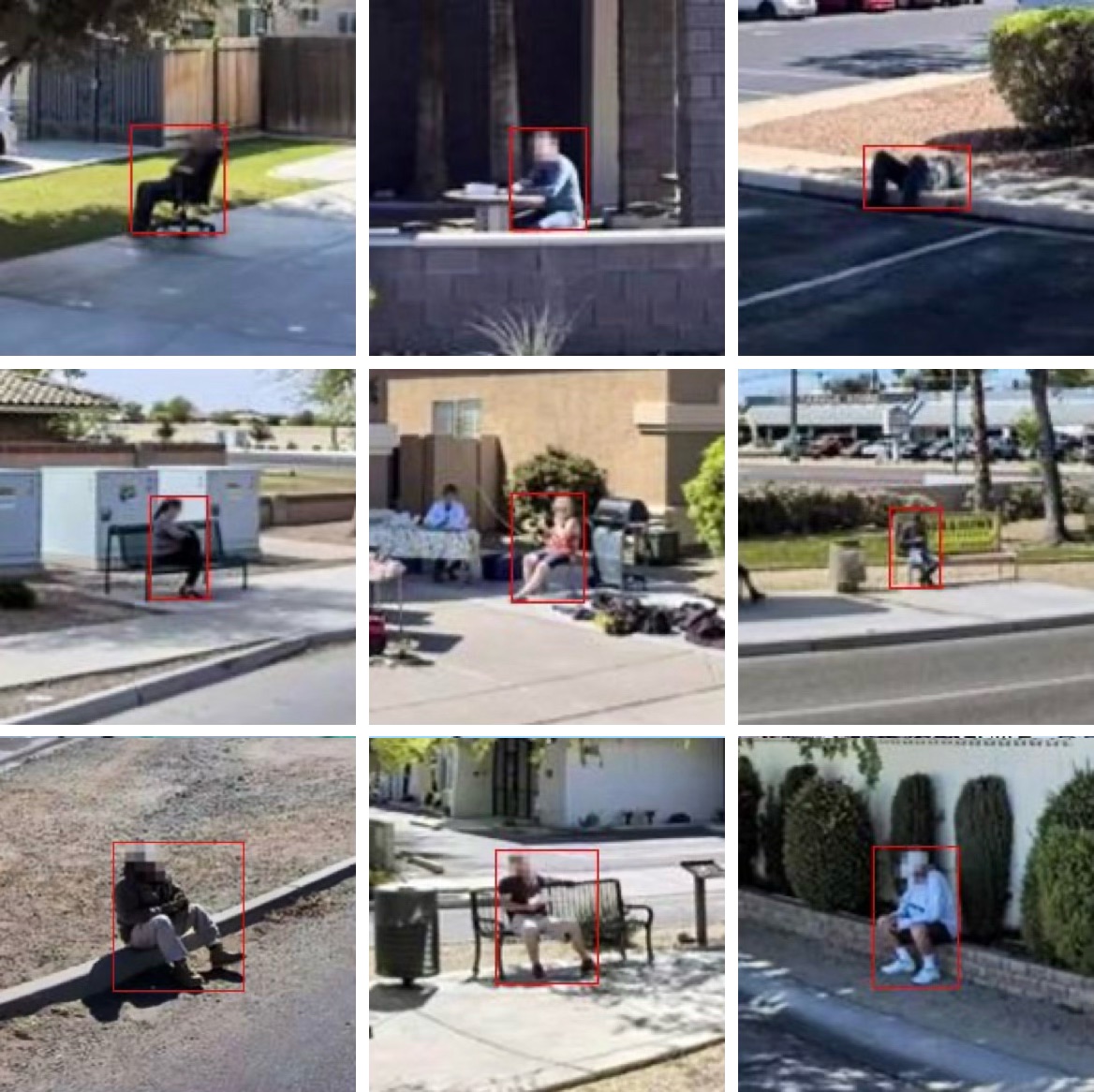}
        \caption{Resting.}
        \label{fig:paf-resting}
    \end{subfigure}
    \hfill
    \begin{subfigure}[t]{0.32\textwidth}
        \includegraphics[width=\linewidth]{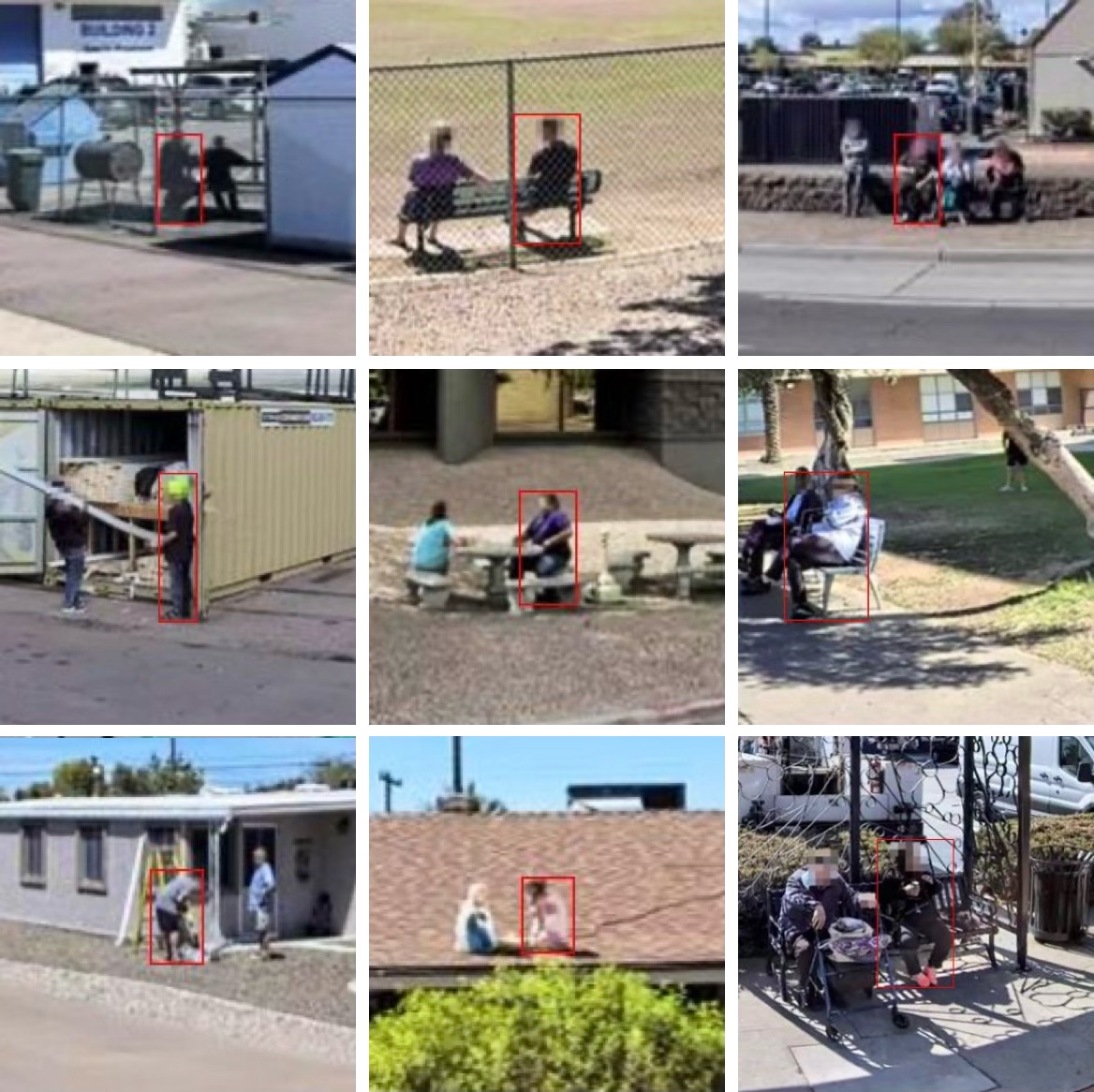}
        \caption{Chatting.}
        \label{fig:paf-chatting}
    \end{subfigure}
    \hfill
    \begin{subfigure}[t]{0.32\textwidth}
        \includegraphics[width=\linewidth]{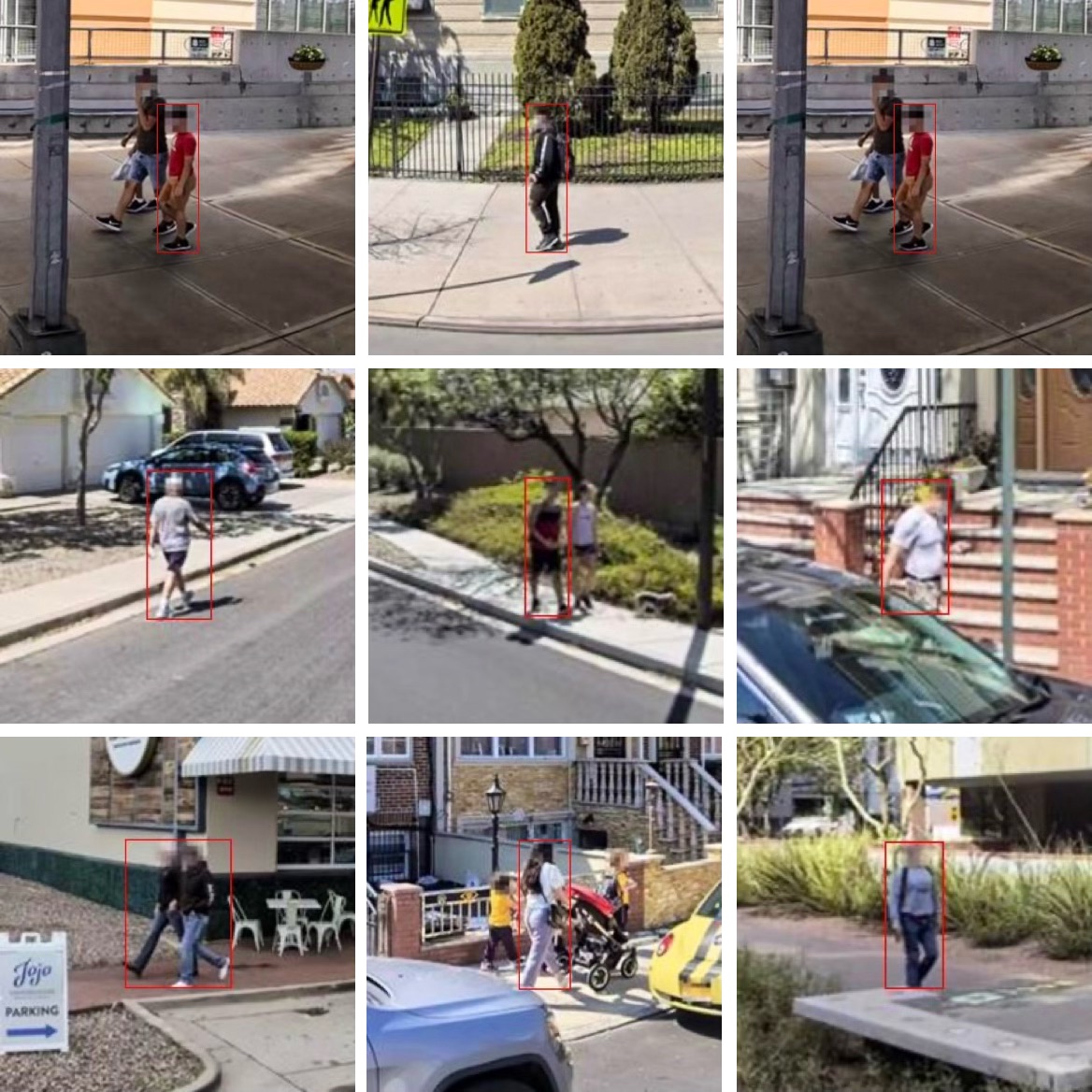}
        \caption{Passing.}
        \label{fig:paf-passing}
    \end{subfigure}
    \caption{Representative street-level examples of the twelve Primary Activity Label categories, arranged in descending priority order. Each crop shows a single detected person assigned to the indicated label based on the hierarchical assignment procedure described in Table~\ref{tab:paf}.}
    \label{fig:paf-examples}
\end{minipage}
\end{figure}

Each detected person receives exactly one label via a hierarchical assignment procedure. When a person satisfies conditions for multiple categories, the higher-priority label takes precedence. The hierarchy ranks object-anchored behaviors above posture-based or zone-inferred states, on the grounds that object detection is more reliable than posture or locomotion coding in street-level imagery. The label \texttt{passing} serves as the broad baseline for upright civilians in motion or at rest without a more specific activity signal; \texttt{other} is the residual fallback for persons who satisfy none of the above conditions. Table~\ref{tab:paf} lists the full label set with detection conditions in priority order; Figure~\ref{fig:paf-examples} shows a representative example for each category.

\begin{table}[htbp]
\renewcommand{\arraystretch}{1}%
\setlength{\tabcolsep}{4pt}%
\scriptsize
\caption{Primary Activity Label assignment conditions, listed in descending priority order. When a person satisfies multiple conditions, the highest-priority label is assigned. All labels require \texttt{uniform\_presence} $=$ \texttt{none} and that the bounding box exceeds the minimum pixel area threshold.}
\label{tab:paf}
\centering
\begin{tabular}{p{0.5cm} p{3.0cm} p{2.3cm} p{6.2cm}}
\toprule
\textbf{} & \textbf{Flag} & \textbf{Variable} &
\textbf{Composite Conditions} \\
\midrule
1 & Maintenance Working
  & \texttt{maintenance}
  & \texttt{uniform\_presence} $=$ \texttt{maintenance\_worker}\newline
    \texttt{perceived\_age\_group} $\neq$ \texttt{child} \\[4pt]
2 & Cycling
  & \texttt{cycling}
  & \texttt{object\_relation} $=$ \texttt{riding}\newline
    \texttt{object} $=$ \texttt{bicycle} \\[4pt]
3 & Dog Walking
  & \texttt{dog\_walking}
  & \texttt{object} $=$ \texttt{animal}\newline
    \texttt{object\_relation} $=$ \texttt{holding} \\[4pt]
4 & Sports
  & \texttt{sports}
  & \texttt{object} $=$ \texttt{sports\_equipment}\newline
    \texttt{object\_relation} $\neq$ \texttt{riding} \\[4pt]
5 & Running
  & \texttt{running}
  & \texttt{locomotion} $=$ \texttt{running} \\[4pt]
6 & Scooter/Skateboard Riding
  & \texttt{riding}
  & \texttt{object\_relation} $=$ \texttt{riding}\newline
    \texttt{object} $\in$ \{\texttt{scooter}, \texttt{skateboard}\}\newline
    \texttt{zone\_type} $\neq$ \texttt{roadway}\newline
    \texttt{perceived\_age\_group} $\neq$ \texttt{older\_adult} \\[4pt]
7 & Eating
  & \texttt{eating}
  & \texttt{posture} $\in$ \{\texttt{seated}, \texttt{squatting},
    \texttt{lying}\}\newline
    \texttt{object\_relation} $\neq$ \texttt{riding}\newline
    \texttt{object} $=$ \texttt{food\_or\_drink}\newline
    \texttt{object\_relation} $\in$ \{\texttt{holding},
    \texttt{carrying}\} \\[4pt]
8 & Phone Using
  & \texttt{phone\_use}
  & \texttt{object} $=$ \texttt{phone}\newline
    \texttt{object\_relation} $=$ \texttt{holding}\newline
    \texttt{locomotion} $=$ \texttt{stationary}\newline
    \texttt{posture} $\in$ \{\texttt{seated}, \texttt{squatting},
    \texttt{lying}, \texttt{upright}\} \\[4pt]
9 & Working with Tools
  & \texttt{working}
  & \texttt{object} $\in$ \{\texttt{tool}, \texttt{cart},
    \texttt{trash\_bag}\}\newline
    \texttt{object\_relation} $\in$ \{\texttt{holding},
    \texttt{pushing}\} \\[4pt]
10 & Resting
  & \texttt{resting}
  & \texttt{posture} $\in$ \{\texttt{seated}, \texttt{lying}\}\newline
    \texttt{object\_relation} $\neq$ \texttt{riding}\newline
    \texttt{group\_size} $= 1$\newline
    \texttt{locomotion} $=$ \texttt{stationary}\newline
    \texttt{object} $\in$ \{\texttt{none}, \texttt{unclear}\} \\[4pt]
11 & Chatting
  & \texttt{chatting}
  & \texttt{group\_size} $\geq 2$\newline
    cluster contains $\geq 1$ non-upright member\newline
    all cluster members \texttt{stationary}\newline
    \texttt{object} $\in$ \{\texttt{none}, \texttt{unclear}\} \\[4pt]
12 & Passing
  & \texttt{passing}
  & \texttt{posture} $=$ \texttt{upright}\newline
    \texttt{locomotion} $\in$ \{\texttt{moving},
    \texttt{stationary}\} \\
\bottomrule
\end{tabular}
\end{table}

\textbf{Spatial Context Flag} is developed, because where a person is located within the street environment carries independent analytical meaning beyond what they are doing. A person crossing a street, lingering near a building entrance, or seated on steps occupies a qualitatively different position in the social ecology of the sidewalk than one simply passing through, even if their posture and locomotion are identical. Spatial Context Flags capture this locational dimension as a set of independent binary indicators that complement the primary activity rather than replace it: a person labeled \texttt{passing} may simultaneously carry \texttt{is\_crosswalk} $= 1$, and a person labeled \texttt{chatting} may carry \texttt{is\_building\_entrance} $= 1$, with each dimension contributing distinct information about the social use of space.

\begin{figure}[htbp]
\centering
\begin{minipage}{0.85\textwidth}
    \begin{subfigure}[t]{0.32\textwidth}
    \includegraphics[width=\linewidth]{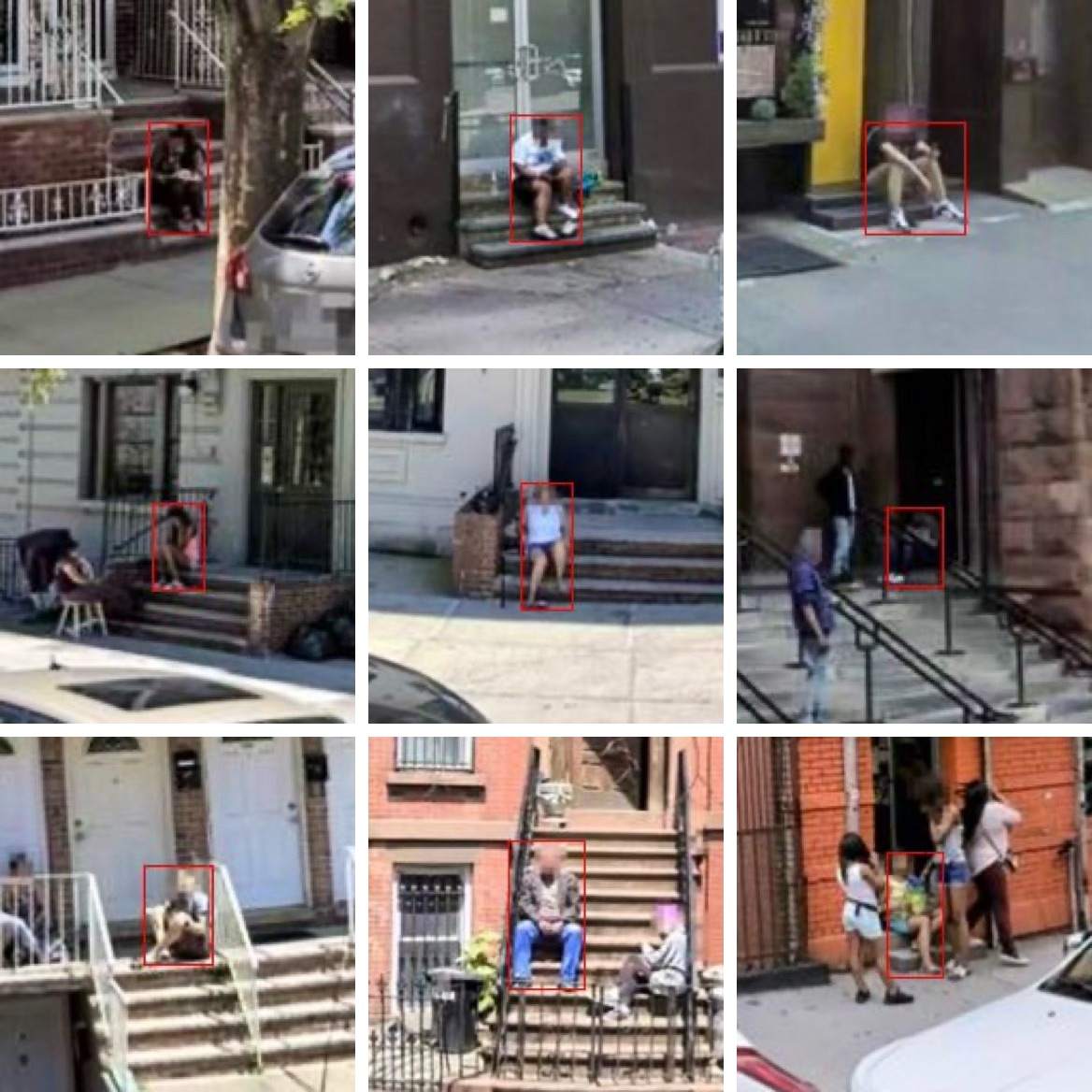}
    \caption{Step Sitting.}
    \label{fig:scf-step-sitting}
    \end{subfigure}
    \hfill
    \begin{subfigure}[t]{0.32\textwidth}
        \includegraphics[width=\linewidth]{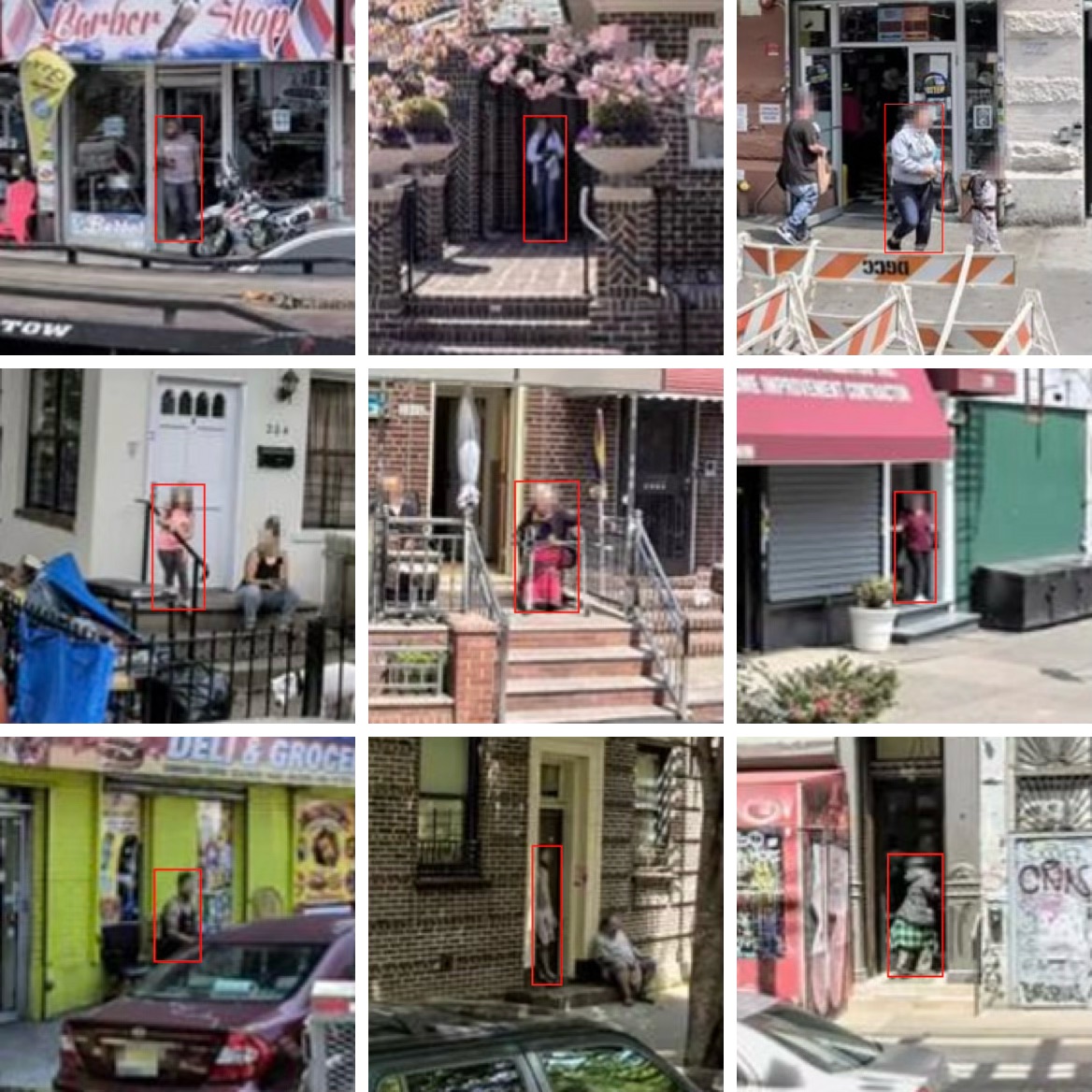}
        \caption{Entrance-area Activity.}
        \label{fig:scf-entrance-area-activity}
    \end{subfigure}
    \hfill
    \begin{subfigure}[t]{0.32\textwidth}
        \includegraphics[width=\linewidth]{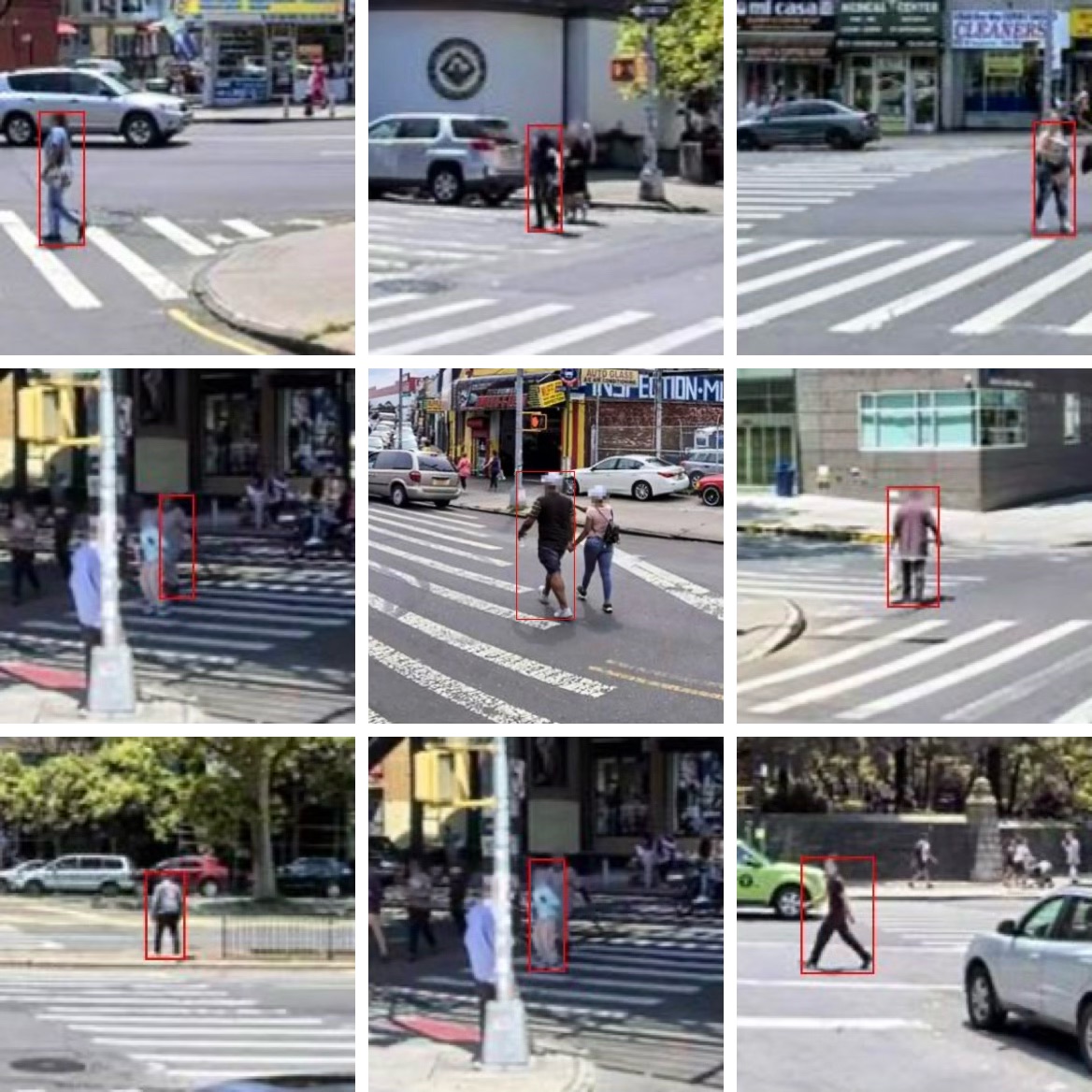}
        \caption{Street Crossing.}
        \label{fig:scf-street-crossing}
    \end{subfigure}
    \medskip
    \begin{subfigure}[t]{0.32\textwidth}
        \includegraphics[width=\linewidth]{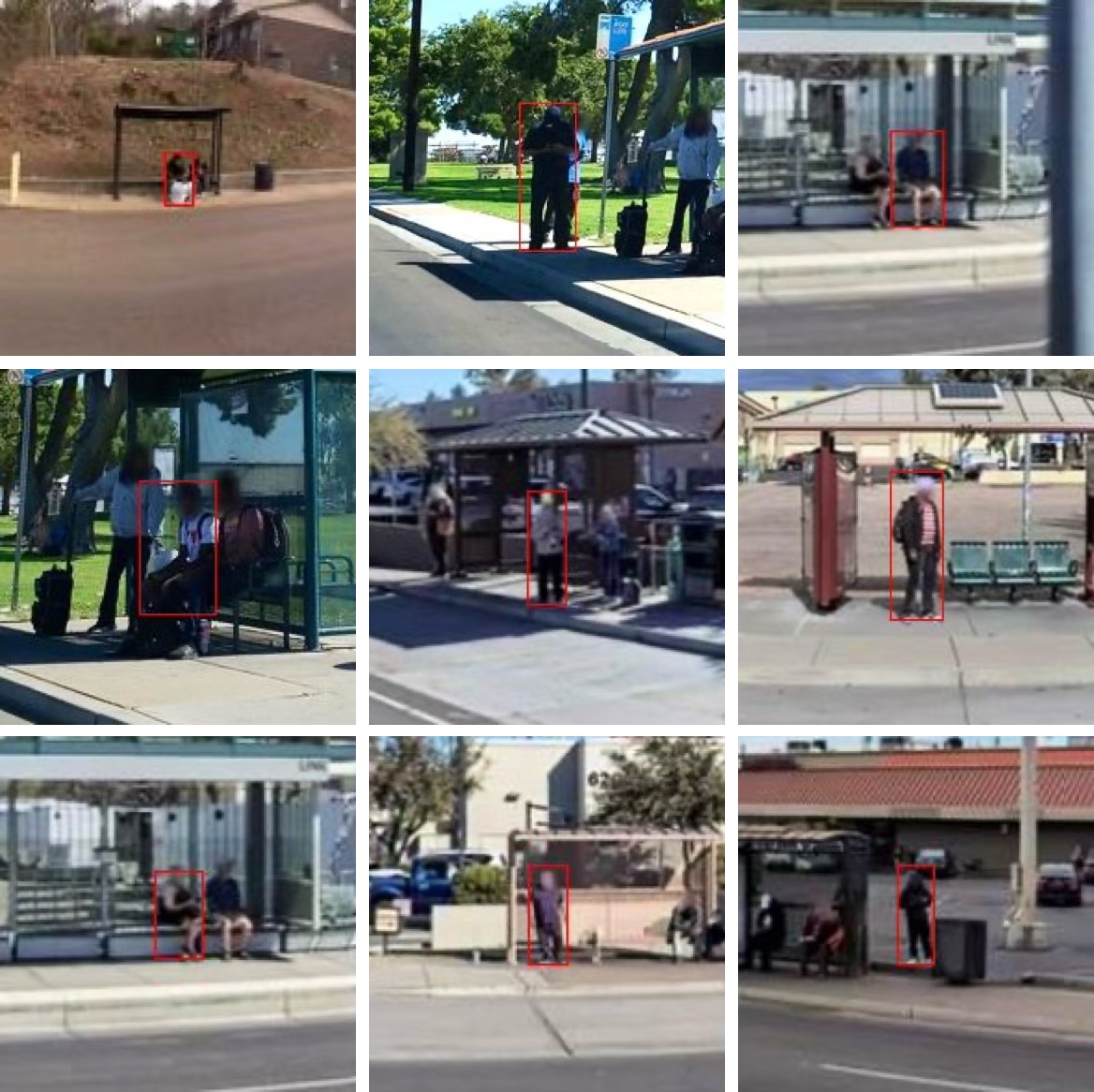}
        \caption{Waiting for Public Transport.}
        \label{fig:scf-waiting-for-public-transport}
    \end{subfigure}
    \hfill
    \begin{subfigure}[t]{0.32\textwidth}
        \includegraphics[width=\linewidth]{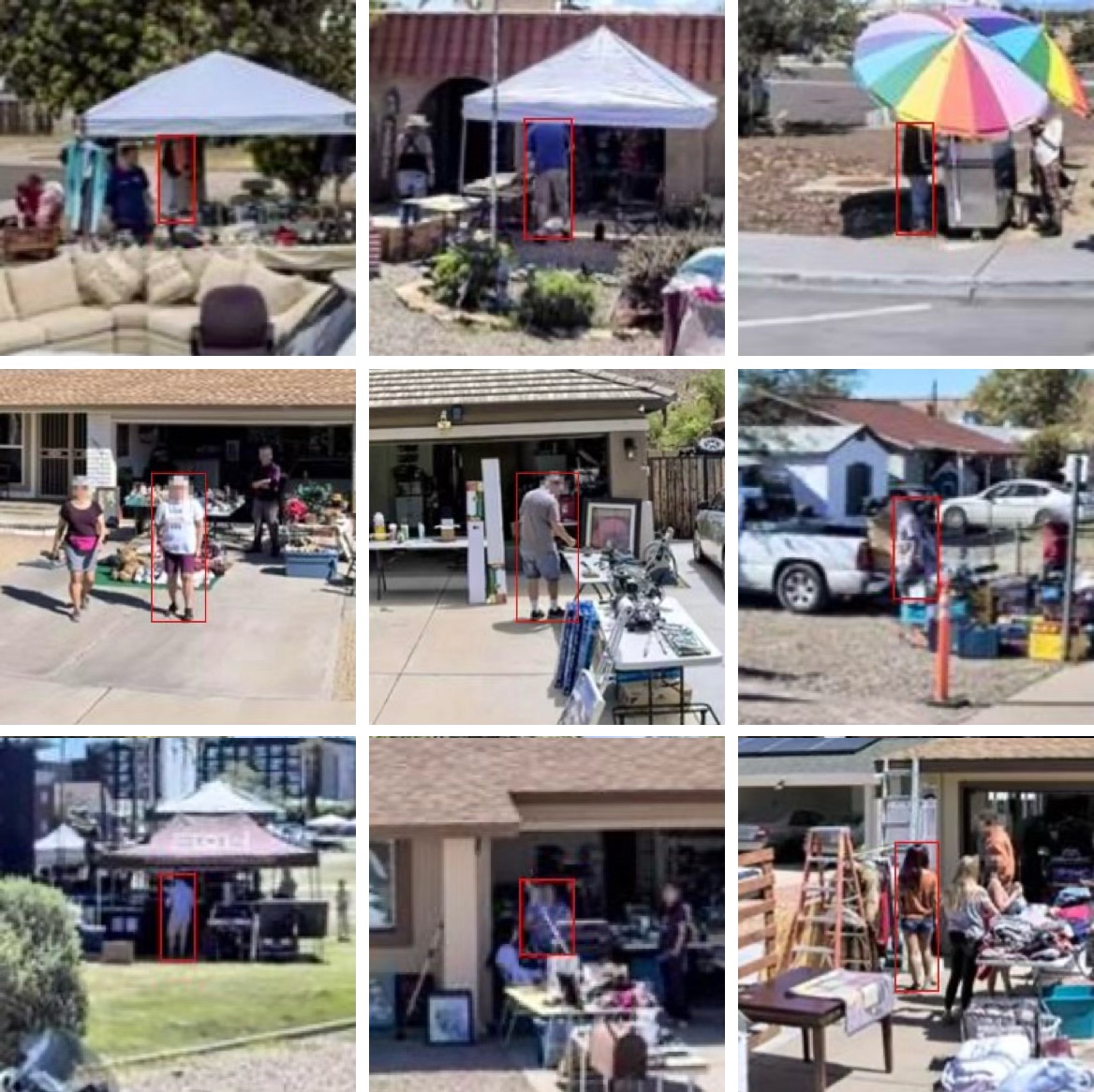}
        \caption{Vending.}
        \label{fig:scf-vending}
    \end{subfigure}
    \hfill
    \begin{subfigure}[t]{0.32\textwidth}
    \mbox{} 
    \end{subfigure}
    \caption{Representative street-level examples of the five Spatial Context Flag categories, arranged to match the detection conditions listed in Table~\ref{tab:scf}. Top row: step sitting, entrance-area activity, street crossing. Bottom row: waiting for public transport, vending. Each crop corresponds to a single bounding box detection passed to the VLM for attribute coding.}
    \label{fig:scf-examples}
\end{minipage}
\end{figure}

These flags exploit a particular strength of VLM-based coding. Identifying the spatial element a person occupies, whether a crosswalk, a building threshold, or a set of steps, requires reading the broader scene context around the person rather than the person's own body attributes alone. This is precisely the kind of holistic scene interpretation that vision-language models handle well, whereas earlier object-detection pipelines trained on bounding-box labels could not reliably produce. The \texttt{zone\_type} attribute coded by the VLM makes these spatial distinctions directly available as structured outputs, without requiring separate scene segmentation or spatial reasoning modules. Five flags are defined. Table~\ref{tab:scf} lists their detection conditions, and Figure~\ref{fig:scf-examples} shows a representative example of each.

\begin{table}[htbp]
\renewcommand{\arraystretch}{1}%
\setlength{\tabcolsep}{4pt}%
\scriptsize
\caption{Spatial Context Flag detection conditions. Flags are binary and independent; a person may satisfy more than one simultaneously. All flags require that the bounding box exceed the minimum-pixel-area threshold.}
\label{tab:scf}
\centering
\begin{tabular}{p{0.4cm} p{2.6cm} p{3.2cm} p{6.4cm}}
\toprule
\textbf{Pri.} & \textbf{Indicator} & \textbf{Variable} &
\textbf{Conditions} \\
\midrule
1 & Step Sitting
  & \texttt{is\_seated\_steps}
  & \texttt{posture} $=$ \texttt{seated}\newline
    \texttt{object\_relation} $\neq$ \texttt{riding}\newline
    \texttt{support\_surface} $=$ \texttt{steps} \\[4pt]
2 & Entrance-Area Activity
  & \texttt{is\_building\_entrance}
  & \texttt{zone\_type} $=$ \texttt{building\_entrance} \\[4pt]
3 & Street Crossing
  & \texttt{is\_crosswalk}
  & \texttt{zone\_type} $=$ \texttt{crosswalk} \\[4pt]
4 & Waiting for Public Transport
  & \texttt{is\_waiting}
  & \texttt{zone\_type} $=$ \texttt{bus\_stop}\newline
    \texttt{locomotion} $=$ \texttt{stationary}\newline
    \texttt{object\_relation} $\neq$ \texttt{riding}\newline
    \texttt{uniform\_presence} $=$ \texttt{none} \\[4pt]
5 & Vending
  & \texttt{is\_vending}
  & \texttt{zone\_type} $=$ \texttt{vendor\_stall}\newline
    \texttt{uniform\_presence} $=$ \texttt{none} \\
\bottomrule
\end{tabular}
\end{table}

\section{Mapping SDI and Activity Entropy in NYC}\label{sec:case-nyc}

We demonstrate the framework at city scale using 102,514 Apple Lookaround sideviews collected in New York City during midday (12:00--14:00) and evening (17:00--19:00) peak hours (see Appendix~\ref{ssec:data-coverage} for the spatial distribution of image locations).

\subsection{SDI and Pedestrian Volume}\label{sec:sdi-people} 


SDI is based on a simple premise: pedestrian volume alone does not capture the social character of sidewalk activity. High pedestrian counts can reflect two very different conditions. In one, \textit{n} people are gathered in groups or engaged in what Gehl has called ``optional'' and ``social'' activities. In the other, the same \textit{n} individuals are simply moving through a scene, without stopping or interacting. Raw pedestrian counts conflate the two situations, whereas SDI is intended to distinguish them.

Table~\ref{tab:sdi-count-correlations} provides a first empirical check of this distinction in the NYC corpus. We find that pedestrian count is only weakly correlated with SDI ($r = 0.168$), indicating that SDI cannot be reduced to a simple measure of crowd size. Count is somewhat more related to WGI ($r = 0.264$) than to WDI ($r = -0.007$), suggesting that busier scenes are slightly more likely to contain groups, but not necessarily stationary dwelling behavior. Meanwhile, WGI and WDI remain nearly orthogonal ($r = -0.017$), supporting their interpretation as distinct and complementary dimensions of sidewalk social activity. Taken together, these relationships suggest that SDI captures a social dimension of sidewalk use that pedestrian volume alone misses.


\begin{table}[h]
    \centering
    \footnotesize
    \caption{Pairwise Pearson correlations among pedestrian count and SDI components (NYC corpus, $n = 102{,}514$ sideviews).}\label{tab:sdi-count-correlations}
    \renewcommand{\arraystretch}{0.8} 
    {\setlength{\tabcolsep}{16pt}
    \begin{tabular}{lcccc}
        \toprule
         & Count & SDI & WGI & WDI \\
        \midrule
        Count & 1.000 & & & \\
        SDI   & 0.168 & 1.000 & & \\
        WGI   & 0.264 & 0.658 & 1.000 & \\
        WDI   & $-$0.007 & 0.696 & $-$0.017 & 1.000 \\
        \bottomrule
    \end{tabular}}
\end{table}

\begin{figure}[htbp]
\centering
\begin{minipage}{0.85\textwidth}
    \begin{subfigure}[t]{0.49\textwidth}
        \includegraphics[width=\linewidth]{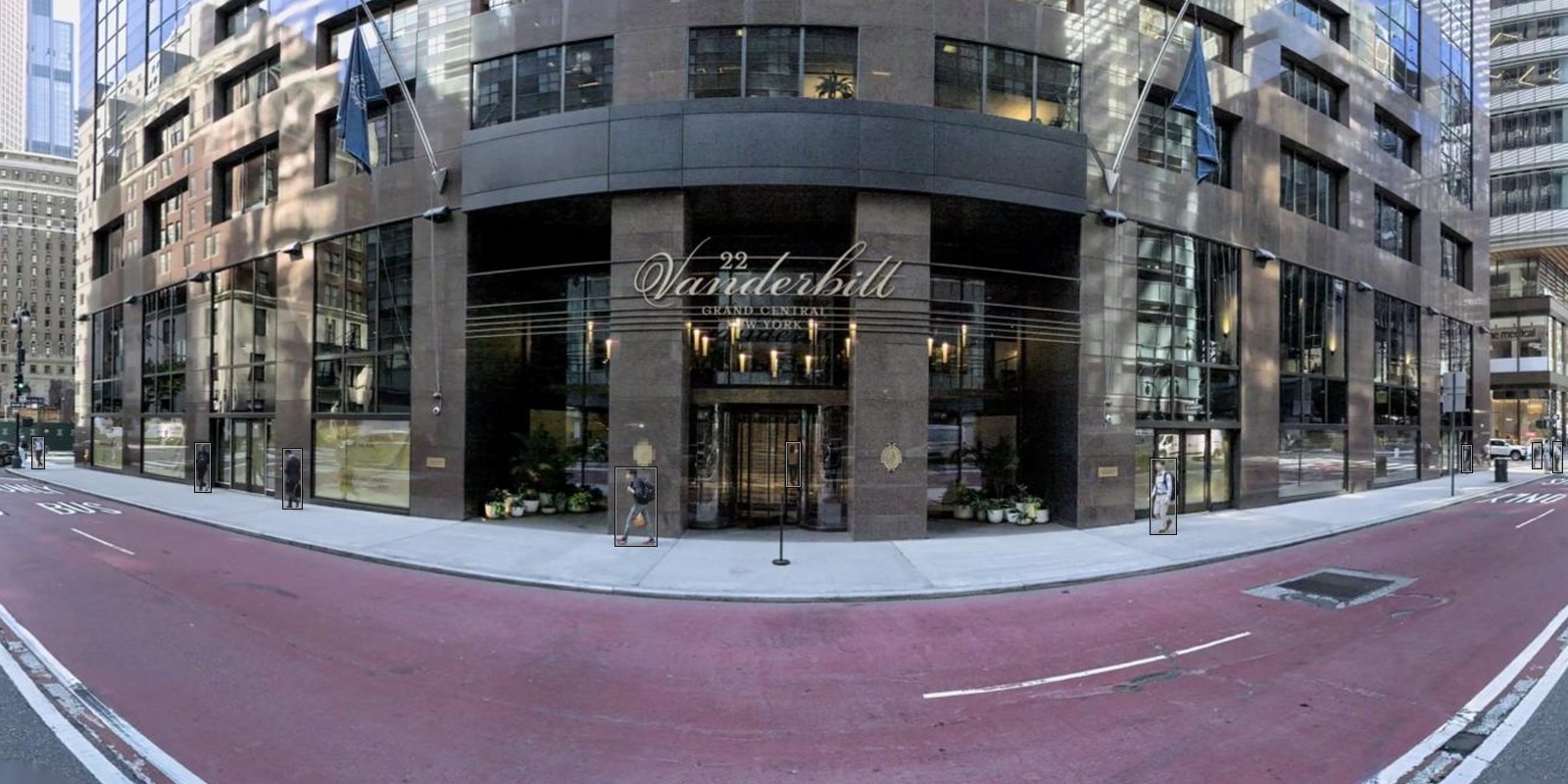}
        \caption{$N = 9$, WGI $= 1.00$, WDI $= 1.11$, SDI $= 1.11$. All nine people are solo pedestrians in active locomotion.}
        \label{fig:sdi-low-1}
    \end{subfigure}
    \hfill
    \begin{subfigure}[t]{0.49\textwidth}
        \includegraphics[width=\linewidth]{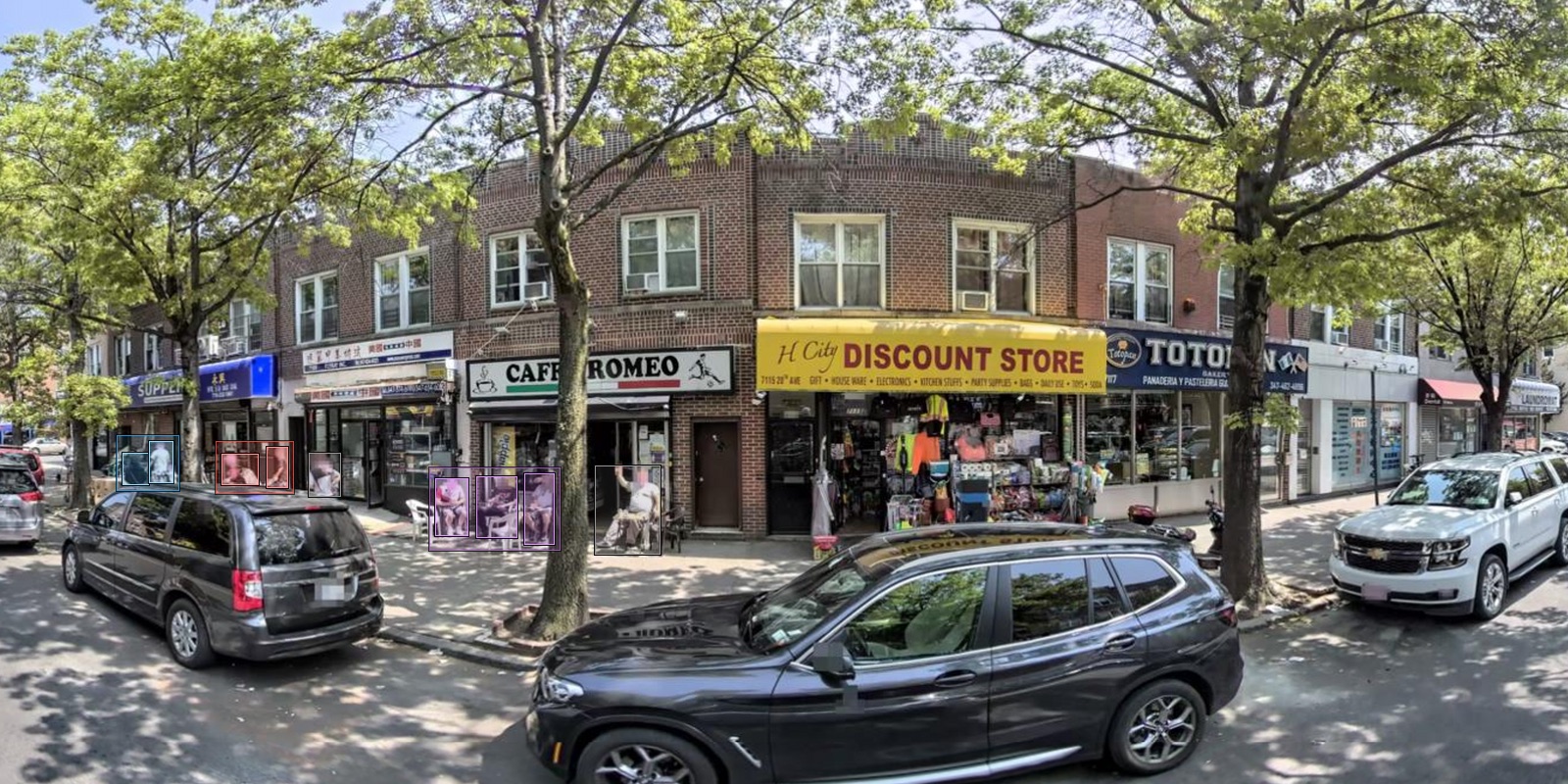}
        \caption{$N = 9$, WGI $= 2.11$, WDI $= 2.11$, SDI $= 4.89$. People are distributed across grouping tiers and include stationary and seated individuals.}
        \label{fig:sdi-high-1}
    \end{subfigure}

    \medskip

    \begin{subfigure}[t]{0.49\textwidth}
        \includegraphics[width=\linewidth]{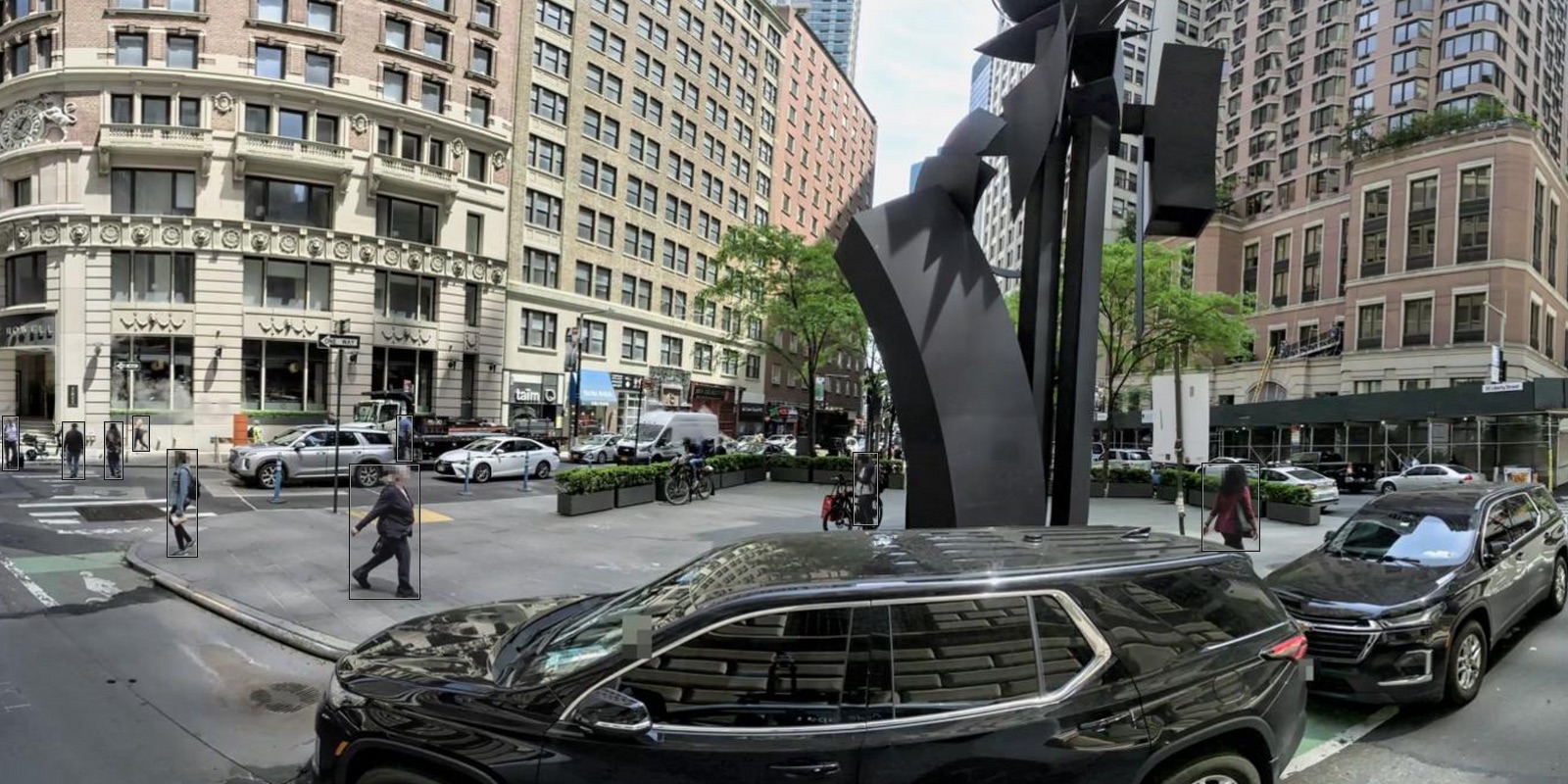}
        \caption{$N = 9$, WGI $= 1.00$, WDI $= 1.11$, SDI $= 1.11$. All nine people are solo pedestrians in active locomotion.}
        \label{fig:sdi-low-2}
    \end{subfigure}
    \hfill
    \begin{subfigure}[t]{0.49\textwidth}
        \includegraphics[width=\linewidth]{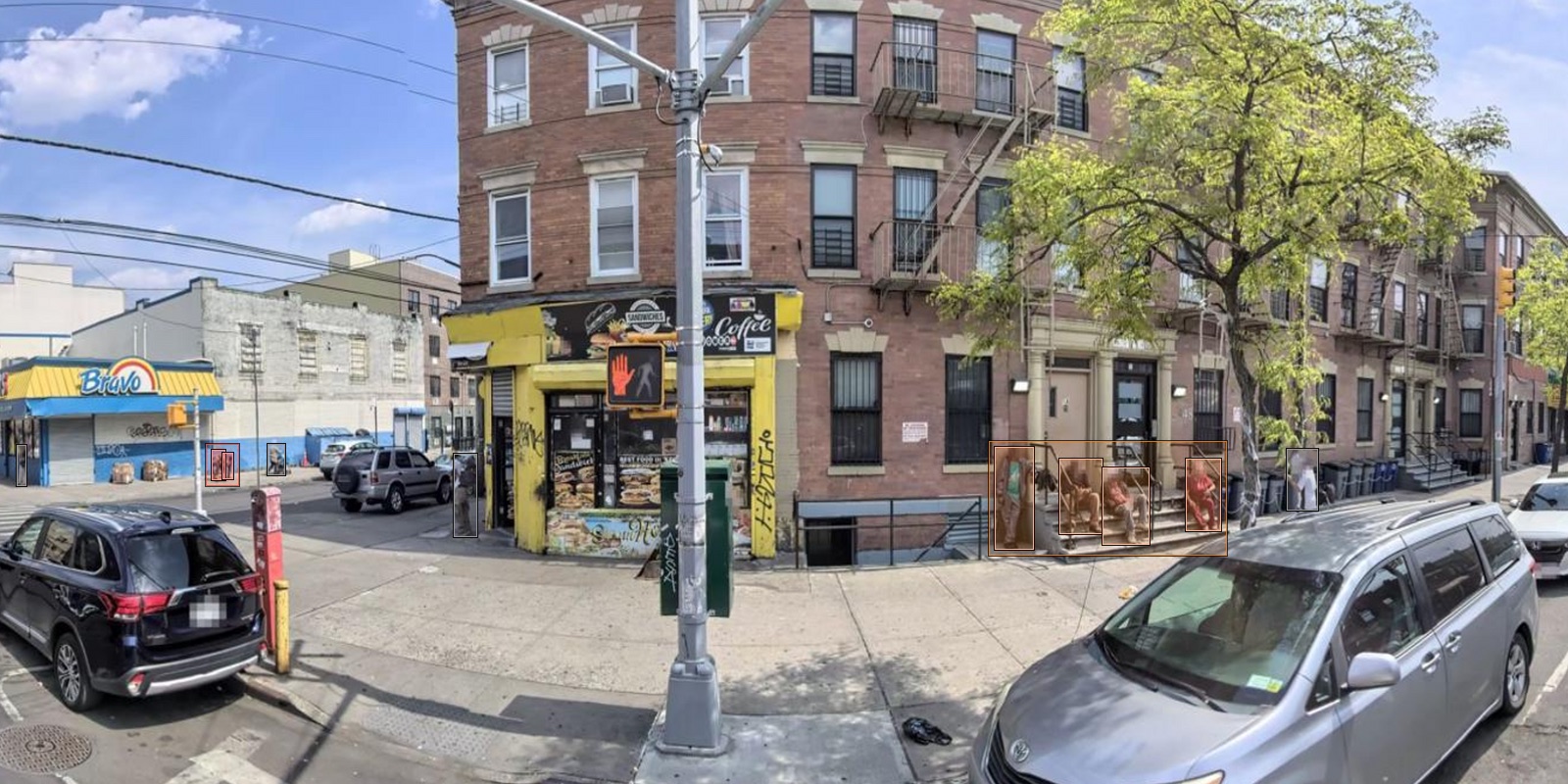}
        \caption{$N = 10$, WGI $= 2.00$, WDI $= 2.00$, SDI $= 4.40$. People include a group of four, a dyad, and four solo pedestrians.}
        \label{fig:sdi-high-2}
    \end{subfigure}

    \medskip

    \begin{subfigure}[t]{0.49\textwidth}
        \includegraphics[width=\linewidth]{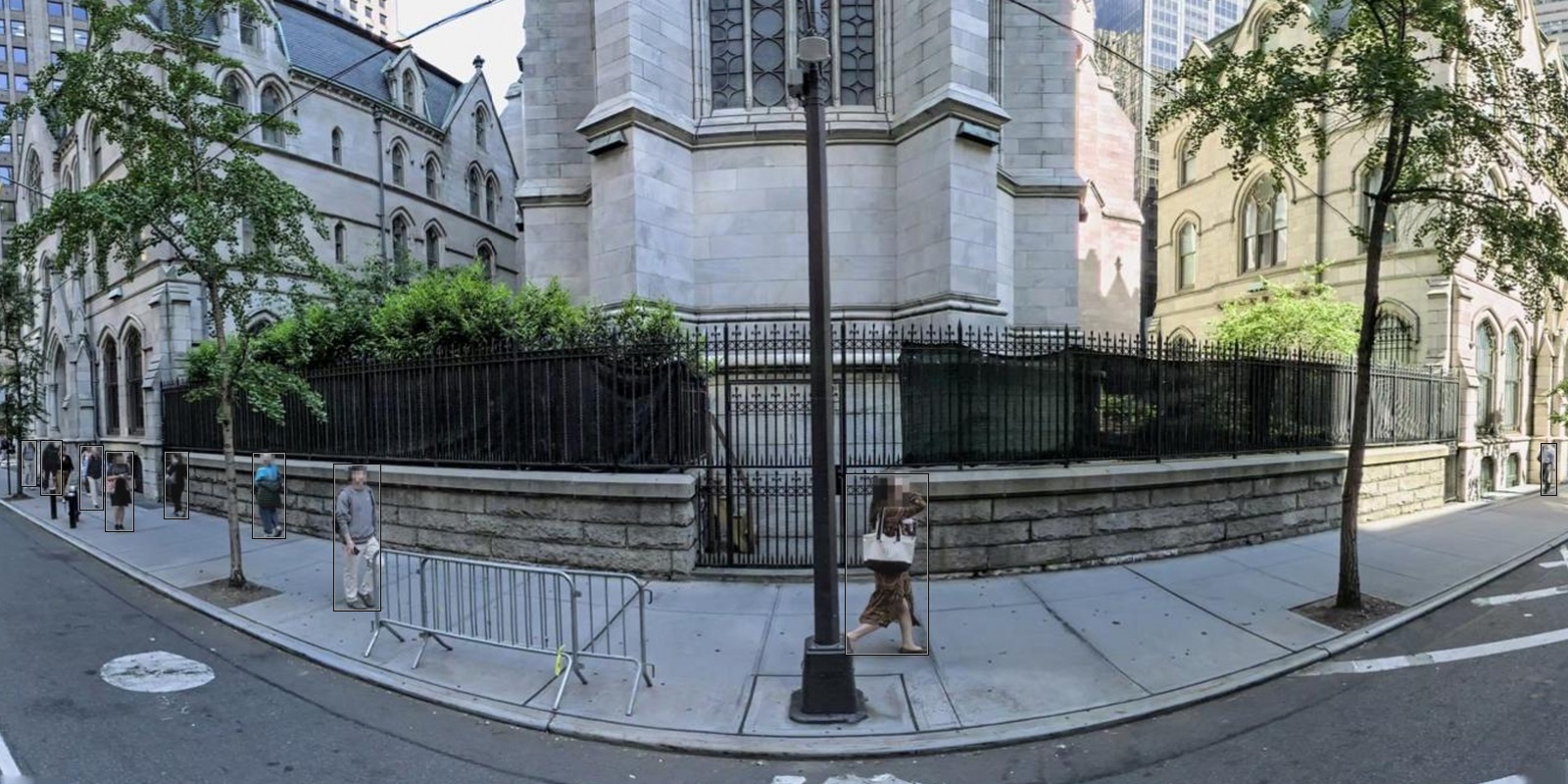}
        \caption{$N = 9$, WGI $= 1.00$, WDI $= 1.11$, SDI $= 1.11$. All nine people are solo pedestrians in active locomotion.}
        \label{fig:sdi-low-3}
    \end{subfigure}
    \hfill
    \begin{subfigure}[t]{0.49\textwidth}
        \includegraphics[width=\linewidth]{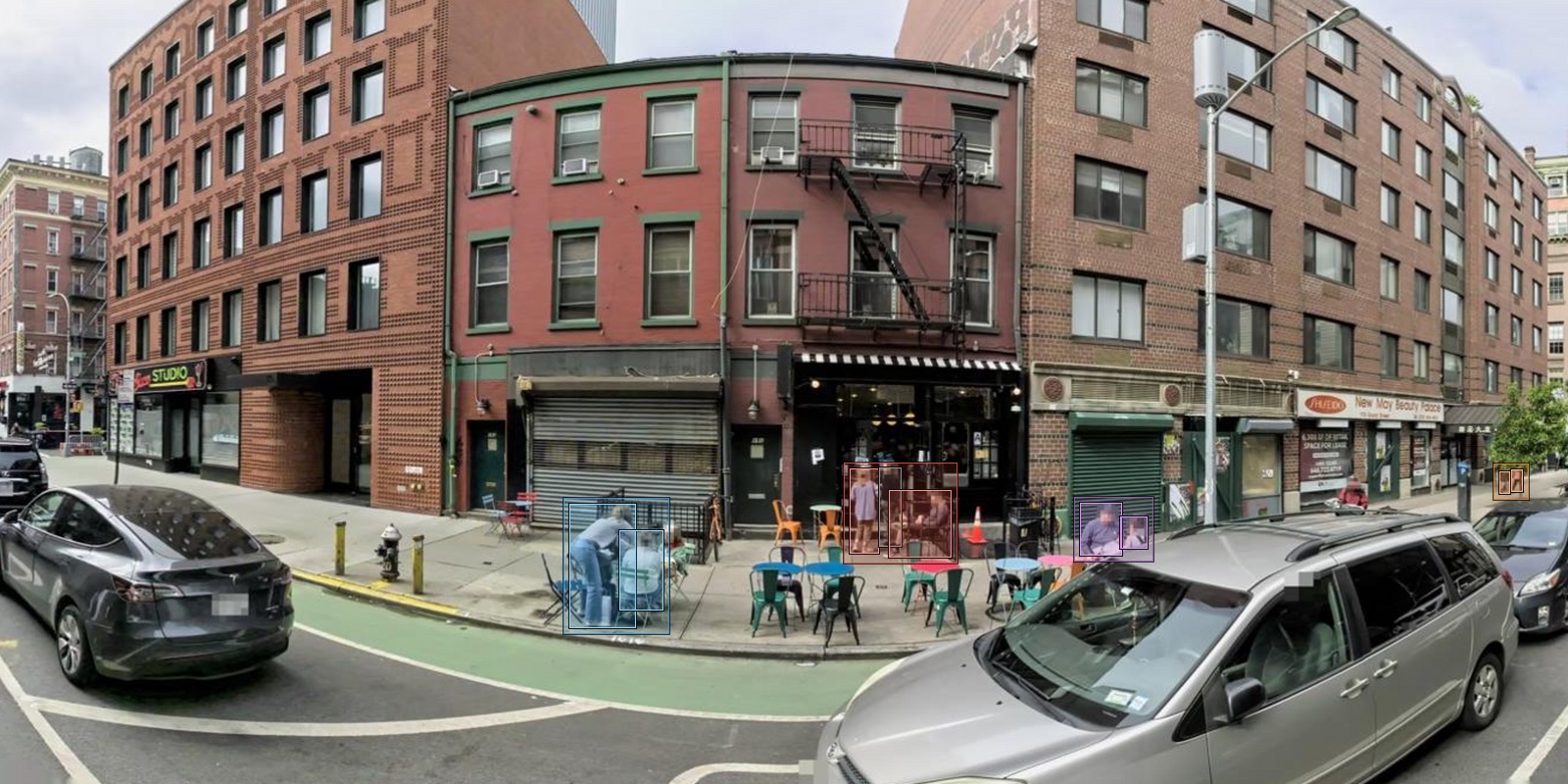}
        \caption{$N = 10$, WGI $= 1.89$, WDI $= 2.22$, SDI $= 4.22$. People form multiple dyads; most are stationary.}
        \label{fig:sdi-high-3}
    \end{subfigure}
    \caption{Three pairs of sideview images matched on pedestrian count ($N \approx 9$--10) but with contrasting SDI values. Left column: low-SDI scenes in which all people are solo pedestrians in active locomotion (SDI $= 1.11$). Right column: high-SDI scenes of similar headcount containing grouped and dwelling individuals (SDI $= 4.22$--$4.89$). The consistent fourfold difference illustrates that pedestrian volume alone does not capture the social character of sidewalk activity.}
    \label{fig:people-count-sdi}
\end{minipage}
\end{figure}

This decoupling between pedestrian count and SDI is visually illustrated in Fig.~\ref{fig:people-count-sdi}, which presents three pairs of street-level images matched on pedestrian count ($N \approx 9$--10) but with different SDI values. In each pair, the low-SDI image contains solo pedestrians in active locomotion (SDI $= 1.11$), whereas the high-SDI image of similar headcount contains grouped and dwelling individuals (SDI $= 4.22$--$4.89$). The roughly fourfold difference in SDI at near-constant $N$ demonstrates that pedestrian volume alone is an inadequate proxy for the social character of sidewalk activity. Segments with high pedestrian counts but low SDI, exemplified by the left column of Fig.~\ref{fig:people-count-sdi}, are locations that are dense in bodies but thin in observed sidewalk sociability.

\subsection{Mapping Decoupling Between SDI and Pedestrian Count in NYC}\label{ssec:nyc-decoupling}

\begin{figure}[htbp]
\centering
\begin{minipage}{0.85\textwidth}
    \includegraphics[width=1\columnwidth]{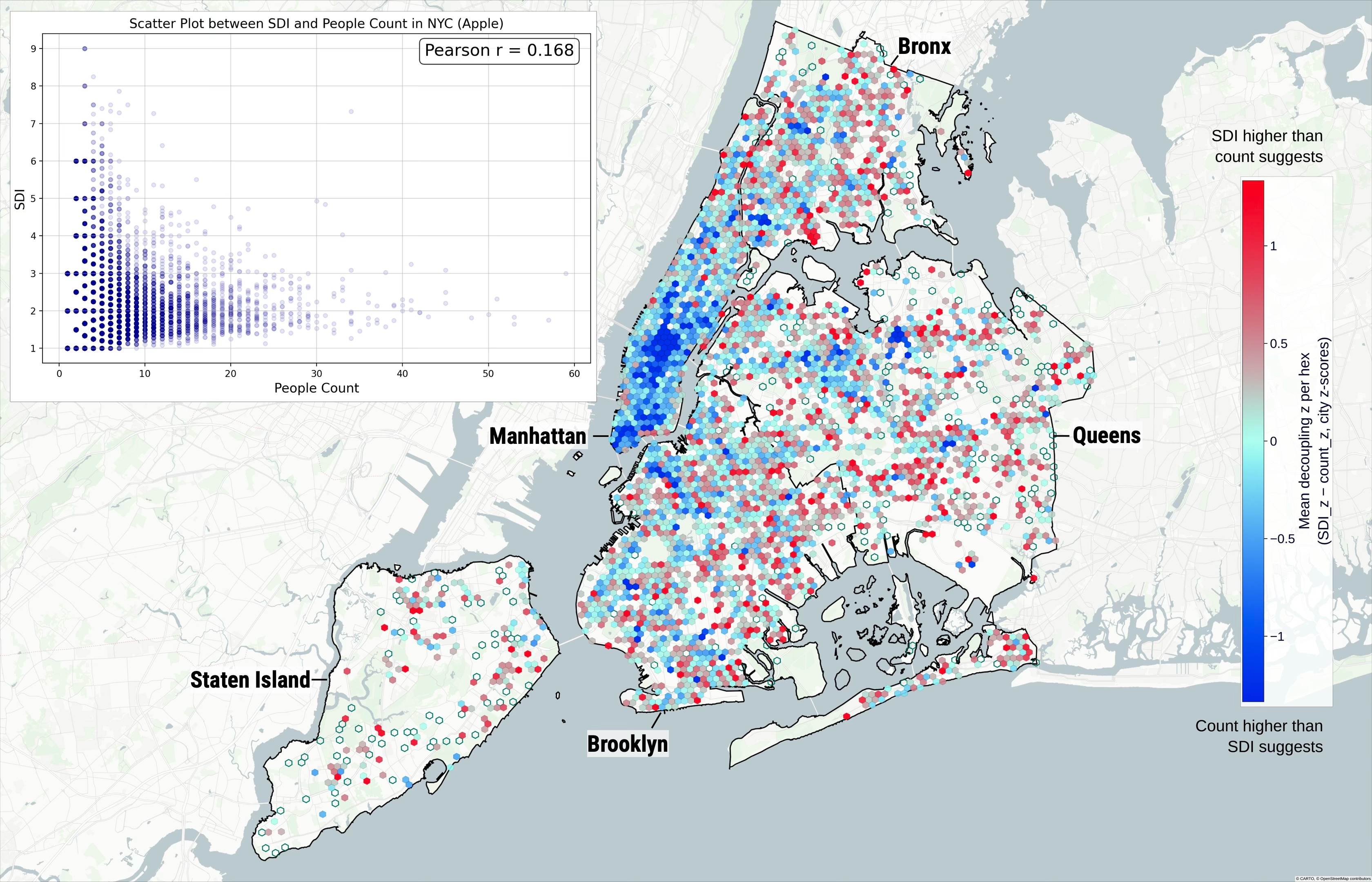}
    \caption{Spatial distribution of decoupling between SDI and pedestrian count across NYC (Apple Lookaround, $n = 102{,}514$     sideviews). Each image is assigned to an H3 hexagonal cell; cell color shows the mean of $z_{\text{SDI}} - z_{\text{count}}$, where both z-scores are computed city-wide within the same image batch. Red cells indicate locations where SDI exceeds the pedestrian volume; blue cells indicate the reverse. The colorscale is symmetrically clipped at the 2nd and 98th percentiles. The inset scatter plot shows the relationship between image-level SDI and person count (Pearson $r = 0.168$). Hollow circles indicate hexes below the minimum sample threshold.}
    \label{fig:nyc-decoupling-map}
\end{minipage}
\end{figure}

Having established the statistical decoupling at the image level, we next examine whether it is spatially structured. Figure~\ref{fig:nyc-decoupling-map} maps this decoupling across the full NYC sample. For each image, both \texttt{person\_count} and \texttt{sdi} are standardized relative to the city-wide mean and standard deviation of the same image batch; the decoupling score is then defined as $z_{\text{SDI}} - z_{\text{count}}$, where a positive value indicates that social activity, as measured by SDI, exceeds what pedestrian volume alone would predict, and a negative value indicates the reverse. Scores are aggregated to H3 hexagonal cells by arithmetic mean. The colorscale is symmetrically clipped at the 2nd and 98th percentiles, with zero mapped to the neutral midpoint. The inset scatter plot shows the image-level relationship between SDI and person count across the full NYC sample, with a Pearson correlation of 0.168, confirming that the two measures are largely independent. The map shows that the divergence between pedestrian volume and SDI is not randomly distributed but forms visible spatial clusters, with high-SDI, low-count areas concentrated in northern Manhattan (Harlem) and selected neighborhoods throughout other boroughs. High-count, low-SDI areas are found in major transit corridors, particularly in the bustling, employment-heavy areas of midtown and downtown Manhattan. This pattern suggests that pedestrian volume and sidewalk social activity capture related but distinct aspects of the street environment. Social and dwelling activities are more common per person in more residential neighborhoods throughout the city.

\subsection{Mapping Activity Entropy in NYC}\label{ssec:nyc-entropy}

\begin{figure}[htbp]
\centering
\begin{minipage}{0.85\textwidth}
    \includegraphics[width=1\columnwidth]{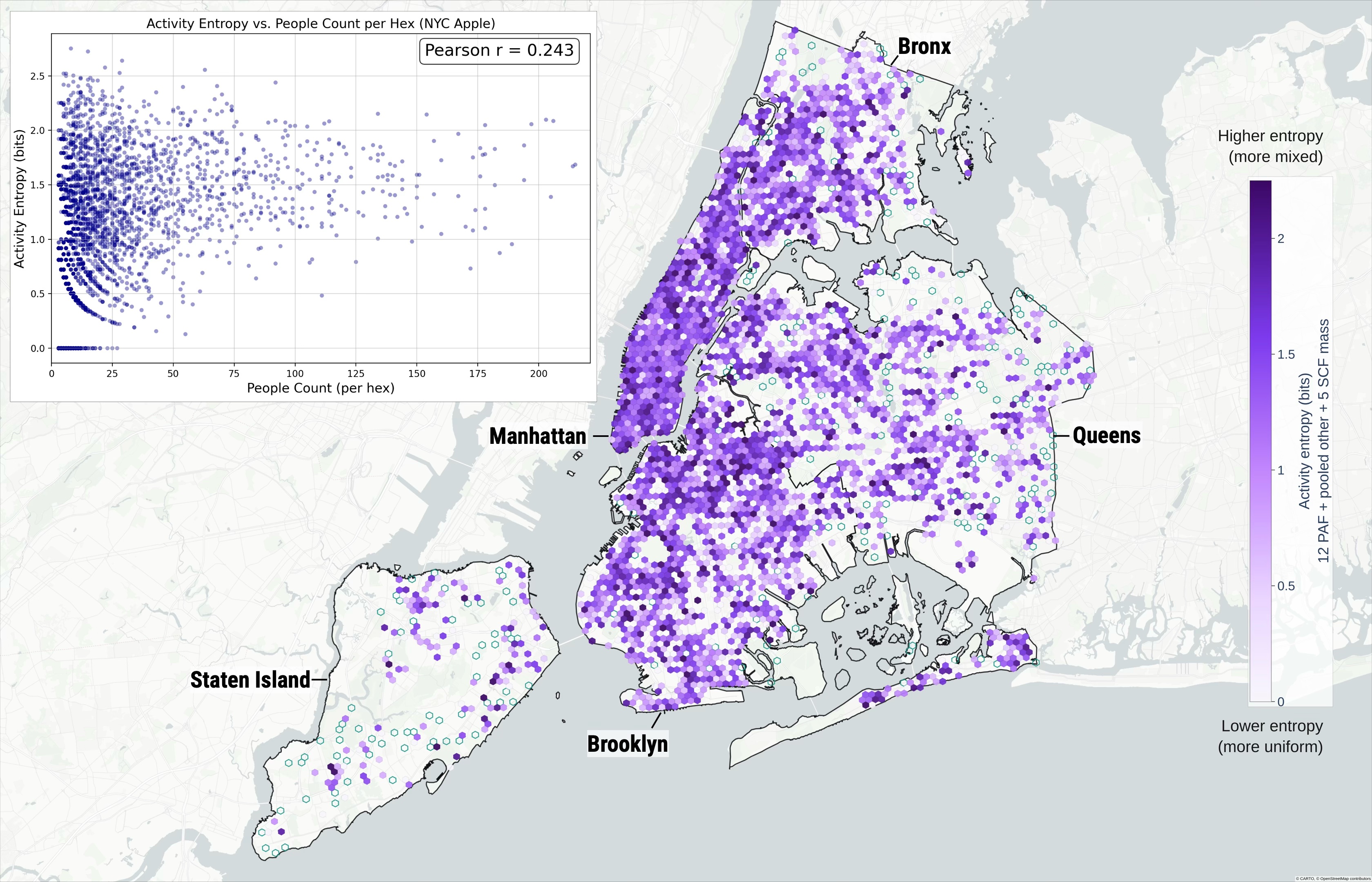}
    \caption{Spatial distribution of activity entropy across NYC (Apple Lookaround, $n = 102{,}514$ sideviews). For each H3 hexagonal cell, detected persons are aggregated into an 18-bin mass vector comprising 12 Primary Activity Flag categories, one pooled residual (\texttt{other}), and 5 Spatial Context Flag counts; Shannon entropy (bits) is computed over the normalized vector. SCF flags are additive and may co-occur with PAF labels for the same person. The colorscale runs from light (lower entropy, more uniform activity mix) to dark purple (higher entropy, more mixed). Hollow circles indicate hexes below the minimum sample threshold. The inset scatter plot shows the hex-level relationship between pedestrian count and activity entropy (Pearson $r = 0.24$); each point represents one H3 hexagonal cell.}
    \label{fig:nyc-entropy-map}
\end{minipage}
\end{figure}

The activity flag system produces not only individual-level labels but also an aggregate measure of behavioral diversity at the neighborhood scale. Figure~\ref{fig:nyc-entropy-map} maps Shannon entropy computed from the joint distribution of Primary Activity Flags and Spatial Context Flags across the same 102,514 NYC sideviews. For each H3 hexagonal cell, all detected persons are pooled into an 18-bin mass vector: 12 Primary Activity Flag categories, one pooled residual category for persons assigned \texttt{other}, and 5 Spatial Context Flag counts. Because SCF flags are not mutually exclusive with PAF labels, a single person may contribute to both a PAF bin and one or more SCF bins simultaneously. Shannon entropy is then computed over the normalized mass vector in bits (see Appendix~\ref{sec:app-entropy} for the formal specification). Higher entropy indicates a more even distribution of activity types within a cell; lower entropy indicates dominance by a single activity type.

The map reveals a pronounced center--periphery gradient. Overall, Manhattan and the inner portions of Brooklyn and Queens show consistently higher entropy, reflecting the co-presence of multiple activity types within the same street environment. However, considerable variation can be seen within Manhattan — higher entropy in the Upper East Side than in the Upper West Side around Central Park, for instance. Peripheral areas, particularly Staten Island and eastern Queens, have fewer observation points but also show a range of activity entropy values. These spatial patterns suggest that activity diversity, as measured by the flag system, captures a dimension of street life that is geographically structured and not reducible to pedestrian volume alone.

\section{Discussion}\label{sec:discussion}

\paragraph{Limitations}

Several limitations should be noted. First, the framework operates on static street-level imagery and therefore captures sidewalk conditions at a single moment in time rather than continuous social activity. The resulting indicators should be interpreted as observational snapshots rather than direct measures of duration, frequency, or temporal dynamics.

Second, all indicators inherit uncertainty from the underlying VLM-based observable query process. Although the held-out evaluation demonstrates strong overall reliability, some observable dimensions and derived activity categories remain more difficult to distinguish under conditions of occlusion, small person scale, motion blur, or image degradation. These uncertainties propagate into the composite indicators and should be interpreted accordingly.

Third, image-level activity composition is inherently sparse in many street scenes because a substantial share of sideviews contain only a small number of detected pedestrians. As a result, fine-grained activity distributions at the single-image level may be unstable, whereas aggregation across street segments or larger spatial units produces more robust patterns.

More broadly, the framework measures observable sidewalk presence rather than internal states, motivations, or social meaning. The indicators should therefore be interpreted as observational measures derived from visible street activity, not as direct measurements of social experience itself.

\paragraph{Directions for future research}

The social measurement system introduced here can be extended in several directions. First, joint distributions of activity types and demographic signals (whether certain activities are systematically associated with particular age compositions, or whether activity diversity correlates with demographic mixing) are addressable with the existing data structure but were not examined here. Second, the \texttt{zone\_type} field provides a direct link between behavioral coding and immediate spatial context; a systematic analysis of how activity profiles vary across zone types would constitute a distinct contribution, particularly once missing categories such as sports courts are added. Third, the sideview structure (each panorama yields two images corresponding to opposite sidewalk faces) enables street-side comparisons not possible with count data, raising questions about whether and under what urban-form conditions the two sides of the same street differ in social activity. Fourth, applying the framework across repeated time windows at the same locations would enable examination of how the indices vary by time of day, day of week, and season, and identification of locations where social activity is structurally stable versus situationally contingent.

The broader implication is methodological. The observational tradition of Whyte, Gehl, and Jacobs produced the conceptual vocabulary for understanding sidewalk social life, but the methods it relied on were necessarily local and labor-intensive. This work produces a different kind of evidence, drawn from the same visual record of the street environment at a scale that fieldwork cannot reach, and whether the two forms of evidence converge or diverge on specific questions is itself a productive research question that this approach makes tractable.
\section{Conclusion}\label{sec:conclusion}

This paper has presented a framework for extracting social indicators from street-level imagery, replacing pedestrian counting with observational measurement of how people use sidewalks. The work makes three contributions.

First, building on MINGLE \citep{liu2026mingle}, an integrated sidewalk social detection framework combines revised social group detection with per-person observable query using vision-language models, operating on timestamped sidewalk-facing sideviews that enable temporal stratification across time windows unavailable to imagery sources without capture timestamps.

Second, a benchmarked activity detection system identifies and resolves a systematic failure mode in applying VLMs to urban behavioral coding. When prompted with high-level social categories, VLMs conflate directly observable physical states with contextual inferences, producing cascading misclassifications. A redesigned taxonomy of ten independent observational dimensions, each restricted to what is directly visible in the image, eliminates this failure mode; a comparative benchmark across six models identifies Qwen3-VL-30B-A3B-Instruct as the most suitable engine for city-scale deployment.

Third, the social indicator system translates VLM-coded observable dimensions into three complementary outputs: the Social Dwelling Index, which captures grouping and dwelling intensity jointly at the image level; binary Vulnerable Street User Flags for accessibility-sensitive populations; and Primary Activity and Spatial Context Flags that document behavioral and spatial diversity.

Applied to 102,514 sideviews in New York City, the framework demonstrates that pedestrian volume and SDI are weakly associated ($r = 0.168$), and that activity entropy is similarly decoupled from foot traffic ($r = 0.24$). Streets with high pedestrian counts are not where people most often linger, gather, or exhibit the greatest behavioral diversity. The spatial distribution of this decoupling identifies a class of street segments that pedestrian counts alone cannot distinguish: locations that are dense in bodies but thin in social life. These are the muted areas that the observational traditions of Whyte, Gehl, and Jacobs described qualitatively, and that prior observational traditions also described qualitatively, but that can now be examined computationally at the city scale.

\section*{Acknowledgements}
This research was supported by the Sagalyn-Hack Dissertation Research Grant at the Massachusetts Institute of Technology Department of Urban Studies and Planning. The authors also acknowledge the MIT--Hasso Plattner Institute Research Partnership. This work used the Jetstream2 GPU and NCSA Delta GPU resources through allocation CIS250088 from the Advanced Cyberinfrastructure Coordination Ecosystem: Services \& Support (ACCESS) program, supported by the U.S. National Science Foundation (NSF) under grants \#2138259, \#2138286, \#2138307, \#2137603, and \#2138296.

\section*{Ethical Statement}
This study did not involve human subjects research requiring institutional review board approval. The analysis used street-level imagery obtained from third-party providers; identifiable facial information was blurred by the providers prior to access. Images were used solely for observational analysis of public-space activity, in compliance with the terms of service of each imagery platform.

\section*{Declaration of Competing Interests}

The authors declare that they have no known competing financial interests
or personal relationships that could have appeared to influence the work
reported in this paper.

\section*{Data Availability}
The framework described in this study is independent of any specific imagery dataset and can be applied to street-level imagery from any source that provides panoramic views with capture timestamps. The source imagery used in this study is owned by third parties and is subject to their respective terms of service. Apple Lookaround, Bing Streetside, and Mapillary imagery were accessed programmatically through the streetlevel library and can be retrieved by other researchers using the same procedure described in Appendix~\ref{sec:data-processing}. The authors are not authorized to redistribute the source imagery. The derived data tables for the New York City case study, including person-level bounding boxes, ten-dimensional activity codes, group assignments, image-level WGI/WDI/SDI values, and binary flag outputs across 102,514 sideviews, will be deposited in a public repository upon publication. Additional intermediate data, including model outputs and annotation files used for VLM benchmarking, are available from the corresponding author upon reasonable request.

\section*{Code Availability}
The code used to generate the results in this study is available from the corresponding author upon reasonable request. A public repository for the framework implementation is being prepared and will be made available upon publication.

\section*{Author Contributions}
L.L. conceived the study, designed and implemented the framework, performed the analysis, and drafted the manuscript. A.S. contributed to the conceptual development of the study, supervised the research, and revised the manuscript. Both authors interpreted the results and approved the final manuscript.

\bibliographystyle{elsarticle-harv}
\bibliography{egbib}

\clearpage

\appendix 

\titleformat{\section}
  {\normalfont\large\bfseries}
  {\thesection}
  {1em}
  {}
  
\renewcommand{\thesection}{A\arabic{section}}
\setcounter{section}{0}
\renewcommand{\thefigure}{A\arabic{figure}} 
\setcounter{figure}{0}
\renewcommand{\thetable}{A\arabic{table}}
\setcounter{table}{0}

\newpage
\noindent {\Large \textbf{APPENDIX}} \\
\noindent {\large \myTitle}

\vspace{1em}

\section{Sideview Extraction Pipeline}\label{sec:data-processing}

\subsection{Source Requirements and Sampling} \label{ssec:data-source}

The framework requires timestamped panoramic street-level imagery. Google Street View, the most widely used source in prior urban research, does not expose capture timestamps in its API and is therefore unsuitable for temporal stratification. Pedestrian activity is highly sensitive to time-of-day and day-of-week rhythms: whether people are lingering or in transit, and the social character of a space, differ substantially across typical work, social, and recreational schedules. Apple Lookaround is used as the illustrative example throughout this appendix; other platforms that provide capture timestamps (including Bing Streetside and Mapillary) are similarly compatible with the pipeline described below.

Images are sampled during four periods aligned with peak travel and social activity \citep{USBureauofLaborStatistics2022}: weekday midday (12:00--14:00), weekday evening (17:00--19:00), and the same windows on weekends. This temporal structure targets the most likely activity periods and enables comparisons that go beyond static pedestrian counts.

\subsection{Image Availability and Spatial Coverage}\label{ssec:data-coverage}

\begin{figure}[htbp]
    \centering
    \includegraphics[width=0.85\linewidth]{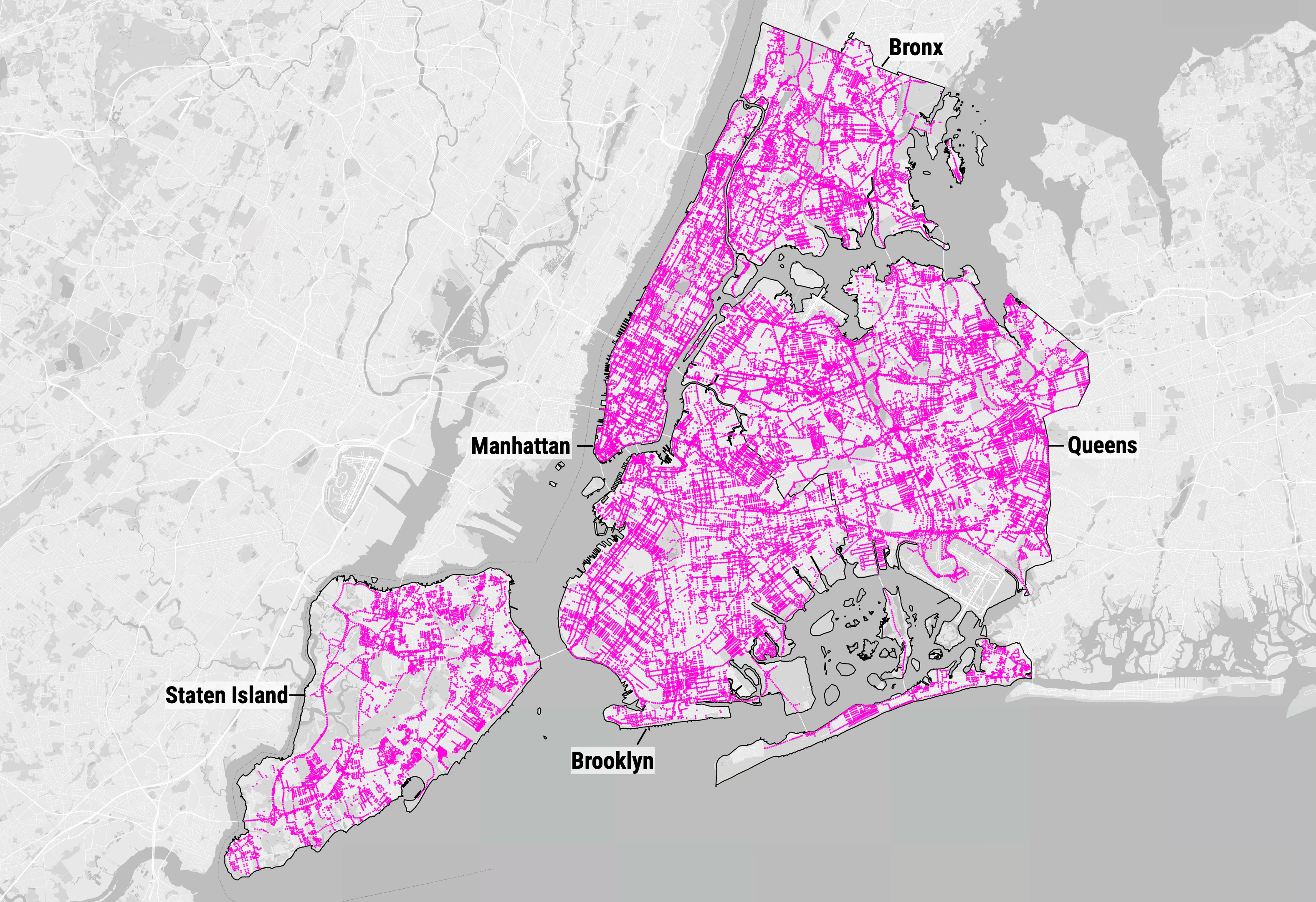}
    \caption{Spatial distribution of Apple Lookaround sideview locations across New York City ($n = 102{,}514$). Each point represents one sideview location.}
    \label{fig:sideview-coverage-nyc}
\end{figure}

Figure~\ref{fig:sideview-coverage-nyc} shows the spatial distribution of all 102,514 Apple Lookaround sideview locations used in the NYC case study, collected during midday and evening peak hours. Because sampling is restricted to these two time windows, coverage follows the street network but is not exhaustive and remains somewhat patchy across the city.

\subsection{Sideview Projection} \label{ssec:data-sideview}

Raw panoramic imagery (360$^\circ$) contains substantial noise (sky, road surfaces, and building rooftops) that can degrade VLM performance. To focus on the pedestrian zone, each panorama is reprojected into two rectilinear sideview images (left and right) facing the sidewalks.

For Apple Lookaround, raw imagery is delivered as six-face tiles. We use the \texttt{streetlevel} library \citep{streetlevel} to assemble these tiles into a unified equirectangular panorama before reprojecting to a perspective view. Other sources follow an analogous procedure with their respective native formats.

\begin{figure}[htbp]
    \centering
    \includegraphics[width=\linewidth]{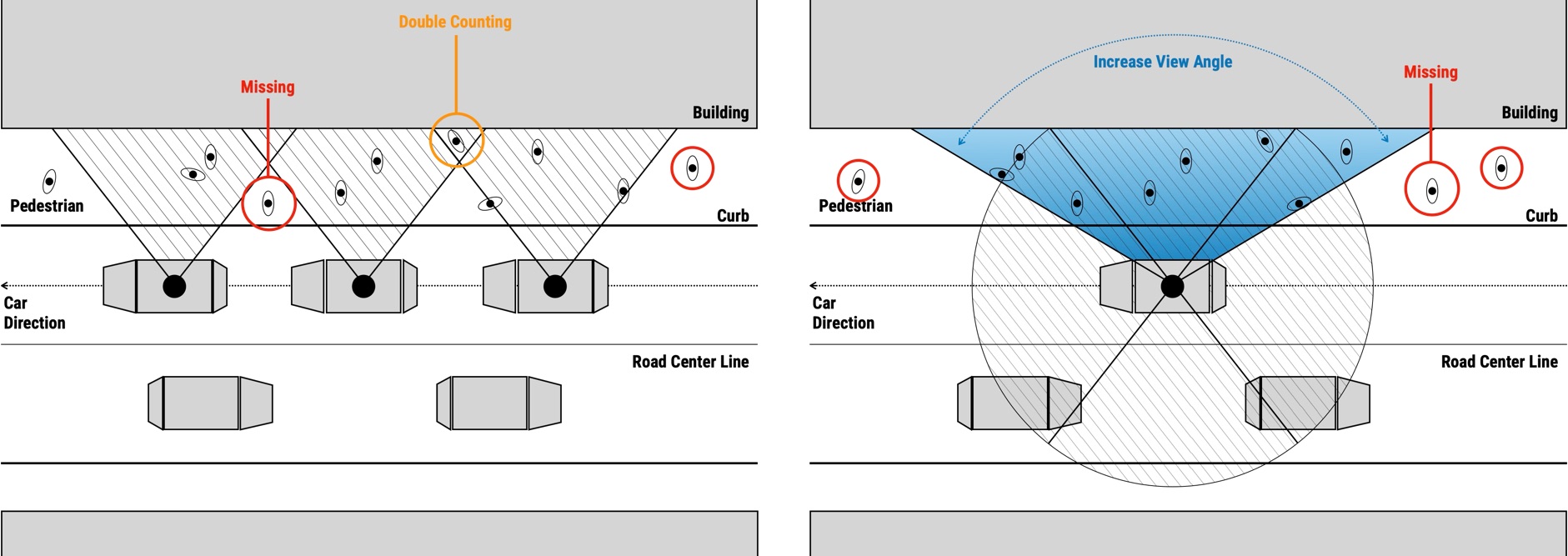}
    \caption{Sideview coverage geometry. The left panel shows the limited spatial coverage and potential overlap at frame boundaries under a narrow field of view. The right panel shows how a 150$^\circ$ FOV expands coverage to the full pedestrian zone on each side without boundary overlap.}
    \label{fig:sideview-coverage}
\end{figure}

\begin{figure}[htbp]
    \centering
    \includegraphics[width=\linewidth]{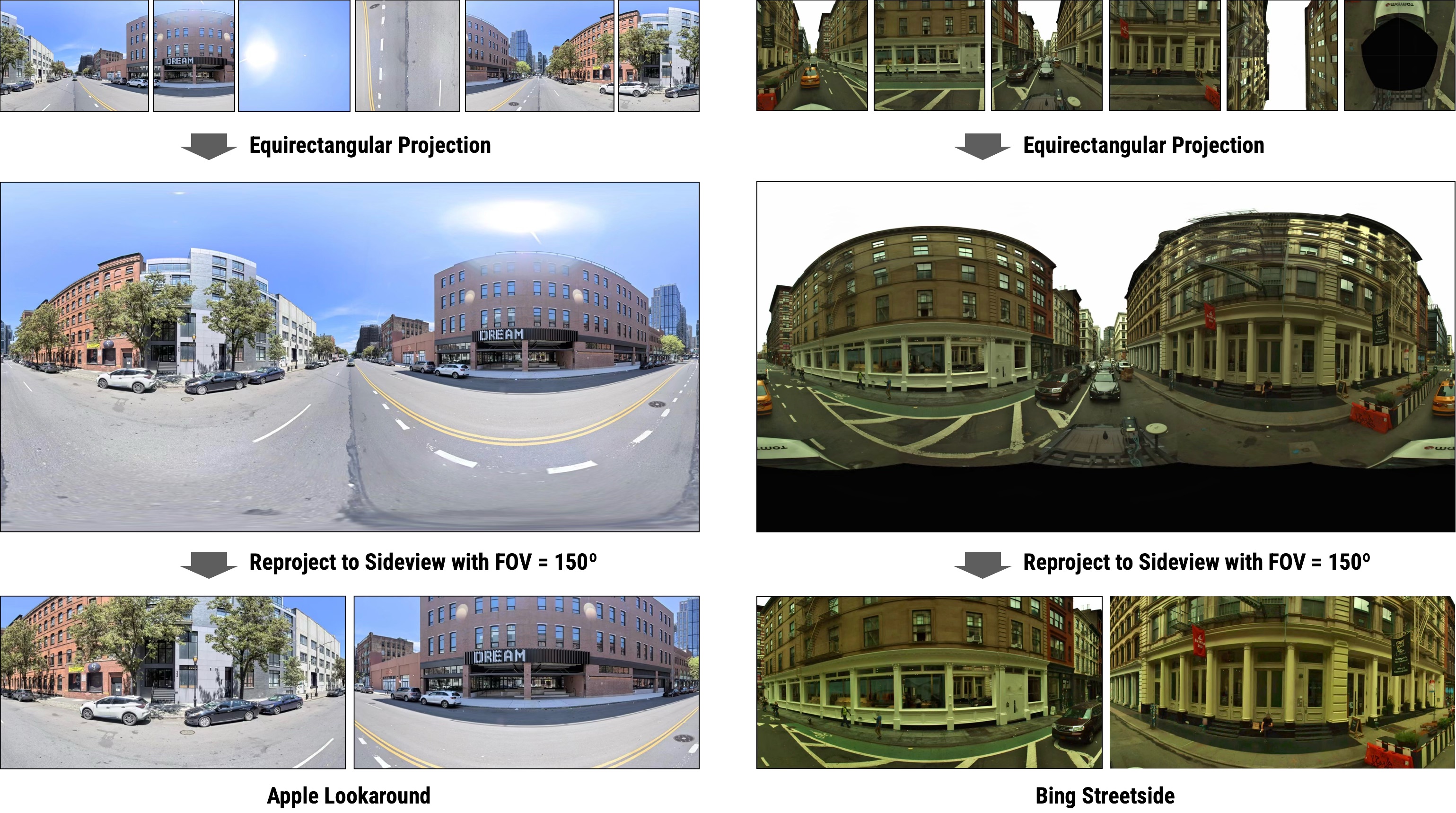}
    \caption{Reprojection pipeline illustrated for Apple Lookaround (left column) and, for comparison, Bing Streetside (right column). Raw tiles from each source are assembled into an equirectangular panorama and reprojected to a rectilinear sideview at $\text{FOV} = 150^\circ$. The bottom row shows the resulting left and right sideview images.}
    \label{fig:sideview-pipeline}
\end{figure}

We use a field of view (FOV) of 150$^\circ$. As shown in Fig.~\ref{fig:sideview-coverage}, this FOV is chosen so that adjacent capture points spaced at 100~m intervals produce continuous lateral coverage of the pedestrian zone without gaps. At a typical perpendicular distance of approximately 15~m from the road centerline to the building face, a $150^\circ$ FOV captures a lateral extent of $2 \times 15 \times \tan(75^\circ) \approx 112$~m, slightly exceeding the 100~m sampling interval. Wider angles were tested but introduced excessive peripheral distortion that degraded VLM coding accuracy. A vertical crop is then applied to remove sky and road regions, producing a 1600$\times$800 pixel image centered on the pedestrian zone. Fig.~\ref{fig:sideview-pipeline} illustrates the complete pipeline.

\section{Supplementary Figures: Revised Group Detection Framework}\label{app:pipeline-revisions}

The following figures supplement the revised social group detection methodology described in Section~\ref{ssec:social-group-detection} and could not be accommodated in the main text.

\begin{figure}[htbp]
    \centering
    \includegraphics[width=1\columnwidth]{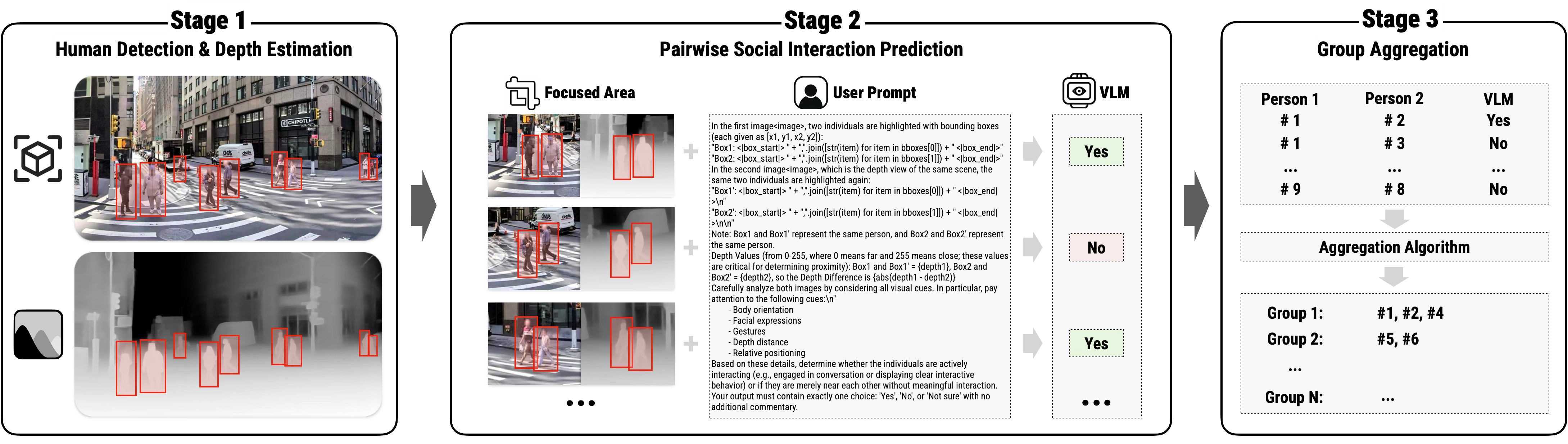}
    \caption{
        Illustration of the three-stage pipeline for social group detection \citep{liu2026mingle}. Stage~1 detects individual persons; Stage~2 evaluates pairwise social relationships between detected persons; Stage~3 integrates local pairwise judgments via graph-based clustering to determine the global group structure.
    }
    \label{fig:three-stage-pipeline}
\end{figure}

\begin{figure}[htbp]
    \centering
    \includegraphics[width=1\columnwidth]{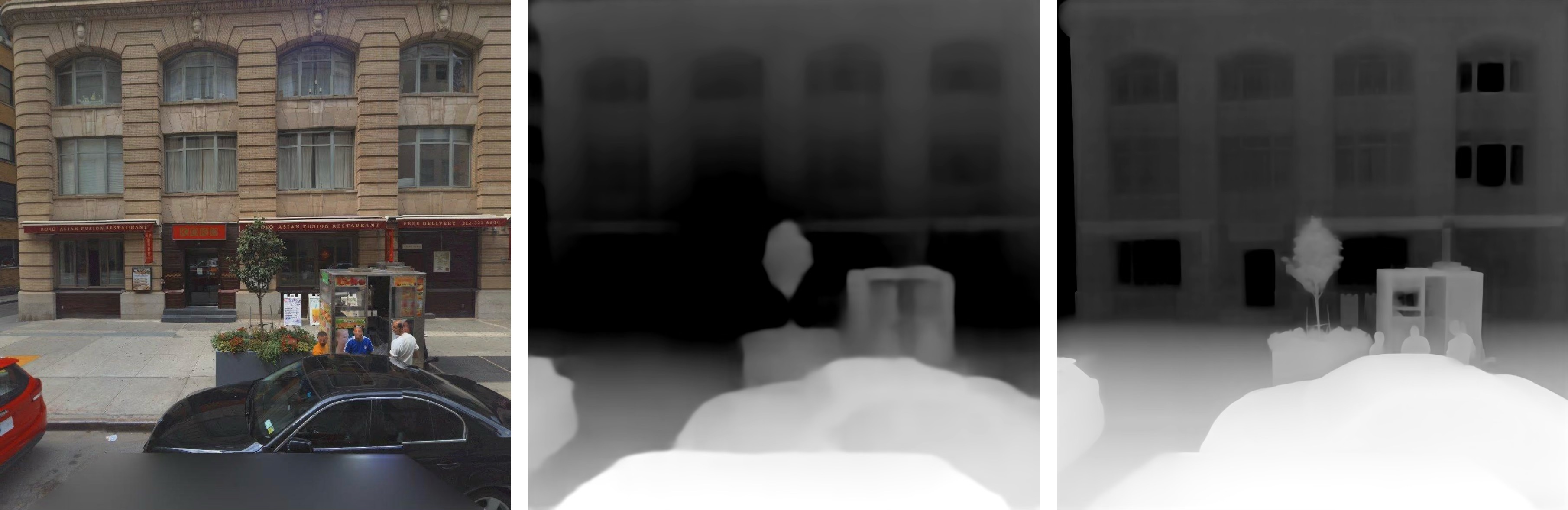}
    \caption{
        Comparison of depth estimation methods.
        Left: Original image;
        Middle: EVP method \citep{lavreniuk2024evp};
        Right: PatchFusion \citep{li2024patchfusion}.
        PatchFusion captures finer structural details and maintains superior continuity across the streetscape.
    }
    \label{fig:depth-comparison}
\end{figure}

\begin{figure}[htbp]
    \centering
    \includegraphics[width=1\columnwidth]{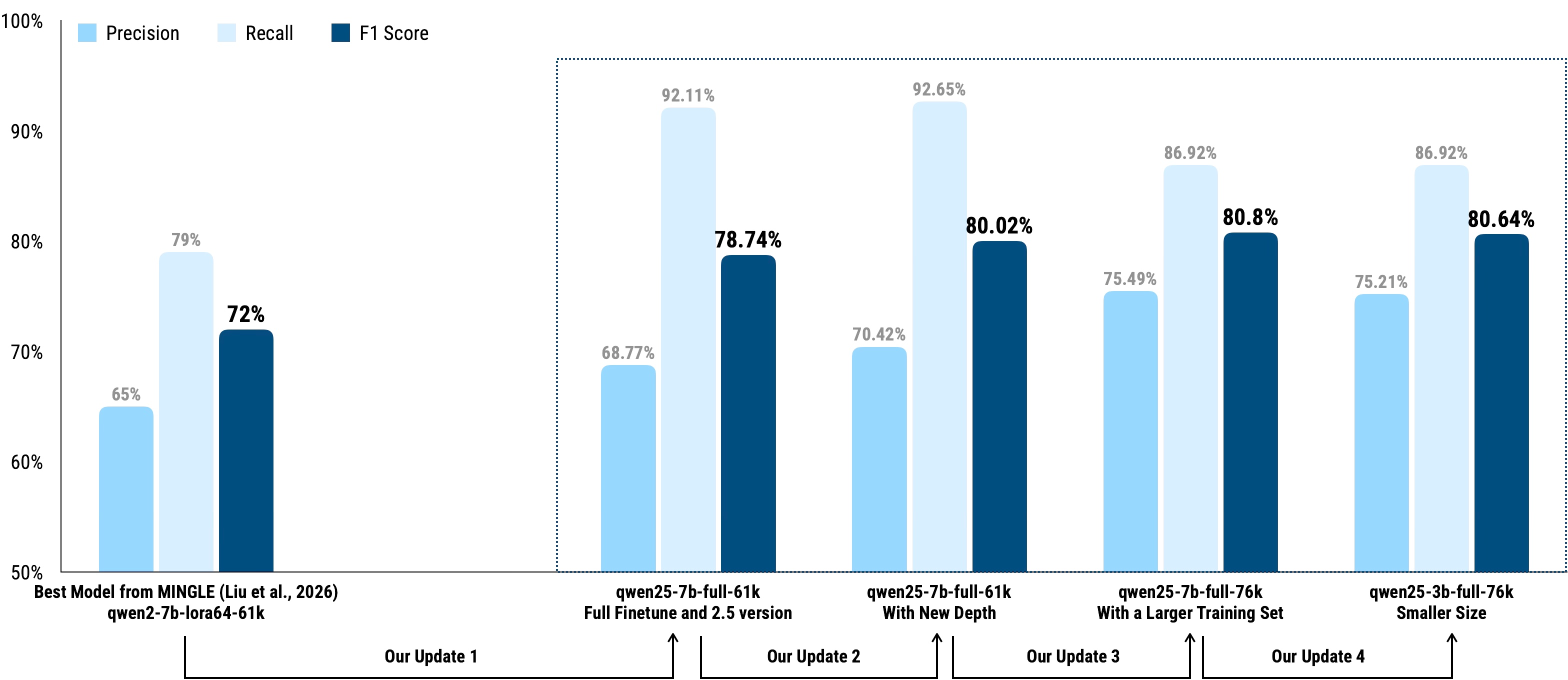}
    \caption{
        F1 scores, precision, and recall across four iterative training stages for the pairwise social group detection model. The MINGLE baseline \citep{liu2026mingle} (Qwen2-VL-7B, LoRA fine-tuning, 61k training images) achieved F1\,=\,72\%. Update~1 upgraded the base architecture to Qwen2.5-VL-7B, switched to full fine-tuning, and simplified the prompt, raising F1 to 78.74\%. Update~2 replaced EVP depth maps with PatchFusion, further improving F1 to 80.02\%. Update~3 expanded the training corpus from 61k to 76k images through additional manual annotation, reaching F1\,=\,80.8\% while reducing the precision--recall imbalance (precision: 70.4\%\,$\to$\,75.5\%; recall: 92.7\%\,$\to$\,86.9\%). Update~4 trained a 3B-parameter variant of the same architecture on the 76k corpus, achieving F1\,=\,80.64\% at substantially lower inference cost. The Update~4 model is used for all pairwise social interaction judgments in this paper.
    }
    \label{fig:updated-f1}
\end{figure}

\begin{figure}[htbp]
    \centering
    \begin{subfigure}[t]{0.48\linewidth}
        \centering
        \includegraphics[width=\linewidth]{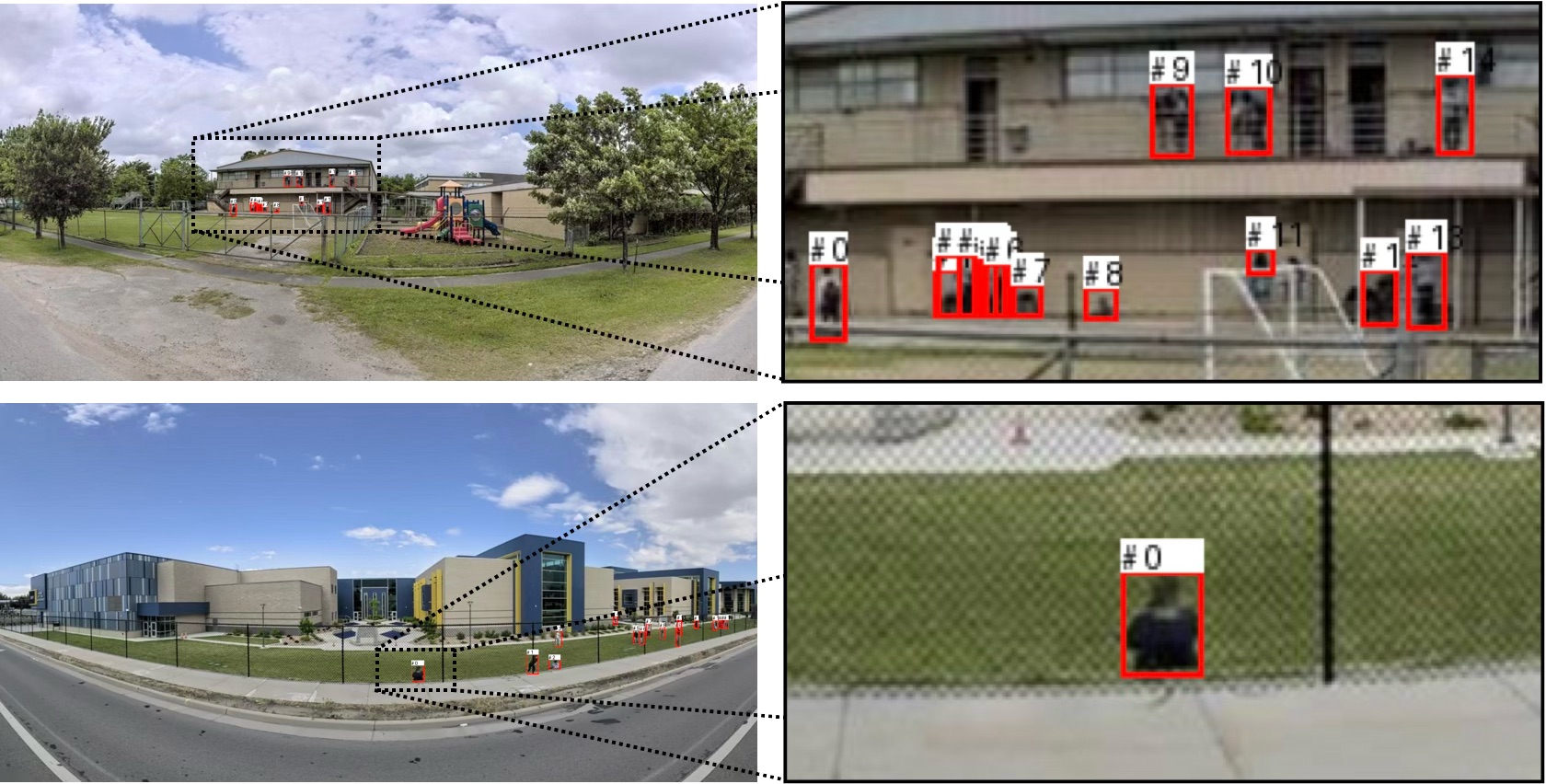}
        \caption{SAM3 recovers individuals that the ATSS-Swin-L detector
        missed, including small-scale figures at a distance and persons partially
        occluded by fencing or background clutter.}
    \end{subfigure}
    \hfill
    \begin{subfigure}[t]{0.48\linewidth}
        \centering
        \includegraphics[width=\linewidth]{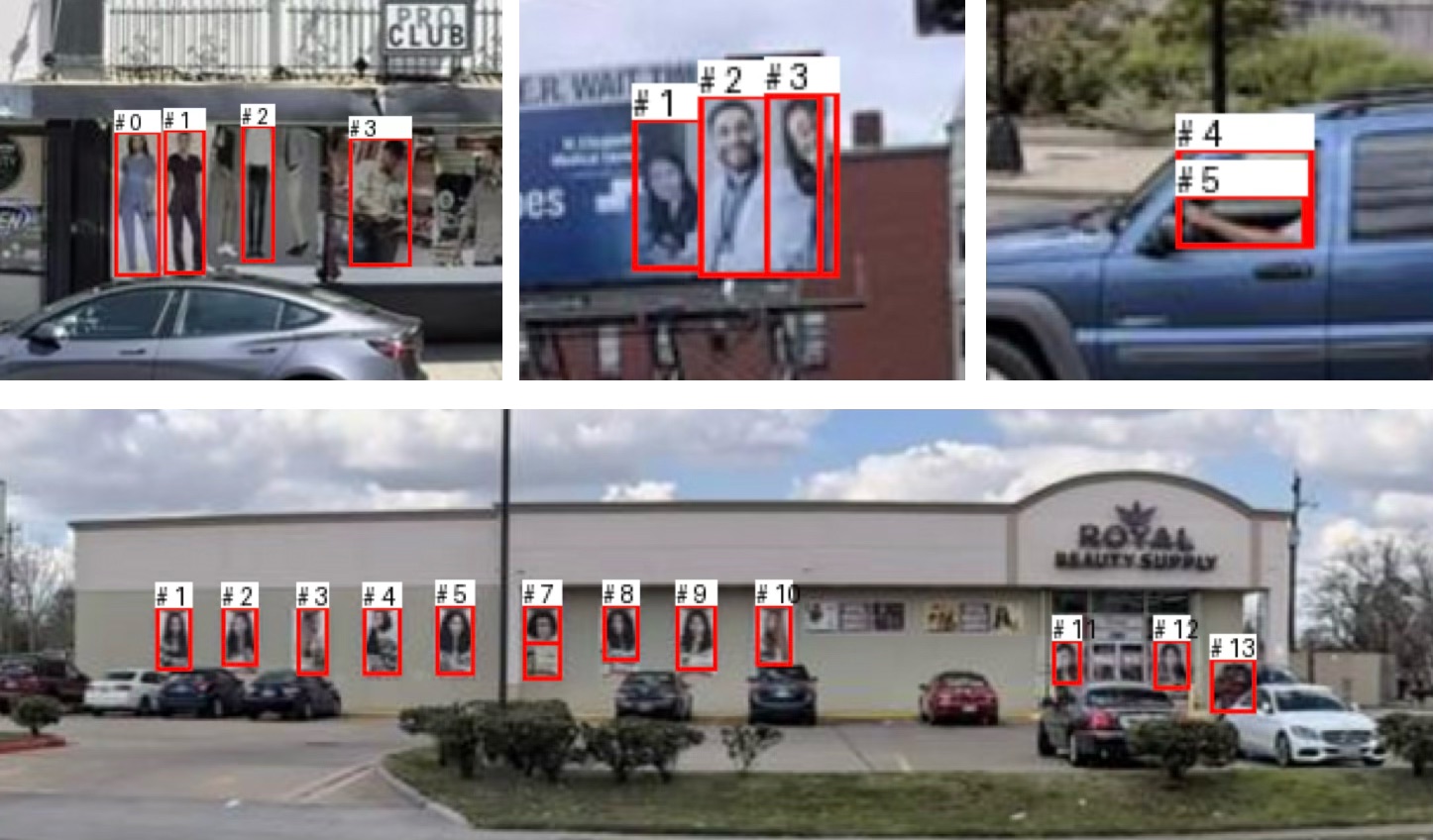}
        \caption{False positives excluded by VLM-based validity filtering. Figures on advertisement displays and vehicle occupants are classified as \texttt{advertisement\_or\_image} and \texttt{inside\_vehicle} respectively, and excluded before the pairwise social interaction stage.}
    \end{subfigure}
    \caption{Refined person detection: higher recall through SAM3 (left) and
    false positive removal through activity detection (Section~\ref{ssec:activity-detection}) (right).}
    \label{fig:refined-person-detection}
\end{figure}

\section{Adaptive Individual Cropping}\label{sec:app-cropping}

\begin{figure}[htbp]
    \centering
    \includegraphics[width=1\columnwidth]{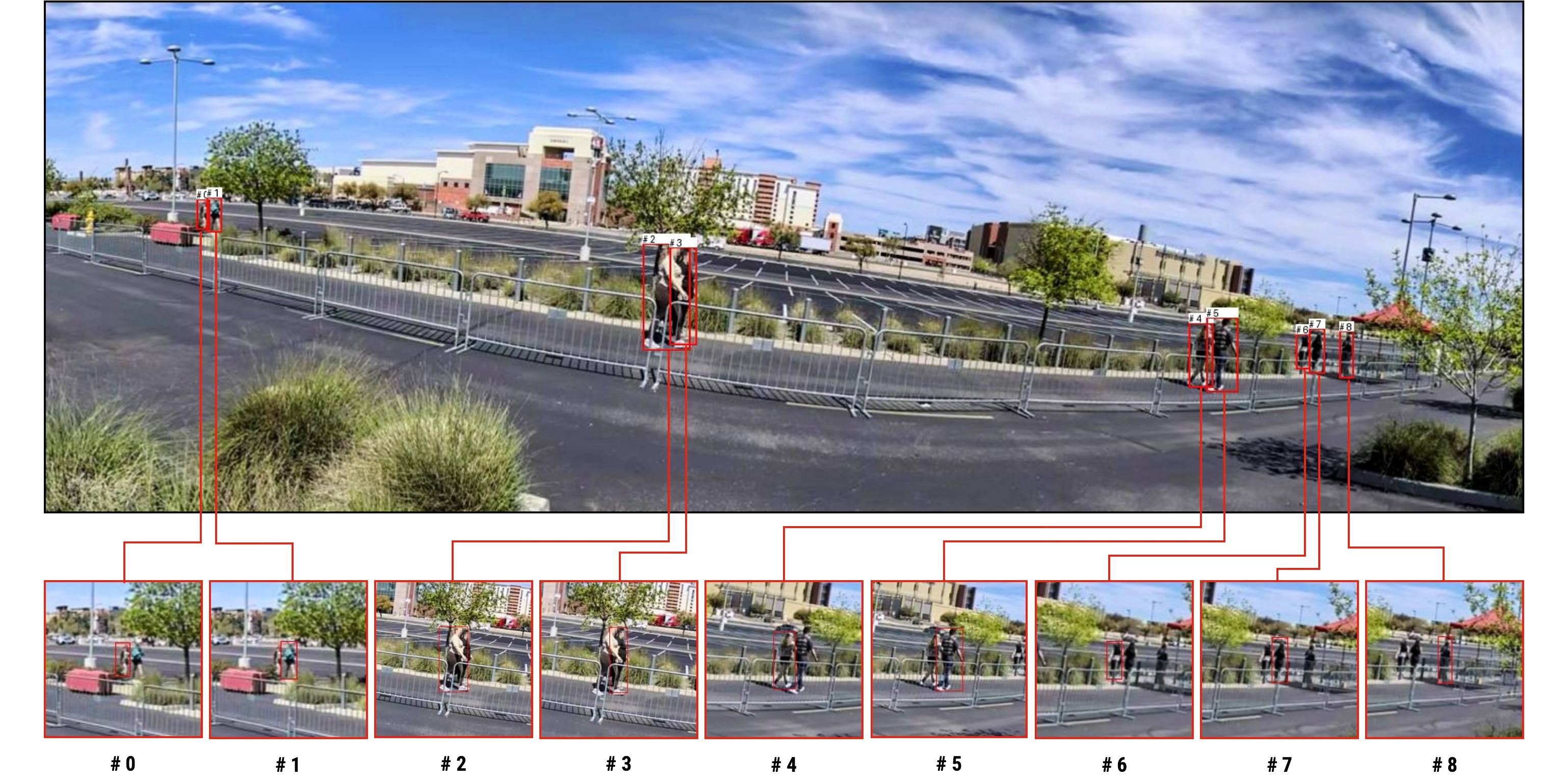}
    \caption{
        Visual demonstration of the dynamic focusing mechanism. Top: The original wide-angle panoramic source image containing multiple detected entities (from \#0 person to \#8 person). Bottom: A sequence of standardized, square focal patches generated via the \texttt{focus} operation. Note how the adaptive padding $P$ maintains contextual information around each target while the square constraint and boundary-aware shifting ensure consistent 1:1 aspect ratios even for objects near the image periphery.
    }
    \label{fig:focus-crop}
\end{figure}

To enhance the visual context of specific objects of interest, we implement a dynamic focusing mechanism that adaptively crops the image based on a subset of bounding boxes $\mathcal{I} \subseteq \{1, \dots, n\}$. For a given set of boxes $\{[x_{i,1}, y_{i,1}, x_{i,2}, y_{i,2}]\}_{i \in \mathcal{I}}$, we first define the tightest axis-aligned bounding box coordinates as $X_{\min} = \min_{i \in \mathcal{I}}(x_{i,1})$, $Y_{\min} = \min_{i \in \mathcal{I}}(y_{i,1})$, $X_{\max} = \max_{i \in \mathcal{I}}(x_{i,2})$, and $Y_{\max} = \max_{i \in \mathcal{I}}(y_{i,2})$. To incorporate sufficient background context, an adaptive padding $P$ is computed such that $P = \max(\max(W_t, H_t) \cdot \rho, P_{\min})$, where $W_t = X_{\max} - X_{\min}$ and $H_t = Y_{\max} - Y_{\min}$ represent the dimensions of the tightest box, $\rho$ denotes the padding percentage (e.g., $0.7$), and $P_{\min}$ is the minimum pixel threshold. The initial crop boundaries $[x_1, y_1, x_2, y_2]$ are derived by extending the tightest box by $P$ and clamping the results within the image dimensions $[0, W_{img}]$ and $[0, H_{img}]$. When the square constraint is enabled, the region is re-centered around $(c_x, c_y) = (\frac{x_1+x_2}{2}, \frac{y_1+y_2}{2})$ with a uniform side length $S = \max(x_2-x_1, y_2-y_1)$, followed by a boundary-aware shifting logic to ensure the square remains within the image manifold without truncation. Finally, the internal object coordinates are translated to the new local coordinate system via the spatial transformation $x' = x - x_1$ and $y' = y - y_1$, yielding a normalized focal representation of the target entities. The empirical results of this adaptive focusing and cropping procedure are illustrated in Fig.~\ref{fig:focus-crop}.

\section{Prompt Design for Individual Activity Detection}\label{sec:app-act-prompt}

The decomposed prompt described in Section~\ref{ssec:activity-detection} improves reliability in part by surfacing behavioral detail that the original posture-action vocabulary suppressed. Seated individuals, for example, are now recorded with both \textit{posture: seated} and \textit{support\_surface}, which distinguishes sitting on a bench from sitting on steps, a ledge, or the ground---a distinction relevant to understanding the affordances a space provides, as documented by Whyte~\citeyearpar{whyte1980social} and Gehl~\citeyearpar{Gehl1987LifeSpace}. The \textit{validity} dimension serves an additional filtering function: each detected crop is classified as \texttt{valid\_person}, \texttt{advertisement\_or\_image}, \texttt{inside\_vehicle}, \texttt{inside\_building}, or \texttt{unclear}, removing a class of false positives that person detectors routinely produce without requiring a separate filtering pass.

\ref{box:posture-action-prompt} shows the original posture-action prompt that exhibited the failure modes described in Section~\ref{ssec:activity-detection}. \ref{box:decomposed-prompt} shows the redesigned decomposed prompt adopted in this paper.

\refstepcounter{promptbox}\label{box:posture-action-prompt}
\begin{tcolorbox}[
    enhanced,
    breakable,
    sharp corners,
    boxrule=0.5pt,
    colback=gray!5,
    colframe=gray!80,
    fontupper=\footnotesize\ttfamily,
    title=Box~\arabic{promptbox}: Posture-Action Prompt for Individual Activity Detection,
    colbacktitle=gray!20,
    coltitle=black,
    attach boxed title to top left={yshift=-2mm, xshift=2mm},
    boxed title style={sharp corners, boxrule=0.5pt},
    pad at break=2mm,
]
\begin{Verbatim}[breaklines=true, breakanywhere=true, fontsize=\scriptsize, baselinestretch=0.8]
Analyze the person **within the red bounding box** in the image
and output the result ONLY as a JSON object.

### Task 1: Reality Check
First, determine if the subject is a real person.

If the subject is NOT a real person (e.g., a picture, statue,
figure on a billboard, etc.):
Return the JSON object below:
{
    "is_real_person": false
}

If the subject IS a real person, proceed to Task 2.

### Task 2: Posture and Actions Analysis
Before making any final choices, **think step by step** and explain
your reasoning based on what you observe, including body orientation,
hand position, gaze direction, and any interactions with surrounding
objects.

1. Reasoning: Describe your thought process and key visual evidence
   that supports your later decisions.

2. Description: Provide a brief, objective description of the person
   (e.g., 'man in suit carrying bag').

3. Posture (Single Choice):
Choose the single best option from the list below:
standing, sitting, lying down, bending, kneeling, squatting

4. Actions (One or More Choices):
Choose ALL relevant actions from the list below. You MUST select
more than one if multiple actions are clearly visible or implied.
Pay special attention to interactions with surrounding objects
(e.g., animals, carts, phones, etc.) to infer actions like
'petting a dog' or 'carrying an object'. If only one action is
visible, select just one.

waiting for public transport; walking on the sidewalk;
crossing the street; walking a pet; using a crutch;
jogging or running; smoking; bicycling; skateboarding;
riding a scooter; vending; interacting with a pet;
pushing a cart; pushing a stroller; pushing a wheelchair;
talking with other(s); talking on a phone;
taking a photo or selfie; reading a book or newspaper;
reading a phone or tablet; window shopping; eating; drinking;
holding a billboard; performing as a street artist;
cleaning and maintaining shop front or sidewalk;
kissing or other intimate actions; handing out flyers; dancing;
sunbathing; begging or panhandling; greeting others on the street;
observing other people and activities; protesting; sleeping;
playing; performing mime, clowns, or juggle; painting;
sports activities; carrying an object;
playing a musical instrument; something else

**Final Output Format (Real Person)**:
Return ***ONLY*** the JSON object below. ***Do not include any
surrounding text, markdown formatting (like ```json), or special
characters***:
{
    "is_real_person": true,
    "description": "<brief_objective_description>",
    "posture": "<chosen_posture_name>",
    "actions": [
        "<chosen_action_1>",
        "<chosen_action_2>",
        ...
    ],
    "reasoning": "<explanation_for_posture_and_actions>"
}
\end{Verbatim}
\end{tcolorbox}

Below is the restructured prompt, which decomposes the Posture-Action Prompt into 10 dimensions. We adopted this prompt for this research.

\refstepcounter{promptbox}\label{box:decomposed-prompt}
\begin{tcolorbox}[
    enhanced,
    breakable,
    sharp corners,
    boxrule=0.5pt,
    colback=gray!5,
    colframe=gray!80,
    fontupper=\footnotesize\ttfamily,
    title=Box~\arabic{promptbox}: Decomposed Prompt for Individual Activity Detection,
    colbacktitle=gray!20,
    coltitle=black,
    attach boxed title to top left={yshift=-2mm, xshift=2mm},
    boxed title style={sharp corners, boxrule=0.5pt},
    pad at break=2mm,
]
\begin{Verbatim}[breaklines=true, breakanywhere=true, fontsize=\scriptsize, baselinestretch=0.8]
You must follow ALL rules below exactly.

1. Analyze ONLY the person inside the red bounding box (the "target person").
2. Use only visible and directly observable information.
3. Do NOT infer high-level intent or social meaning.
4. An object may extend outside the box ONLY if it is clearly connected to the target person (e.g., directly held, worn, attached, or continuous physical contact).
5. Do NOT attribute nearby objects without clear physical connection.
6. If object attribution is ambiguous, set:
   - "object": "none" or "unclear"
   - "object_relation": "none" or "unclear"
7. If "object" is "none", then "object_relation" MUST be "none".
8. For any field, if the required information is not clearly visible, select "unclear".

Return ONLY a JSON object with the following fields.

All values MUST be chosen exactly from the allowed options.
Do NOT invent new labels.
Do NOT include explanations.

Fields and allowed values:

validity:
- valid_person
- advertisement_or_image
- inside_vehicle
- inside_building
- unclear

posture:
- upright
- seated
- lying
- squatting
- unclear

locomotion:
- stationary
- moving
- running
- unclear

object:
- none
- animal
- phone
- tablet
- book_or_newspaper
- smoking_item
- sports_equipment
- bag
- trash_bag
- box
- tool
- stroller
- cart
- wheelchair
- walker
- bicycle
- scooter
- skateboard
- musical_instrument
- sign_or_billboard
- food_or_drink
- flyer
- luggage
- unclear

object_relation:
- none
- holding
- carrying
- pushing
- riding
- unclear

support_surface:
- ground
- bench_or_chair
- steps
- elevated_surface
- unclear

zone_type:
- vendor_stall
- roadway
- crosswalk
- sidewalk
- park
- building_entrance
- bus_stop
- unclear

uniform_presence:
- none
- police
- maintenance_worker
- unclear

perceived_age_group:
- child
- adult
- older_adult
- unclear

perceived_sex:
- male
- female
- unclear

Return ONLY:
{
  "validity": "",
  "posture": "",
  "locomotion": "",
  "object": "",
  "object_relation": "",
  "support_surface": "",
  "zone_type": "",
  "uniform_presence": "",
  "perceived_age_group": "",
  "perceived_sex": "",
}
\end{Verbatim}
\end{tcolorbox}

\section{Model Selection and Validation} \label{sec:app-activity-model}

In this study, we primarily select two mainstream open-source Vision-Language Model (VLM) families for evaluation: the \textbf{InternVL3.5} series and the \textbf{Qwen3-VL} series. Specifically, the evaluated models include \textit{InternVL3\_5-30B-A3B-Instruct}, \textit{InternVL3\_5-GPT-OSS-20B-A4B-Preview}, \textit{Qwen3-VL-30B-A3B-Instruct}, \textit{Qwen3-VL-32B-Instruct}, \textit{Qwen3-VL-4B-Instruct}, and \textit{Qwen3-VL-8B-Instruct}.

It is worth noting that we excluded several models from the final comparison based on preliminary testing. For instance, smaller models such as the InternVL3.5-Instruct series (14B and below) and the Qwen3-VL-Instruct series (2B and below) exhibited significant deficiencies in instruction-following under our specific prompts. These models frequently failed to generate parsable JSON outputs or to produce answers within the predefined options. Furthermore, larger-scale models such as \textit{InternVL3.5-38B-Instruct}, \textit{InternVL3.5-241B-A28B-Instruct}, and \textit{Qwen3-VL-235B-A22B-Instruct} were also omitted due to the excessive computational overhead and memory pressure associated with their deployment.

To identify the most suitable model for this decomposed framework, we conducted a comparative analysis of prediction performance across a suite of candidate models. These performance metrics were validated against a manually annotated dataset comprising 999 individual pedestrian samples. Based on the evaluation results, \textbf{Qwen3-VL-30B-A3B-Instruct} was selected as the primary engine for activity detection and annotation in this study, as it demonstrated the most robust instruction-following capabilities and classification accuracy. The detailed performance metrics across various dimensions are summarized in Table~\ref{tab:model-performance}.

\begin{table}[htbp]
    \centering
    \caption{Performance Evaluation of Selected VLM Models Across Different Dimensions}
    \label{tab:model-performance}
    \resizebox{\textwidth}{!}{%
        \begin{tabular}{lcccccccccc|c}
            \toprule
            \textbf{Model} & \textbf{Val.} & \textbf{Pos.} & \textbf{Loc.} & \textbf{Obj.} & \textbf{Rel.} & \textbf{Surf.} & \textbf{Zone} & \textbf{Uni.} & \textbf{Age} & \textbf{Sex} & \textbf{Mean} \\
            \midrule
            InternVL3\_5-30B-A3B & 91.3\% & 64.7\% & 56.3\% & 74.2\% & 75.7\% & 77.0\% & 77.0\% & 95.2\% & 85.0\% & 60.7\% & 75.7\% \\
            InternVL3\_5-GPT-OSS-20B & 95.0\% & 81.4\% & 69.9\% & 73.8\% & 70.7\% & 80.0\% & 78.8\% & 96.2\% & 76.9\% & 62.0\% & 78.4\% \\
            Qwen3-VL-30B-A3B & 98.2\% & 95.6\% & 82.6\% & 77.1\% & 77.2\% & 95.5\% & 89.8\% & 99.0\% & 93.9\% & 76.0\% & 88.5\% \\
            Qwen3-VL-32B-Instruct & 98.2\% & 95.4\% & 82.4\% & 71.1\% & 70.7\% & 74.5\% & 86.0\% & 98.8\% & 94.6\% & 72.4\% & 84.4\% \\
            Qwen3-VL-4B-Instruct & 97.9\% & 92.8\% & 80.9\% & 80.4\% & 79.8\% & 50.7\% & 84.2\% & 67.7\% & 73.8\% & 72.2\% & 78.0\% \\
            Qwen3-VL-8B-Instruct & 98.4\% & 91.4\% & 82.5\% & 81.1\% & 81.8\% & 94.4\% & 84.3\% & 89.3\% & 92.5\% & 71.5\% & 86.7\% \\
            \bottomrule
        \end{tabular}%
    }
    \vspace{1mm}
    \begin{flushleft}
        \footnotesize \textit{Note: Abbreviations represent Validity, Posture, Locomotion, Object, Object\_relation, Support\_surface, Zone\_type, Uniform\_presence, Perceived\_age\_group, and Perceived\_sex respectively.}
        \end{flushleft}
\end{table}


\section{Shannon Entropy over Activity Labels}\label{sec:app-entropy}

For each H3 hexagonal cell $c$, all $N_c$ detected persons are pooled into an 18-dimensional mass vector $\mathbf{m}_c$ partitioned into two independent components. The first 13 entries record PAF mass: each person contributes one unit to the bin corresponding to their assigned PAF label, or to a pooled \texttt{other} bin if no named category applies:

\begin{equation}
    m^{\mathrm{PAF}}_{c,k} = \bigl|\{i \in c : \mathrm{PAF}_i = k\}\bigr|, \quad
    k \in \{\texttt{maintenance},\, \texttt{cycling},\, \ldots,\, \texttt{passing},\, \texttt{other}\},
    \label{eq:paf-mass}
\end{equation}

with $\sum_{k=1}^{13} m^{\mathrm{PAF}}_{c,k} = N_c$. The remaining 5 entries record SCF mass; each person contributes independently to each flag:

\begin{equation}
    m^{\mathrm{SCF}}_{c,j} = \sum_{i \in c} \mathbf{1}[\mathrm{SCF}_{i,j} = 1], \quad
    j \in \{\texttt{seated-steps},\, \texttt{building-entrance},\, \texttt{crosswalk},\, \texttt{waiting},\, \texttt{vending}\}.
    \label{eq:scf-mass}
\end{equation}

Because SCF flags are not mutually exclusive with PAF labels, a single person may contribute to both a PAF bin and one or more SCF bins, so the total mass $M_c = \sum_k m^{\mathrm{PAF}}_{c,k} + \sum_j m^{\mathrm{SCF}}_{c,j} \geq N_c$. The normalized probability vector is $p_{c,\ell} = m_{c,\ell} / M_c$ for $\ell = 1, \ldots, 18$. Shannon entropy in bits is then:

\begin{equation}
    H_c = -\sum_{\ell=1}^{18} p_{c,\ell} \log_2 p_{c,\ell},
    \label{eq:entropy}
\end{equation}

with terms where $p_{c,\ell} = 0$ taken as zero by convention. The maximum possible entropy under this scheme is $\log_2 18 \approx 4.17$ bits; a cell dominated by a single activity type approaches $H_c = 0$.

\section{Sensitivity of the Social Dwelling Index to Alternative Weighting Schemes}\label{sec:app-sensitivity}

The grouping weights $\alpha$ and dwelling-intensity weights $\beta$ that define the SDI are theory-informed ordinal values rather than empirically estimated structural parameters. The grouping tiers 1:2:3 correspond directly to representative group sizes (Section~\ref{ssec:sdi}); the dwelling-intensity tiers 3:2:1 encode an ordinal distinction between stronger and weaker forms of observed dwelling, broadly consistent with Gehl's \citeyearpar{Gehl1987LifeSpace} emphasis on lingering versus pass-through activity, with seated and stationary states assigned higher weights than active locomotion. Because these values are fixed by theoretical convention rather than calibration, their precise magnitudes carry inherent uncertainty. This appendix examines whether the main empirical finding (the weak association between SDI and pedestrian volume documented in Section~\ref{ssec:nyc-decoupling}) is stable under plausible alternative weighting schemes.

The baseline specification assigns grouping weights $(\alpha = 1, 2, 3)$ to the solo, dyad, and group-of-3+ tiers and dwelling-intensity weights $(\beta = 3, 2, 1)$ to the seated, stationary, and active-locomotion tiers, respectively. Three alternative specifications are evaluated. The \textit{flattened} specification compresses both gradients uniformly: grouping $(1, 1.5, 2)$ and dwelling-intensity $(2, 1.5, 1)$. The \textit{steep} specification amplifies both: grouping $(1, 2, 4)$ and dwelling-intensity $(4, 2, 1)$. The \textit{dwelling-adjusted} specification retains the baseline grouping weights but applies a compressed dwelling-intensity gradient $(2, 1.5, 1)$, shifting relative emphasis from dwelling-intensity toward grouping.

\begin{figure}[htbp]
\centering
\begin{minipage}{0.90\textwidth}
    \begin{subfigure}[t]{0.32\textwidth}
        \includegraphics[width=\linewidth]{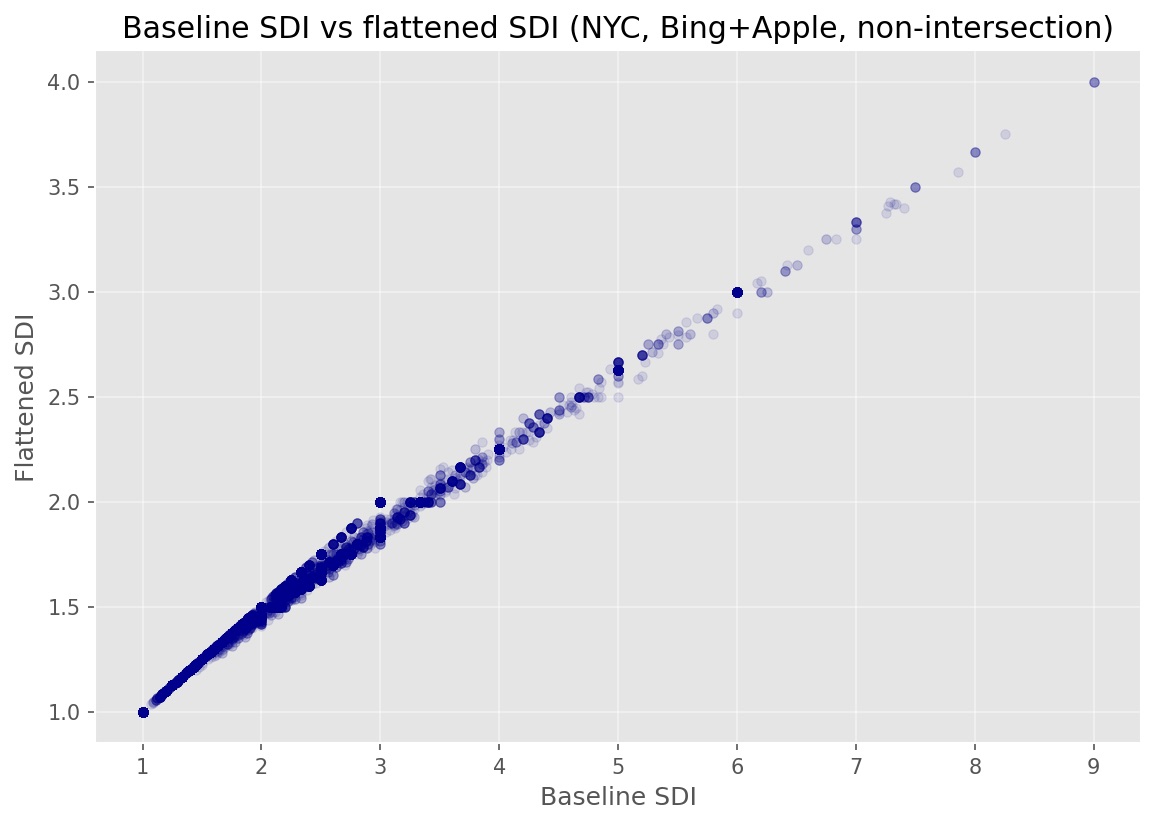}
        \caption{Baseline vs.\ flattened\\($\rho = 0.998$).}
        \label{fig:sdi-robustness-flattened}
    \end{subfigure}
    \hfill
    \begin{subfigure}[t]{0.32\textwidth}
        \includegraphics[width=\linewidth]{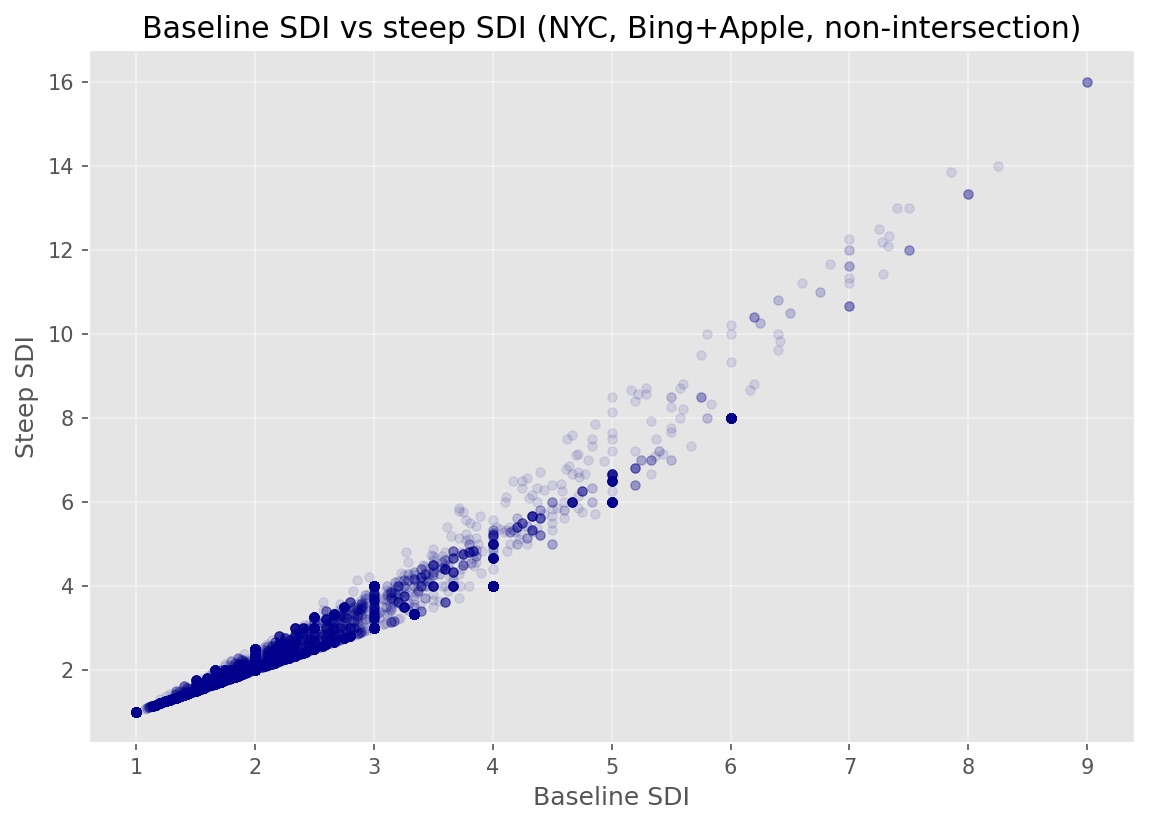}
        \caption{Baseline vs.\ steep\\($\rho = 0.992$).}
        \label{fig:sdi-robustness-steep}
    \end{subfigure}
    \hfill
    \begin{subfigure}[t]{0.32\textwidth}
        \includegraphics[width=\linewidth]{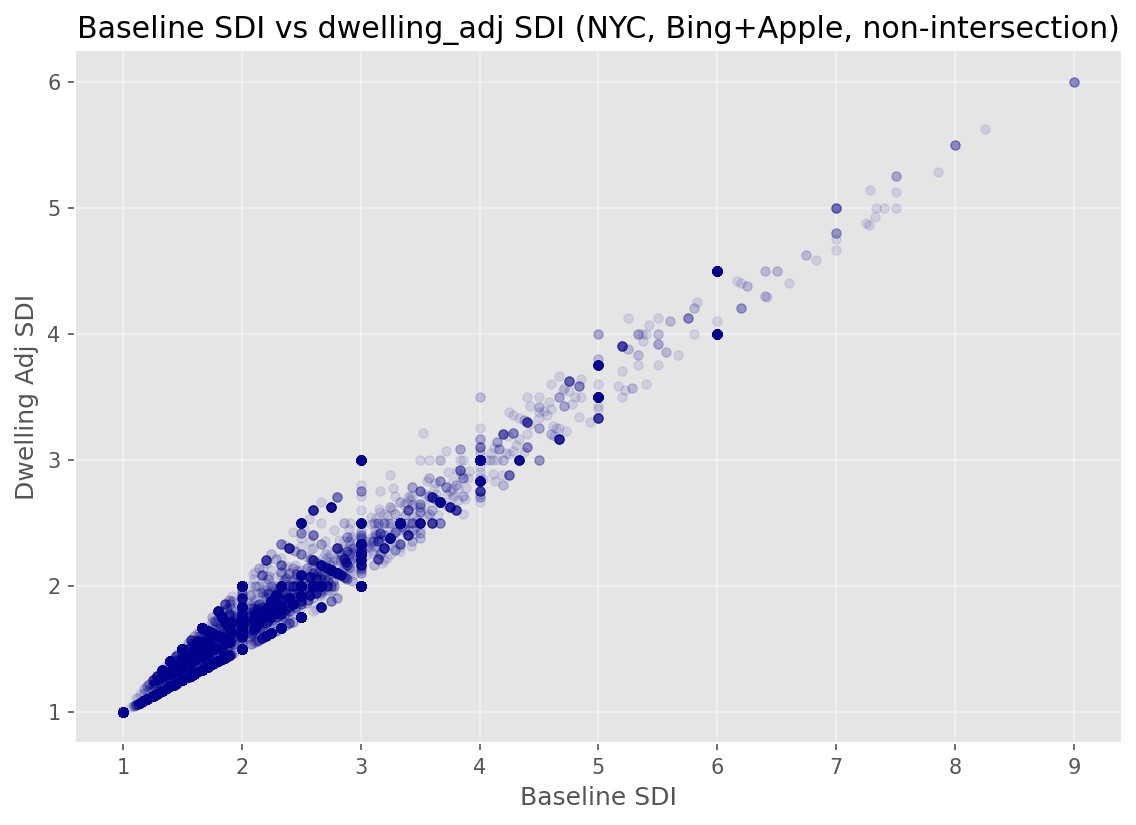}
        \caption{Baseline vs.\ dwelling-adjusted ($\rho = 0.949$).}
        \label{fig:sdi-robustness-dwelling}
    \end{subfigure}
    \caption{Image-level SDI under the baseline specification versus three alternative weighting schemes (NYC Apple Lookaround, non-intersection images, $n = 102{,}514$ sideviews). Each point represents one sideview. Greater dispersion in panel (c) reflects the structural shift in the relative contribution of WGI and WDI under the dwelling-adjusted specification.}
    \label{fig:sdi-robustness}
\end{minipage}
\end{figure}

All analyses use the same sample as the main empirical section: $n = 102{,}514$ sideviews from Apple Lookaround in NYC (non-intersection images only). Table~\ref{tab:sdi-robustness-summary} reports pairwise Pearson correlations among SDI, WGI, WDI, and pedestrian count under each specification. Table~\ref{tab:sdi-robustness-rank-stability} reports the Spearman rank correlation between baseline SDI and each alternative at the image level. Figure~\ref{fig:sdi-robustness} plots baseline SDI against each alternative.

Rank stability relative to the baseline is high under all three alternatives: $\rho = 0.998$, $0.992$, and $0.949$ for the flattened, steep, and dwelling-adjusted specifications, respectively (Table~\ref{tab:sdi-robustness-rank-stability}; Fig.~\ref{fig:sdi-robustness}). The dwelling-adjusted specification shows the greatest departure, consistent with its structural shift in the relative contribution of the WGI and WDI components. Across all specifications, the association between SDI and pedestrian count remains weak: $r = 0.163$, $0.165$, and $0.212$ for the flattened, steep, and dwelling-adjusted alternatives, compared with $r = 0.168$ at baseline (Table~\ref{tab:sdi-robustness-summary}). The correlation structure between SDI and its components shifts notably under the dwelling-adjusted specification: SDI correlates more strongly with WGI ($r = 0.822$ versus $0.658$ at baseline) and less strongly with WDI ($r = 0.521$ versus $0.696$), reflecting the reduced weight assigned to dwelling intensity in that specification. These results indicate that the substantive conclusion that SDI captures a social dimension of sidewalk use not reducible to pedestrian volume is robust to plausible alternative weight specifications.

\begin{table}[htbp]
\renewcommand{\arraystretch}{1.2}
\setlength{\tabcolsep}{8pt}
\footnotesize
\caption{Pairwise Pearson correlations among pedestrian count, SDI, WGI, and WDI under the baseline and three alternative weighting specifications. Sample: NYC Apple Lookaround, non-intersection images, $n = 102{,}514$ sideviews.}
\label{tab:sdi-robustness-summary}
\centering
\begin{tabular}{lcccc}
\toprule
\textbf{Specification} & \textbf{Count--SDI} & \textbf{SDI--WGI} & \textbf{SDI--WDI} & \textbf{WGI--WDI} \\
\midrule
Baseline          & 0.168 & 0.658 & 0.696 & $-$0.017 \\
Flattened         & 0.163 & 0.634 & 0.747 & $-$0.017 \\
Steep             & 0.165 & 0.669 & 0.688 &    0.016 \\
Dwelling-adjusted & 0.212 & 0.822 & 0.521 & $-$0.017 \\
\bottomrule
\end{tabular}
\end{table}

\begin{table}[htbp]
\renewcommand{\arraystretch}{1.2}
\setlength{\tabcolsep}{10pt}
\footnotesize
\caption{Spearman rank correlation between baseline SDI and each alternative weighting specification at the image level. Higher values indicate closer agreement in the ordering of sideviews by social dwelling intensity. Sample: NYC Apple Lookaround, non-intersection images.}
\label{tab:sdi-robustness-rank-stability}
\centering
\begin{tabular}{lcc}
\toprule
\textbf{Specification} & \textbf{$n$} & \textbf{Spearman $\rho$} \\
\midrule
Flattened         & 102{,}514 & 0.998 \\
Steep             & 102{,}514 & 0.992 \\
Dwelling-adjusted & 102{,}514 & 0.949 \\
\bottomrule
\end{tabular}
\end{table}

\end{document}